\documentclass[twocolumn, trackchanges]{aastex701}

\usepackage{url}
\usepackage{amsmath}
\usepackage{gensymb}
\usepackage{amssymb}
\usepackage{natbib}
\usepackage{multirow}
\usepackage{subcaption}
\usepackage{graphicx}
\usepackage{xcolor}
\usepackage{lineno}

\received{}
\revised{}
\accepted{}

\submitjournal{}

\shorttitle{COMPASS First 7}
\shortauthors{Gordon}

\begin{document}


\title{JWST COMPASS: Insights into the Systematic Noise Properties of NIRSpec/G395H From a Uniform Reanalysis of Seven Transmission Spectra}


\correspondingauthor{Tyler A. Gordon}
\email{tygordon@ucsc.edu}

\author[0000-0001-5253-1987]{Tyler A. Gordon}
\affiliation{Department of Astronomy and Astrophysics, University of California, Santa Cruz, CA, USA}
\email{tygordon@ucsc.edu}


\author[0000-0002-7030-9519]{Natalie M. Batalha}
\affiliation{Department of Astronomy and Astrophysics, University of California, Santa Cruz, CA, USA}
\email{}

\author[0000-0003-1240-6844]{Natasha E. Batalha}
\affiliation{NASA Ames Research Center, Moffett Field, CA 94035, USA}
\email{}


\author[0000-0002-8949-5956]{Artyom Aguichine}
\affiliation{Department of Astronomy and Astrophysics, University of California, Santa Cruz, CA, USA}
\email{}

\author[0009-0003-2576-9422]{Anna Gagnebin}
\affiliation{Department of Astronomy and Astrophysics, University of California, Santa Cruz, CA, USA}
\email{}

\author[0000-0002-4207-6615]{James Kirk}
\affiliation{Imperial College London}
\email{}

\author[0000-0003-3204-8183]{Mercedes López-Morales}
\affiliation{Space Telescope Science Institute, 3700 San Martin Drive, Baltimore MD 21218}
\email{}

\author[0000-0002-7500-7173]{Annabella Meech}
\affiliation{Center for Astrophysics, Harvard \& Smithsonian, 60 Garden St, Cambridge, MA 02138, USA}
\email{}

\author[0000-0003-3623-7280]{Nicholas Scarsdale}
\affiliation{Department of Astronomy and Astrophysics, University of California, Santa Cruz, CA 95064, USA}
\email{}

\author[0009-0008-2801-5040]{Johanna Teske}
\affiliation{Earth and Planets Laboratory, Carnegie Institution for Science, 5241 Broad Branch Road, NW, Washington, DC 20015, USA} \affiliation{The Observatories of the Carnegie Institution for Science, 813 Santa Barbara St., Pasadena, CA 91101, USA}
\email{}

\author[0000-0003-0354-0187]{Nicole L. Wallack}
\affiliation{Earth and Planets Laboratory, Carnegie Institution for Science, 5241 Broad Branch Road, NW, Washington, DC 20015, USA}
\email{}

\author[0000-0002-0413-3308]{Nicholas Wogan}
\affiliation{SETI Institute, Mountain View, CA 94043, USA}
\affiliation{NASA Ames Research Center, Moffett Field, CA 94035}
\email{}

\begin{abstract}

JWST has already observed near-infrared transmission spectra of over a dozen super-Earths and sub-Neptunes. While some observations have allowed astronomers to characterize sub-Neptunes in unprecedented detail, small feature amplitudes and poorly-understood systematics have led to ambiguous results for others. Using the first seven targets from the COMPASS program, which will survey 12 small planets using NIRSpec/G395H, we investigate timeseries systematics. We implement a model that uses the principle components of the normalized pixel fluxes to account for variations in the shape and position of the spectral trace. We find that observations with a smaller number of groups-per-integration benefit most profoundly from the use of this model, and that systematics are particularly strong between 2.8 and 3.5 $\mu$m. Despite these systematics, \texttt{pandexo} is a relatively accurate predictor of the precision of the spectra, with real error bars on average 5\% larger in NRS1 and 12\% larger in NRS2 than predicted. We compute new limits on metallicity and opaque pressure level for each target and compare these to previous results from the COMPASS program. Next, we co-add spectra from multiple targets to reduce the effective noise in the combined spectra in hopes of detecting transmission features in common between the targets, but this exercise does not yield compelling evidence any signals. We find that a handful of additional transits are sufficient to break the degeneracy between metallicity and aerosols for the majority of our targets, pointing towards the possibility of unraveling the mysteries of these worlds with future allocations of JWST time.

\end{abstract}

\section{Introduction}

    Sub-Neptunes and super-Earths, planets larger than the Earth but smaller than Neptune's radius of 3.86$R_\oplus$, are a numerous class of exoplanets that have no counterpart in the Solar System. Since the discovery of this population of small planets by the \textit{Kepler} mission \citep{Borucki2011, Batalha2013} the origins and atmospheres of these planets have been a mystery. As super-Earths and sub-Neptunes represent potentially habitable environments \citep{Nixon2021, Luger2015} and make enticing targets for biosignature searches \citep{Rauer2011, Seager2021, Tsai2024, Madhusudhan2021, Phillips2021}, understanding their histories and present-day environments is a key goal for exoplanet scientists. 

    The masses and radii of super-Earths and sub-Neptunes permit a range of possible compositions \citep{Seager2007, Fortney2007, Valencia2007}. The smallest members of this population, the super-Earths, are likely to have a predominantly rocky interior composition and may or may not possess atmospheres \citep{Ginzburg2016}. Throughout this work we use the term ``super-Earth'' to refer to planets below the radius gap \citep{Fulton2017} at about 1.8$R_\oplus$, a region of the mass-radius parameter space with a low occurrence of planets which is thought to separate terrestrial worlds from the lower density sub-Neptunes. The radii of sub-Neptunes, defined in this work as any planets smaller than Neptune but above the radius gap, can be explained by several different classes of interior models including gas dwarfs (a rocky core underneath a thick, low mean molecular weight atmosphere, \citealt{Lopez2014, Rigby2024}), true mini-Neptunes (small rocky cores overlain by some combination of water ice, liquid water, and an H$_2$-dominated atmosphere, \citealt{Rogers2011}), steam worlds (a rocky core underneath a high mean molecular weight steam/supercritical H$_2$O atmosphere, \citealt{Mousis2020, Aguichine2021}), or even Hycean worlds (a rocky core overlain by thick hydrosphere with liquid water at the surface and a thin H$_2$-dominated atmosphere, \citealt{Madhusudhan2021, Madhusudhan2023}). While some studies have attempted to deduce the existence of distinct populations of these planet types by careful examination of their mass-radius distribution, these efforts have been hampered by degeneracies between different interior models and uncertainties in models of atmospheric loss \citep{Otegi2020, Rogers2023, Burn2024}.

    The role of transmission spectroscopy in this landscape is to provide additional constraints to break the degeneracies in mass and radius that permit multiple possible compositions for any given planet. Furthermore, by allowing us to detect molecules and measure their abundances, transmission spectroscopy gives us clues to the origins and subsequent evolution of these worlds. 

    While the NIRSpec instrument has displayed excellent performance and several JWST sub-Neptune observations have allowed for detailed atmospheric characterizations \citep[e.g.][]{Benneke2024, Madhusudhan2023, Schlawin2024}, atmospheric signals for all but the nearest planets with clear, low-metallicity, large scale-height atmospheres still sit near the observed noise floor. This has resulted in a number of more ambiguous observations \citep[e.g.][and the COMPASS papers listed in Table \ref{tbl:observations}]{Ahrer2025, Cadieux2024}. Compounding this challenge is the fact that JWST time-series observations frequently display red noise of uncertain origin on timescales of minutes to hours with amplitudes of tens to hundreds of parts-per-million in the white light curves \citep[e.g.,][]{Luque2024, Wallack2024, Sarkar2024}, degrading the precision of transmission spectra to below what is predicted by tools like \texttt{Pandeia} \citep{Pontoppidan2016} and \texttt{Pandexo} \citep{Batalha2017} (which uses Pandeia as the backend noise simulator). Studies of individual planets observed with NIRSpec have dealt with this noise in a variety of ways, for example by fitting the light curve with a systematics model that incorporates auxiliary timeseries such as the x and y-positions of the trace on the detector as measured during spectral extraction \citep[e.g.][]{Rustamkulov2022, Scarsdale2024} or various engineering parameters associated with the telescope \citep[e.g.,][]{Wallack2024, Kirk2024}. While these techniques show promise, there is still a need to develop a more complete understanding of detector effects as well as a set of best-practices for mitigating red noise and reaching closer to the photon noise limit. A primary aim of this study is to progress towards that goal with respect to NIRSpec. 

    In this work we leverage the uniformity of the COMPASS dataset of sub-Neptune and super-Earth transmission spectra to study the characteristics of the NIRSpec detectors and the noise properties of the light curves. All COMPASS targets were observed using NIRSpec/G395H in the Bright Object Time-Series mode with the SUB2048 subarray, F290LP filter, and S1600A1 slit in the NRSRAPID readout pattern. The number of transits for each target was chosen to achieve a transit SNR of 30 ppm at 4 $\mu$m when binned to a spectral resolution of R=100, according to pre-observation simulations using \texttt{Pandexo}. We carry out uniform reductions of all seven datasets using \texttt{ExoTiC-JEDI} \citep{Alderson2022} holding all reductions procedures and settings identical between all targets, followed by white and spectral light curve fitting using a custom fitting code. We describe both the reduction and the fitting procedure in Section \ref{sec:obs_and_reduction}. When fitting the light curves we apply a newly-developed systematics model, which is introduced in Section \ref{sec:obs_and_reduction} and described in more detail in Section \ref{sec:detector_systematics}. Using this model, we proceed to analyze the instrumental systematics present in NIRSpec/G395H observations and discuss their impacts on transmission spectra. 
    
    Following the analysis of the timeseries systematics we detail our methodology for constraining the atmospheric metallicity and aerosol properties of the planets in Section \ref{sec:spec_analysis}. Our method is similar to that used in previous COMPASS papers, with the difference that we use MCMC to sample the likelihood space whereas previous papers used a grid of models in metallicity and opaque pressure level and compared these to the transmission spectra to find the contour along which these models could be ruled out at $3-\sigma$. In Section \ref{sec:results} we present our lower limits on metallicity and opaque pressure level for each planet. We compare these to the limits obtained in previous papers in order to understand how robust these lower limits are with respect to differences in the reduction and modeling assumptions. In Section \ref{sec:discussion} we draw comparisons between these seven COMPASS targets and the eleven other planets smaller than 3 $R_\oplus$ observed in transmission with JWST in the near-infrared. We then investigate the potential for detecting transmission features and breaking the metallicity-opaque pressure level degeneracy for the seven COMPASS targets with additional observations, before offering some concluding remarks in Section \ref{sec:conclusions}. 
    
\section{Observations and Data Reduction}
    \label{sec:obs_and_reduction}

    The seven planets included in this study are GJ\,357\,b \cite[TOI-562.01,][]{Adams2025}, TOI-836\,b \cite[TOI-836.02,][]{Alderson2024}, TOI-836\,c \cite[TOI-836.01,][]{Wallack2024}, TOI-776\,b \cite[TOI-776.02,][]{Alderson2025}, TOI-776\,c \cite[TOI-776.01,][]{Teske2025}, L\,98-59\,c \cite[TOI-175.01,][]{Scarsdale2024}, and L\,168-9\,b \cite[TOI-134.01,][]{Alam2025}. Details of each of these observations can be found in the associated papers, but to summarize, all observations were made with JWST NIRSpec in the G395H mode, which provides spectra between 2.87 and 5.14 $\mu$m with an average resolution of R=2700. All observations used the NIRSpec Bright Object Time Series mode with the SUB2048 subarray, F290LP filter, and S1600A1 slit in the NRSRAPID readout pattern. NIRSpec/G395H spectra are dispersed across the two NIRSpec detectors, NRS1 and NRS2, resulting in a gap from 3.72 to 3.82 $\mu$m. Table \ref{tbl:observations} summarizes the observations for each target. 

    \begin{deluxetable*}{cccccc}[htb!]
    \tablehead{\colhead{target} & \colhead{$N_\mathrm{transits}$} & \colhead{$N_\mathrm{groups}$} & \colhead{$N_\mathrm{integrations}$} & \colhead{HGA move?} & \colhead{reference}}
    \startdata
        GJ\,357\,b & 1 & 3& 4401 & No & \cite{Adams2025} \\
        TOI-836\,b & 2 & 3 & 5259 & No & \cite{Wallack2024} \\
        TOI-836\,c & 1 & 3 & 6755 & No & \cite{Alderson2024} \\ 
        TOI-776\,b & 2 & 7 & 3288 & Transit 2 & \cite{Alderson2025} \\
        TOI-776\,c & 2 & 7 & 3636 & Transit 2 & \cite{Teske2025} \\
        L\,98-59\,c & 2 & 4 & 3311 & No & \cite{Scarsdale2024} \\
        L\,168-9\,b & 3 & 4 & 3359 & Transit 1 & \cite{Alam2025} 
    \enddata
    \caption{Summary of the observations. For observations affected by high gain antenna movements, data points from one minute prior to five minutes after the antenna movement were masked out and not used for during transit fitting.}
    \label{tbl:observations}
    \end{deluxetable*}

    \subsection{ExoTiC-JEDI Reduction}

        We carried out a homogeneous reduction of all observations using the \texttt{ExoTiC-JEDI} package \citep{Alderson2022}. By homogeneous reduction, we mean that all parameters of the reduction were held constant between observations. For the first stage of the reduction we use the standard \texttt{jwst} pipeline \cite[version 1.18.1, context map 1364;][]{jwst_pipeline}, substituting the custom bias subtraction routine implemented in \texttt{ExoTiC-JEDI} and described in \cite{Alderson2024} and adding in the custom group-level destriping routine from \texttt{ExoTic-JEDI}, which uses the column-by-column median value of the out-of-trace pixels
        to correct for 1/f noise, where the in-trace pixels are identified as all those within $\pm10$ standard deviations of the trace center, with the trace center and width defined by a second-order polynomial fit to the column-by-column medians and standard deviations of a Gaussian fit to the PSF. We then conduct ramp fitting using the routine implemented in the \texttt{jwst} pipeline. 
    
        Operating now at the integration level rather than the group level, we identify pixels with data quality bit values indicating ``do not use'', ``saturated'', ``dead'', ``hot'', ``low quantum efficiency'', and ``no gain value''. These pixels are replaced with the median of the neighboring four pixels. We then use \texttt{ExoTiC-JEDI}'s outlier detection to replace outlier pixels exceeding 10$\sigma$ in the spatial dimension with the median of the nearest 10 pixels, and those exceeding 6$\sigma$ in the time dimension with the median of the nearest 4 pixels. 
        
        We extract the spectrum using the intrapixel aperture extraction routine implemented in \texttt{ExoTiC-JEDI}. The aperture width is defined in units of the full-width half-maximum (FWHM) of the trace and is allowed to vary across the detector as the width of the trace increases at longer wavelengths. Both the center and width of the aperture are smoothed by fitting a fourth-order polynomial to the column-by-column trace centers and widths, themselves determined by a Gaussian fit to the counts in each column. The extraction aperture for each target, visit, and detector is individually selected by carrying out a trial extraction of the last 500 integrations of the observation for candidate aperture widths ranging from 2 to 10 FWHM in units of 0.2 FWHMs. We chose to carry out this trial extraction on the last 500 integrations in order to avoid selecting in-transit points and to avoid any possible effects of instrument settling at the beginning of the observation. Each 500-integration trial lightcurve is then smoothed using a Gaussian filter with a width of 30 integrations to remove red noise and the selected aperture is the one that minimizes the remaining scatter, computed as the standard deviation of the smoothed lightcurve. Table \ref{tbl:extraction_aperture} contains the extraction apertures in units of FWHM and the corresponding aperture width in pixels at 4 $\mu$m for each observation. 

        \begin{deluxetable*}{ccccc}[htb!]
            \tablehead{\colhead{target} & \colhead{visit} & \colhead{detector} & \colhead{aperture (FWHM)} & \colhead{aperture in pixels}}
            \startdata
            TOI-836 c  &  T1  &  NRS1  &  3.8  & 5.30  \\
              &    &  NRS2  &  3.4  &  5.12  \\
            \hline
            GJ 357 b  &  T1  &  NRS1  &  4.0  &  5.68  \\
             &    &  NRS2  &  4.0  &  6.22  \\
             \hline
            TOI-836 b  &  T1  &  NRS1  &  4.0  &  5.60  \\
              &    &  NRS2  &  3.8  &  5.74  \\
              &  T2  &  NRS1  &  3.6  &  5.04  \\
              &    &  NRS2  &  3.6  &  5.44  \\
              \hline
            L 98-59 c  &  T1  &  NRS1  &  4.0  &  5.58  \\
              &    &  NRS2  &  3.6  &  5.48  \\
              &  T2  &  NRS1  &  4.2  &  5.92  \\
              &    &  NRS2  &  4.0  &  6.16  \\
              \hline
            TOI-776 b  &  T1  &  NRS1  &  4.0  &  5.62  \\
              &    &  NRS2  &  4.2  &  6.40  \\
              &  T2  &  NRS1  &  4.2  &  5.92  \\
              &    &  NRS2  &  4.2  &  6.42  \\
              \hline
            L 168-9 b  &  T1  &  NRS1  &  4.0  &  5.60  \\
              &    &  NRS2  &  4.0  &  6.04  \\
              &  T2  &  NRS1  &  4.2  &  5.90  \\
              &    &  NRS2  &  4.0  &  6.10  \\
              &  T3  &  NRS1  &  4.0  &  5.64  \\
              &    &  NRS2  &  3.8  &  5.80  \\
              \hline
            TOI-776 c  &  T1  &  NRS1  &  4.2  &  5.86  \\
              &    &  NRS2  &  4.6  &  6.94  \\
              &  T2  &  NRS1  &  4.0  &  5.62  \\
              &    &  NRS2  &  4.2  &  6.42  \\
            \enddata
            \caption{Aperture width used to extract the stellar spectra. The FWHM value is in reference to the profile of the trace in the cross-dispersion direction. The ``aperture (FWHM)'' column gives the aperture size as a multiple of the FWHM of the trace, which varies with wavelength causing the width of the aperture to vary as well. The last column labeled ``aperture in pixels'' gives the aperture width in pixels at 3.5 $\mu$m for NRS1 or 4 $\mu$m for NRS2. \label{tbl:extraction_aperture}}
        \end{deluxetable*}
        
        In order to maintain communication with ground facilities, JWST's high-gain antenna (HGA) must reorient every 10,000 seconds. For long observations like those of exoplanet transits, those movements occasionally take place during the observation and can cause systematics in the light curve which last for several minutes. In order to limit the impact of these HGA movements we mask the integrations starting one minute prior to the movement and extending to five minutes after. 

    \subsection{White Light curve Analysis}

        \subsubsection{Systematics Model}

            The origin of the correlated noise seen in NIRSpec time-series observations is uncertain, and it may arise from a combination of both astrophysical and instrumental sources. Previous studies have attempted to mitigate red noise by fitting the light curve with a systematics model that incorporates auxiliary timeseries such as the x and y-positions of the trace on the detector as measured during spectral extraction \citep[e.g.][]{Rustamkulov2022, Alderson2023, Scarsdale2024} or various engineering telemetry timeseries associated with the telescope \citep[e.g.,][]{Wallack2024, Kirk2024}. We take a similar approach, but rather than using engineering parameters or trace positions we use basis vectors extracted using a principal component analysis (PCA) of the relative pixel flux (RPF) timeseries. Principal component analysis is a dimension reduction technique that can, for our purposes, be understood as a method for reducing the full set of RPF vectors to a set of $N$ orthogonal vectors that together explain the majority of the variance in the relative pixel fluxes. PCA techniques have been used to reveal detector-level artifacts in \textit{Kepler} \citep{Luger2018} and \textit{Spitzer} observations \citep{Deming2015} as well as for NIRISS/SOSS \citep{Radica2025, Coulombe2023} and NIRSpec observations \citep[][discussed further below]{Luque2024} from JWST.
            
            The RPF is defined as the flux in an individual pixel divided by the total flux on all pixels, including the background pixels. If the pixels are indexed by $i$ in the dispersion direction and $j$ in the cross-dispersion direction, then we have: 
            \begin{equation}
                \mathrm{RPF}(t)_{i, j} = \frac{F(t)_{i, j}}{\sum_{i, j}F(t)_{i, j}}.
            \end{equation} 
            We then compute the principal components of the set of RPF timeseries for each individual visit and detector using the PCA algorithm implemented in \texttt{SciPy} \citep{scipy}. PCA is a method for reducing the dimensionality of a dataset, in this case the collection of RPF timeseries, when the individual members of the dataset can be represented as linear combinations of a smaller set of basis vectors plus some additional noise. By applying this technique, we are able to identify a set of $N_\mathrm{comp}$ principal component timeseries which represent the components of the correlated noise which are shared among the RPF vectors. We construct our systematics model as a linear combination of these principal components, plus a first order polynomial trend in time. The systematics model for the $n^{\mathrm{th}}$ visit is therefore given by 
            \begin{equation}
              S_n(t) = a_n + b_nt + \sum_{j=0}^{N_\mathrm{comp}}c_{n,j} P_n(t)_{j}
            \end{equation}
            where $P_n(t)_{0, 1\dots N_\mathrm{comp}}$ are the first $N_\mathrm{comp}$ principal components of the relative pixel fluxes for the $n^{\mathrm{th}}$ visit. 

            We demonstrate in Section \ref{sec:detector_systematics} that the principal component vectors correspond to small changes in the morphology of the trace on the detector over time. This systematics model thus extends the practice of fitting shifts in the position of the trace to include more complex, higher-order changes in trace morphology. It has the advantage that the basis vectors are extracted directly from pixel fluxes and is therefore independent of the method of spectral extraction, making the technique pipeline-independent. It can therefore be used with pipelines that carry out optimal extraction such as \texttt{Eureka!} \citep{Bell2022} or aperture extraction such as \texttt{ExoTiC-JEDI}. 

            \cite{Luque2024} constructs a similar systematics model for their NIRSpec/G395H observations of TOI-1685 b's phase curve. However, their model differs in that it uses the total flux in each pixel rather than the relative pixel flux. By using the relative pixel flux we remove any astrophysical trends while preserving detector effects such a residual 1/f noise, gain variations, and changes in the shape or position of the trace. This potentially explains our model's success at fitting red noise whereas their model overfit the transit signal and was not found to reduce the red noise. In addition to PCA, \cite{Luque2024} also use Independent Component Analysis, or ICA, to decompose the pixel timeseries and obtain a set of basis vectors. ICA is related to PCA, but rather than ensuring that the resulting components are orthogonal, ICA guarantees that the components will be statistically independent. In developing our systematics model we also experimented with ICA, and found that while ICA and PCA produce distinct basis vectors, the best-fit or maximum-likelihood systematics model is indistinguishable between the two techniques.
            
            The phase curve observations for TOI-1685\,b use 16 groups per integration, which is larger than the three to seven groups used for the COMPASS observations. As we will see in Section \ref{sec:detector_systematics}, the amplitude of the systematics decreases with the number of groups, so that a 16 group observation may not display significant red noise of the type seen in our observations. This is borne out by the fact that the red noise in the TOI-1685\,b observation occurs mainly on timescales of hours, suggesting that it has a different source altogether. 
            
        \subsubsection{White Light curve Fitting Procedure}

            For each target we fit the white lightcurves for each detector separately. We produce three fits for each target and detector using three different systematics models in order to investigate the impact of each model on the final white lightcurves. The first, which we refer to as the ``0-vector'' model, is a two-parameter model in which we fit for the out-of-transit flux and a linear slope in time. Fitting this slope has become standard for NIRSpec lightcurves in order to account for the pronounced downward trend in flux over time that is seen in NRS1, and the smaller slope seen in NRS2 \citep{Espinoza2023}. This 0-vector model provides a baseline against which we can compare the two more complex systematics models. 

            The second model, which we refer to as the ``shifts-only'' model, is a four-parameter model defined by $S(t) = a + bt + cx(t) + dy(t)$ where $x(t)$ and $y(t)$ are the trace positions in the x and y-directions respectively recorded as offsets from the position of the integration-averaged trace position. Systematics models like this one, that account for variations in the telescope pointing over time, have been used in many recent JWST observations including NIRSpec/G395H observations \citep[e.g.][]{Rustamkulov2022, Teske2025, Alderson2023, Scarsdale2024}.

            The third systematics model, which we refer to as the ``6-vector'' model, is an eight-parameter model consisting of the out-of-transit flux and linear slope from the 0-vector model plus coefficients for the first six principal components of the RPF timeseries described above. We choose to fit six components because we find that every observation has at most six components which display correlated noise, with higher order components appearing consistent with white noise when inspected by eye. In the interest of carrying out a uniform analysis, we elect to use the same number of principal components for each observation, and using fewer than six would result in excluding principal components that show correlated noise for some observations, while using more than six would mean including components that display only white noise. 
            
            We parameterize the transit in terms of the period, $P$, planet/star radius ratio, $R_p/R_*$, time of the transit relative to the start of the observation $t_0$, semimajor axis in units of stellar radii $a/R_*$, inclination $i$, eccentricity $e$ and argument of periastron $\omega$. Because the JWST observations do not meaningfully constrain $P$, $i$, $e$, or $\omega$, we place informative Gaussian priors on these parameters that are derived from the reported maximum likelihood values and uncertainties from the studies referenced in Table \ref{tbl:priors}. We place uninformative uniform priors on $R_p/R_*$ (uniform between 0 and 1), $t_0$ (uniform between the beginning and end of the observation), and $a/R_*$ (uniform between $\mu_{t_0}\pm10\sigma_{t_0}$ where $\mu_{t0}$ and $\sigma_{t0}$ are the mean and standard deviation of the reported $t_0$ value from the reference in \ref{tbl:priors}) in order to allow the data to determine the posteriors for these parameters. For GJ\,357\,b and L\,168-9\,b the studies referenced in Table \ref{tbl:priors} assumed a fixed eccentricity of zero. For these cases we also set eccentricity to zero rather than fitting for its value. For the limb-darkening parameters we use informative priors computed with \texttt{ExoTiC-LD} \citep{Grant2024} using the MPS-ATLAS-1 stellar models with the stellar parameters given in Table \ref{tbl:planetary_stellar_params}.

            \begin{deluxetable}{cccc}[b!]
                \tablehead{\colhead{target} & \colhead{reference} & \colhead{notes}}
                \startdata
                  GJ 357 b & \cite{Oddo2023} & eccentricity set to 0\\
                  TOI-836 b & \cite{Hawthorn2023} & \\
                  TOI-836 c & \cite{Hawthorn2023} & \\
                  TOI-776 b & \cite{Fridlund2024} & \\
                  TOI-776 c & \cite{Fridlund2024} & \\
                  L 98-59 c & \cite{Demangeon2021} & \\
                  L 168-9 b & \cite{Hobson2024} & eccentricity set to 0\\
                \enddata
                \caption{References for priors used in white light curve fitting. For each planet the posterior distributions reported in the reference given were adopted directly as the priors for our fitting procedure, with any exceptions given in the notes column.}
                \label{tbl:priors}
            \end{deluxetable}

            \begin{figure}[htb!]
                \centering
                \includegraphics[width=0.5\textwidth]{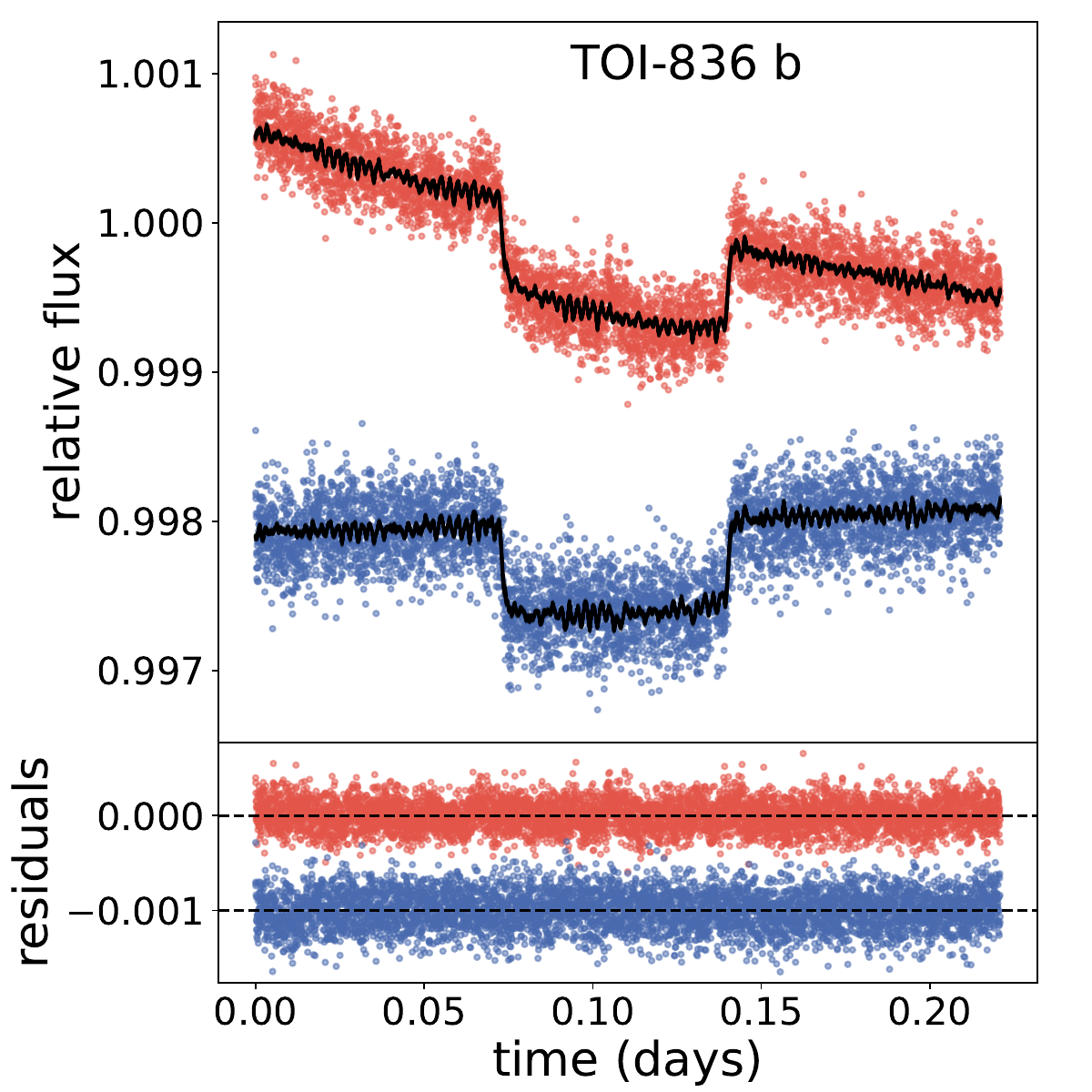}
                \caption{White light curves and models drawn from the posterior distribution for TOI-836\,b. \textbf{Top: } The red points are NRS1 data and the blue points are NRS2. The black lines are the superposition of 100 samples of our 6-vector systematics + transit models drawn from the posterior of the MCMC simulations. \textbf{Bottom:} Residuals after subtraction of the transit + systematics model. The complete figure set (7 images) is available in the online journal.}
                \label{fig:wlc_single}
            \end{figure}

            \begin{figure}[htb!]
                \centering
                \includegraphics[width=0.5 \textwidth]{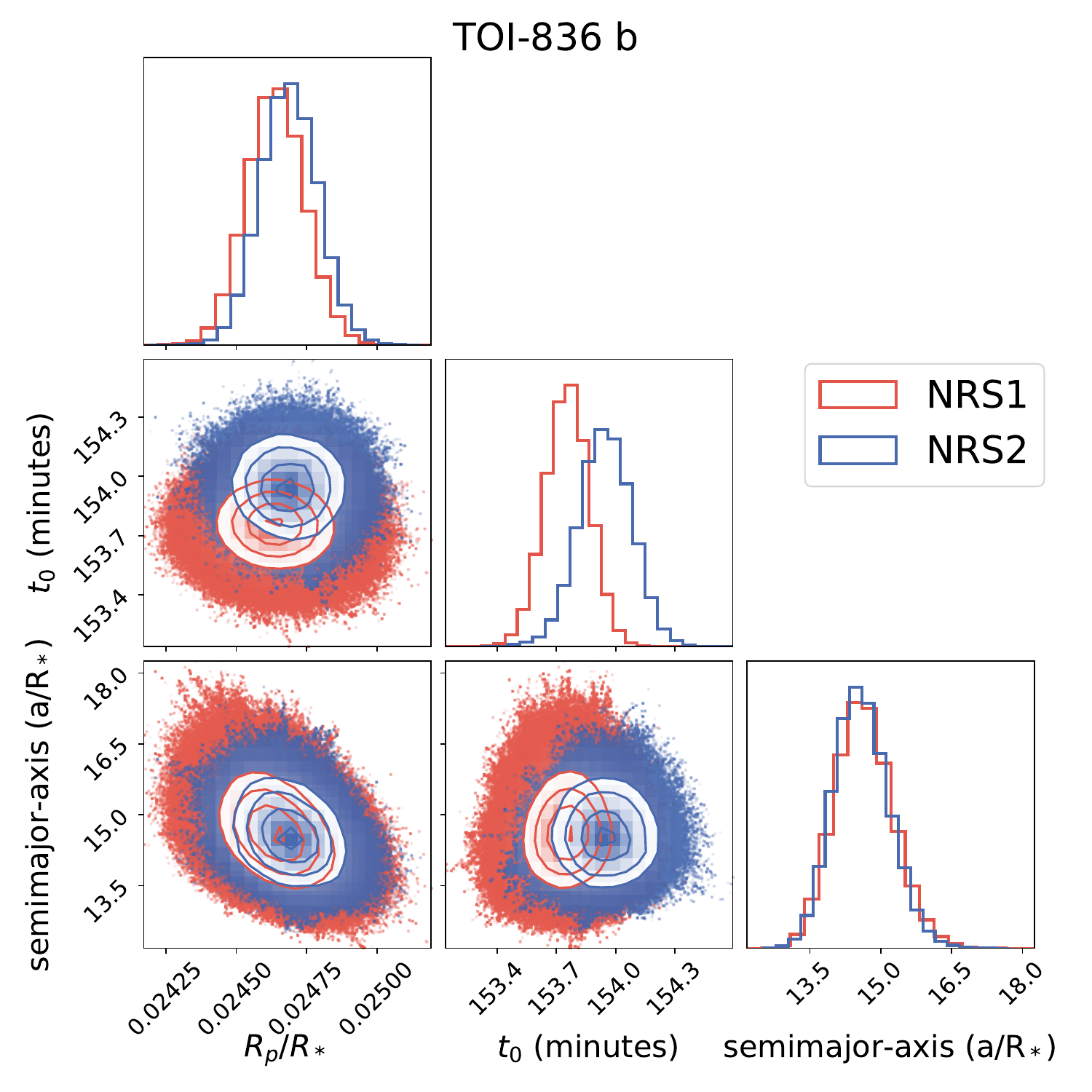}
                \caption{Corner plot showing the posterior distribution of the transit parameters for TOI-836\,b derived from the white light curve. The red distribution corresponds to the NRS1 data and the blue is for NRS2. The complete figure set (7 images) is available in the online journal.}
                \label{fig:corner_single}
            \end{figure}

            \begin{figure*}
                \centering
                \includegraphics[width=0.49\textwidth]{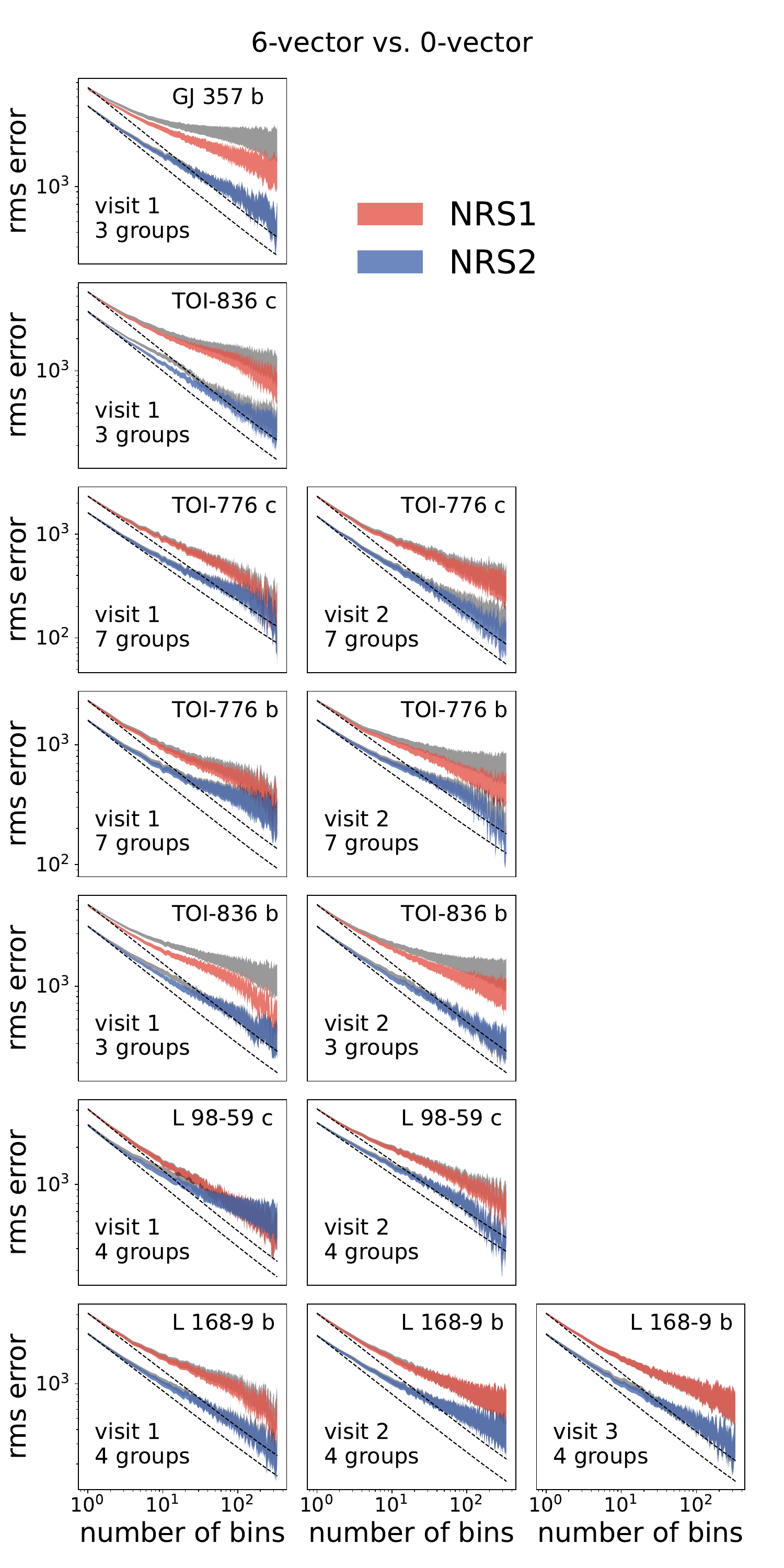}
                \includegraphics[width=0.49\textwidth]{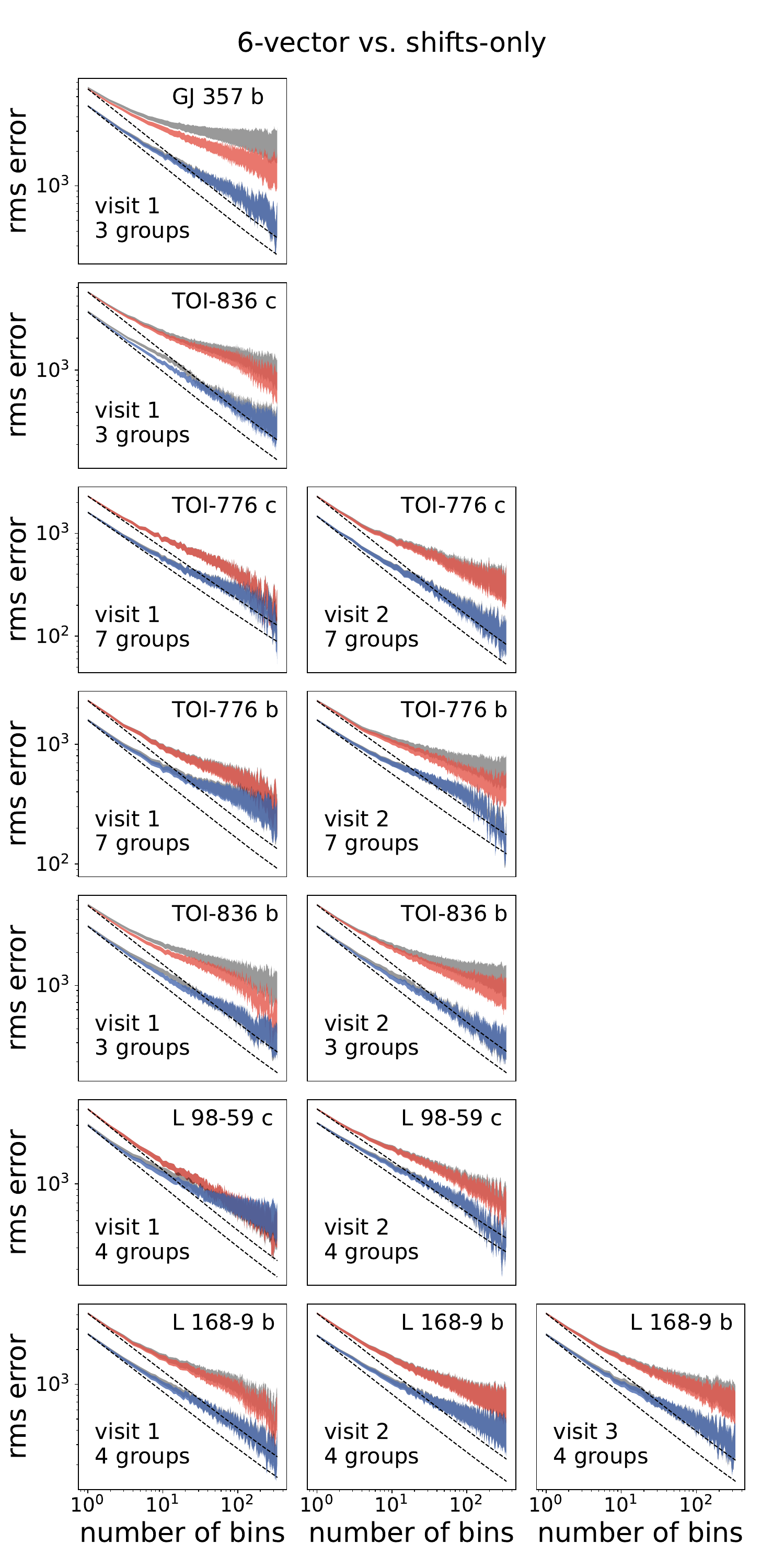}
                \caption{RMS versus bin-size for the residuals to the white light curves after subtraction of the maximum-likelihood transit + systematics model for both detectors and all visits of the seven planets in our sample. \textbf{Left:} Comparison between the 6-vector model (red and blue shaded regions) and the 0-vector model (gray shaded regions). \textbf{Right:} Comparison between the 6-vector model (red and blue shaded regions) and the shifts-only model (gray shaded regions).
                While some degree of red noise remains for all of the datasets, the 6-vector systematics model is more effective at reducing the red noise than either alternative systematics model, particularly for the three-group observations.}
                \label{fig:allan_var_all}
            \end{figure*}

            We use MCMC sampling to infer the parameters of the transit and systematics model. We run 60 chains for 100,000 samples each, discarding the first 5,000 samples of each chain as burn-in, for a total of 5,700,000 samples. We simultaneously model all available visits of each target, but we model the white light curves for the NRS1 and NRS2 detectors separately in order to account for differences in the properties of the detectors, and to assess the agreement in the transit parameters between them. We obtain $>2\times10^7$ effective samples for all parameters and all targets. In general, we find good agreement between the transit parameters for each detector with some observations showing statistically significant offsets in the transit time between NRS1 and NRS2 as has been seen in previous analysis of these targets \citep[e.g.][]{Wallack2024}. Table \ref{tbl:wl_posteriors} summarizes our posterior distributions for $R_p/R_*$, $t_0$, and $a/R_*$. Because the $P$, $i$, $e$, and $\omega$ are not well constrained by these observations, the posteriors are nearly identical to the priors and we elect not to report them in order to avoid confusion. 

            In Figure \ref{fig:wlc_single} we plot 100 randomly drawn samples of the combined transit and systematics model from the posterior distribution for TOI-836\,b, which we have chosen as a representative target. Corresponding plots for the full set of targets are available as a figure set in the HTML version of this article. Note that, while the systematics model primarily fits short timescale variability, and does not capture longer timescale variations in flux such as the pre-transit bump in NRS1. In Section \ref{sec:detector_systematics} we discuss in detail the variability timescales present in the principle component vectors and explore the origins of this variability. Figure \ref{fig:corner_single} shows the posterior distribution for the transit parameters for the white light curve of TOI-836\,b. Corner plots for the full set of targets are also available as a figure set in the HTML version of the article. Figure \ref{fig:allan_var_all} shows the improvement in the degree of red noise in the residuals when using the 6-vector systematics model (colored lines) in comparison to a 0-vector model (gray lines, left side) and the shifts-only model (gray lines, right side). From these plots we can see that use of the 6-vector model results in less red noise than either of the alternative models in the majority of cases. The difference is particularly notable for the NRS1 detector and for the three-group observations (GJ\,357\,b, TOI-836\,b, and TOI-836\,c). The improvement is less significant for the NRS2 detector and for the four-group and seven-group observations (L\,98-59\,c, L\,168-9\,b, TOI-776\,b and TOI-776\,c), though, for unknown reasons, the second visit of TOI-776\,b does show significant improvement in the red noise for the NRS1 white lightcurve.

            \begin{deluxetable*}{ccccccccc}[htb!]
                \tablehead{\colhead{planet name} & \colhead{mass (M$_\oplus$)} & \colhead{T$_\mathrm{eq}$ (K)} & \colhead{radius (R$_\oplus$)} & \colhead{T$_\mathrm{eff}$} & \colhead{R ($R_*$)} & \colhead{$\log(g)$} & \colhead{$\log(m_H)$} & \colhead{reference}}
                \startdata
                    L\,168-9\,b & 4.07 & 998 & 1.58 & 3842 & 0.60 & 4.84 & -0.03 & \cite{Hobson2024}\\
                    L\,98-59\,c & 2.22 & 553 & 1.39 & 3415 & 0.30 & 4.45 & -0.46 & \cite{Demangeon2021}\\
                    GJ\,357\,b & 1.84$^\mathrm{a}$ & 525 & 1.33 & 3505 & 0.34 & 4.94 & -0.12 & \cite{Oddo2023}\\
                    TOI-776\,c & 6.90 & 420 & 2.02 & 3725 & 0.55 & 4.80 & -0.21 & \cite{Fridlund2024}\\
                    TOI-776\,b & 5.00 & 520 & 1.82 & 3725 & 0.55 & 4.80 & -0.21 & \cite{Fridlund2024}\\
                    TOI-836\,b & 4.53 & 871 & 1.79 & 4552 & 0.67 & 4.74 & -0.28 & \cite{Hawthorn2023}\\
                    TOI-836\,c & 9.60 & 665 & 2.50 & 4552 & 0.67 & 4.74 & -0.28 & \cite{Hawthorn2023}\\
                \enddata
                \caption{Planetary parameters used to produce atmospheric forward model grids and stellar parameters used to compute limb-darkening priors with \texttt{ExoTiC-LD}. Masses and equilibrium temperatures are taken from the given reference, while radii are computed as the average $R_p/R_*$ between NRS1 and NRS2 from our analysis, multiplied by the stellar radius given by the listed reference.}
                \footnotesize{(a) \cite{Luque2019}}
                \label{tbl:planetary_stellar_params}
            \end{deluxetable*}

\begin{figure*}[htb!]
            \centering
            \includegraphics[width=\textwidth]{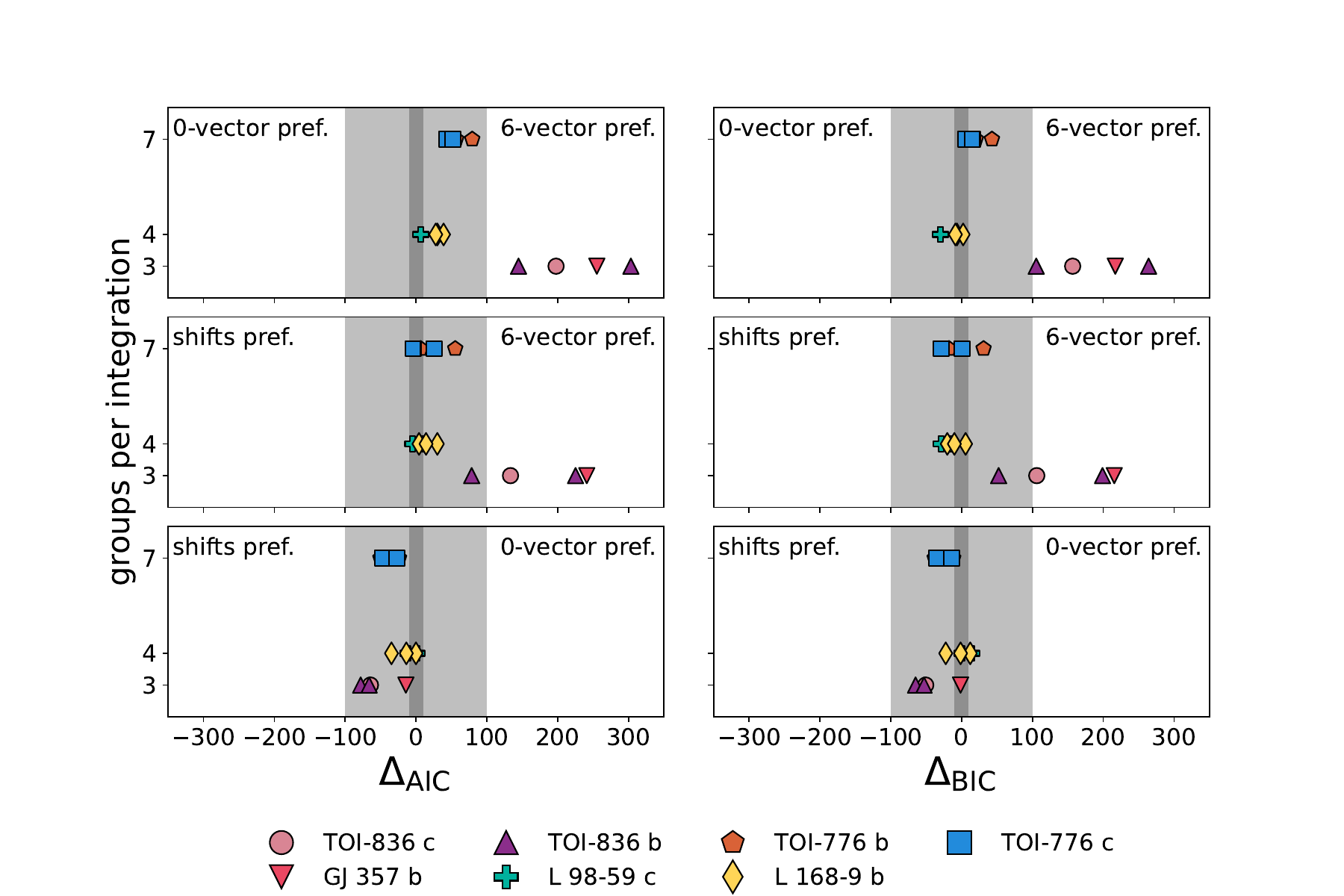}
            \caption{$\Delta_\mathrm{AIC}$ (\textbf{left}) and $\Delta_\mathrm{BIC}$ (\textbf{right}) between each set of systematics models for the NRS1 detector, with each point representing a single visit of the target. The number of groups per integration for the observation is on the y-axis. The dark gray shaded region indicates that there is no clear evidence for one model over the other. For observations falling in the light gray region there is strong evidence for the indicated model, and for observations falling in the unshaded region there is decisive evidence for the indicated model. The thresholds for ``strong'' and ``decisive'' evidence are taken from \cite{Raftery1995}.}
            \label{fig:bic_aic_nrs1}
        \end{figure*}

\begin{figure*}[htb!]
            \centering
            \includegraphics[width=\textwidth]{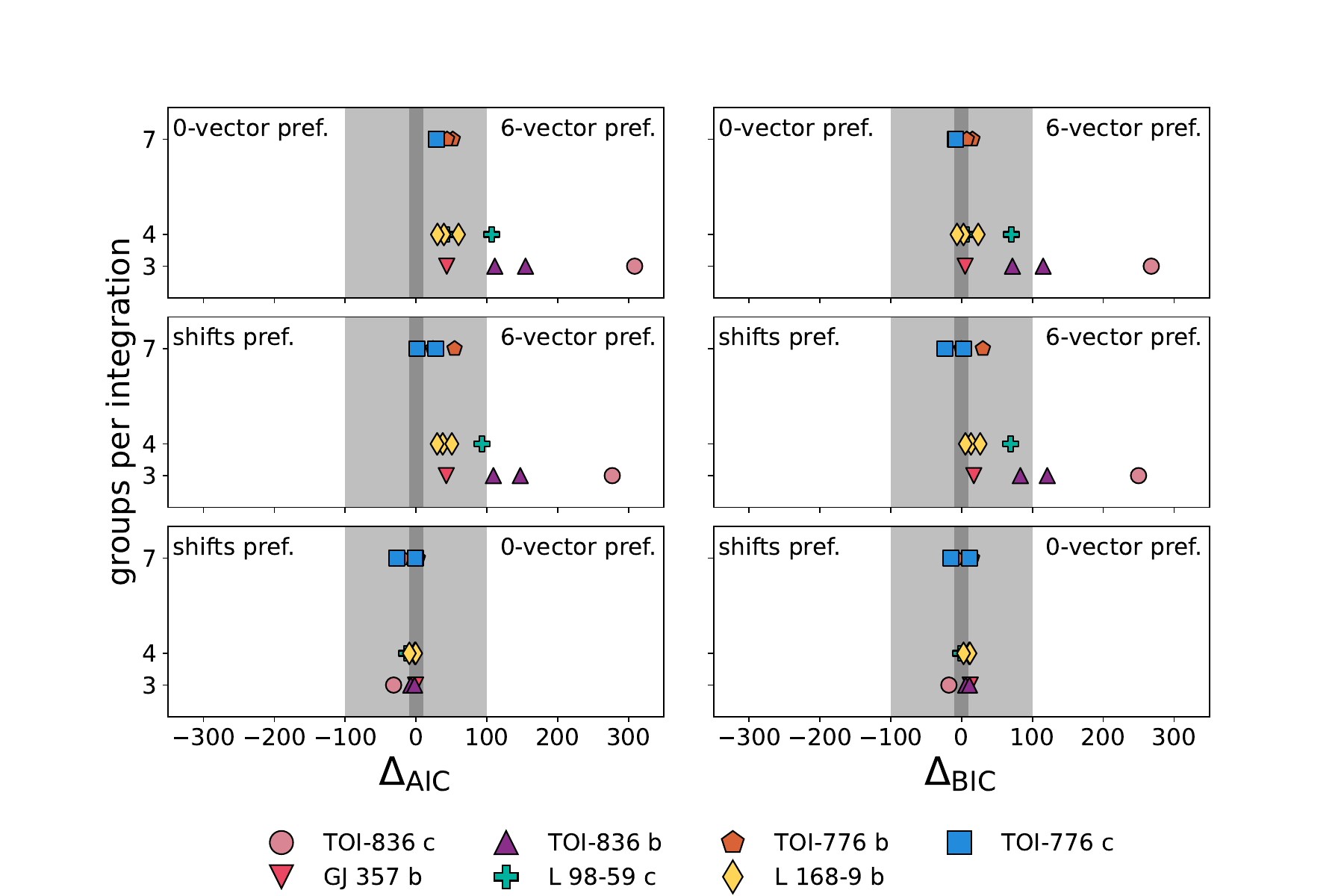}
            \caption{$\Delta_\mathrm{AIC}$ (\textbf{left}) and $\Delta_\mathrm{BIC}$ (\textbf{right}) between each set of systematics models for the NRS2 detector.}
            \label{fig:bic_aic_nrs2}
        \end{figure*}

        In Figures \ref{fig:bic_aic_nrs1} and \ref{fig:bic_aic_nrs2} we plot the difference in the Bayesian Information Criterion ($\Delta_\mathrm{BIC}$) and the Akaike Information Criterion ($\Delta_\mathrm{AIC}$) between the 6-vector model and the 0-vector model, and between the 6-vector model and the shifts-only model. For each comparison a higher value of $\Delta_\mathrm{AIC}$ or $\Delta_\mathrm{BIC}$ indicates a preference for the 6-vector model. Following \cite{Raftery1995}, we consider a $\Delta_\mathrm{BIC}$ or $\Delta_\mathrm{AIC}>10$ to indicate strong evidence for the 6-vector model and $\Delta_\mathrm{BIC}$ or $\Delta_\mathrm{AIC}<-10$ to indicate strong evidence against the 6-vector model, as indicated by the shaded region in the figures. This corresponds roughly to an odds ratio of 150:1 in favor of the selected hypothesis. The difference between the Akaike and Bayesian Information Criterion is in how they penalize model complexity: the Bayesian Information Criterion is smaller by $\log(N)$ for each additional model parameter, where $N$ is the number of samples in the dataset (in this case the number of integrations in the timeseries). In contrast, the Akaike Information Criterion is reduced only by $2$ for each additional model parameter. The result is that the BIC appears less favorable for the more complex 6-vector model compared to the 0-vector model than does the AIC. 

        Focusing first on the NRS1 detector (Figure \ref{fig:bic_aic_nrs1}), we find that both the AIC and BIC indicate strong evidence for the 6-vector model over the both 0-vector model and the model that only accounts for the trace shifts (i.e. the shifts-only model) for every three-group observation. For the four-group observations neither model is strongly preferred, and for the seven-group observations there is slight preference for the 6-vector model over the 0-vector model according to the AIC, less preference for the 6-vector model according to the BIC, and evidence for the 6-vector model over the shifts-only model in some observations but not others according to both the AIC and the BIC. 

        For NRS2 (Figure \ref{fig:bic_aic_nrs2}) the picture is somewhat murkier, but the 6-vector model is still strongly preferred over both the 0-vector model and the shifts-only model for all three-group observations according to the AIC. For the BIC this is true as well, with the exception of the single visit of GJ\,357\,b. For the four-group and seven-group observations both the AIC and BIC indicate a preference for the 6-vector model over the 0-vector and shift-only models for some observations but not others. 

        When we consider the shifts-only model compared to the 0-vector model we see that the shifts-only model is frequently preferred over the 0-vector model for the NRS1 light curves, while there is not strong evidence for one model over the other in the NRS2 light curves. In no case is the evidence for the shifts-only model over the 0-vector model as strong as the evidence for the 6-vector model over the 0-vector model or the shifts-only model. We conclude that, while detrending with the trace shifts may be helpful, especially for NRS1 light curves, this model does not capture nearly as much of the variance as does detrending with principal components, at least for the low to moderate group number observations studied here. 

        \begin{deluxetable*}{ccccccccc}[htb!]
                \tablehead{\colhead{target} & \colhead{detector} & \colhead{$R_p/R_*$} & \colhead{$t_0$ (BMJD)}  & \colhead{$a/R_*$}}
                \startdata
                    GJ 357 b & NRS1 & $0.0313_{-0.0002}^{0.0002}$ & $60282.34125_{-6.0018E-05}^{6.1034E-05}$ & $23.2268_{-1.2085}^{0.8366}$ \\ 
                     & NRS2 & $0.0304_{-0.0002}^{0.0002}$ & $60282.34135_{-6.4022E-05}^{6.3693E-05}$ & $23.5880_{-0.8872}^{0.4565}$ \\ 
                    TOI 836 b & NRS1 & $0.0246_{-0.0002}^{0.0002}$ & $60007.8652_{-0.0002}^{0.0002}$ & $14.6518_{-1.1883}^{1.2298}$ \\ 
                     & NRS2 & $0.0247_{-0.0002}^{0.0002}$ & $60007.8654_{-0.0002}^{0.0002}$ & $14.6085_{-1.1765}^{1.2411}$ \\ 
                    TOI 836 c & NRS1 & $0.0346_{-0.0002}^{0.0002}$ & $59991.7277_{-0.0001}^{0.0001}$ & $23.4682_{-3.3405}^{3.4459}$ \\ 
                     & NRS2 & $0.0343_{-0.0002}^{0.0002}$ & $59991.7280_{-0.0001}^{0.0001}$ & $23.7342_{-3.6525}^{3.8932}$ \\ 
                    TOI 776 b & NRS1 & $0.0301_{-0.0001}^{0.0001}$ & $60088.2926_{-0.0001}^{0.0001}$ & $27.4523_{-2.2128}^{2.0658}$ \\ 
                     & NRS2 & $0.0307_{-0.0002}^{0.0002}$ & $60088.2925_{-0.0001}^{0.0001}$ & $27.0646_{-2.4149}^{2.3556}$ \\ 
                    TOI 776 c & NRS1 & $0.0338_{-0.0001}^{0.0001}$ & $60075.9679_{-0.0001}^{0.0001}$ & $40.2338_{-4.4140}^{5.4070}$ \\ 
                     & NRS2 & $0.0339_{-0.0002}^{0.0002}$ & $60075.9679_{-0.0001}^{0.0001}$ & $39.9822_{-5.0251}^{6.3386}$ \\ 
                    L 98-59 c & NRS1 & $0.0402_{-0.0001}^{0.0001}$ & $60134.60953_{-4.5307E-05}^{4.5809E-05}$ & $23.5017_{-1.6504}^{1.7823}$ \\ 
                     & NRS2 & $0.0402_{-0.0001}^{0.0002}$ & $60134.60949_{-6.2033E-05}^{6.2380E-05}$ & $22.1817_{-1.5992}^{1.7094}$ \\ 
                    L 168-9 b & NRS1 & $0.0238_{-0.0002}^{0.0002}$ & $60098.4580_{-0.0001}^{0.0001}$ & $7.2142_{-0.4748}^{0.4733}$ \\ 
                     & NRS2 & $0.0239_{-0.0002}^{0.0002}$ & $60098.4583_{-0.0001}^{0.0001}$ & $7.2274_{-0.4856}^{0.4740}$ \\
                \enddata
                \caption{Inferred parameters for radius, transit time, and semimajor axis from white light curve fitting. We also fit for inclination, period, eccentricity, and argument of periastron, but because those parameters are poorly constrained by the JWST data the posteriors are nearly identical to the priors. We therefore elect not to summarize the posteriors for those parameters in this table. For targets with multiple visits all visits were fit simultaneously to obtain a single set of parameters. Values are given as the mean of the posterior distribution with upper and lower bounds of the 95\% confidence interval. Note the inconsistency between the time of transit for TOI-836\,c in NRS1 and NRS2, which was also observed in \cite{Wallack2024}.}\label{tbl:wl_posteriors}
            \end{deluxetable*}
        
    \subsection{Spectral Light curve Analysis}

         We independently fit the light curves for each spectral bandpass, holding all transit parameters constant at the maximum-likelihood values from the white lightcurve posteriors except for the planetary radius and the two quadratic limb-darkening parameters. The planetary radius, out-of-transit flux level, slope of the linear trend, and PCA vector coefficients are unconstrained. We fit the spectral light curves in 0.02 micron wide bins, corresponding to approximately 30 pixels-per-bin or a spectral resolution of R=200 at 4 $\mu$m, by running 10,000 steps of MCMC, discarding the first 5,000 as burn-in for each of 36 chains to produce posterior distributions for all parameters. 

         \begin{figure*}[htb!]
            \centering
            \includegraphics[width=1.0\textwidth]{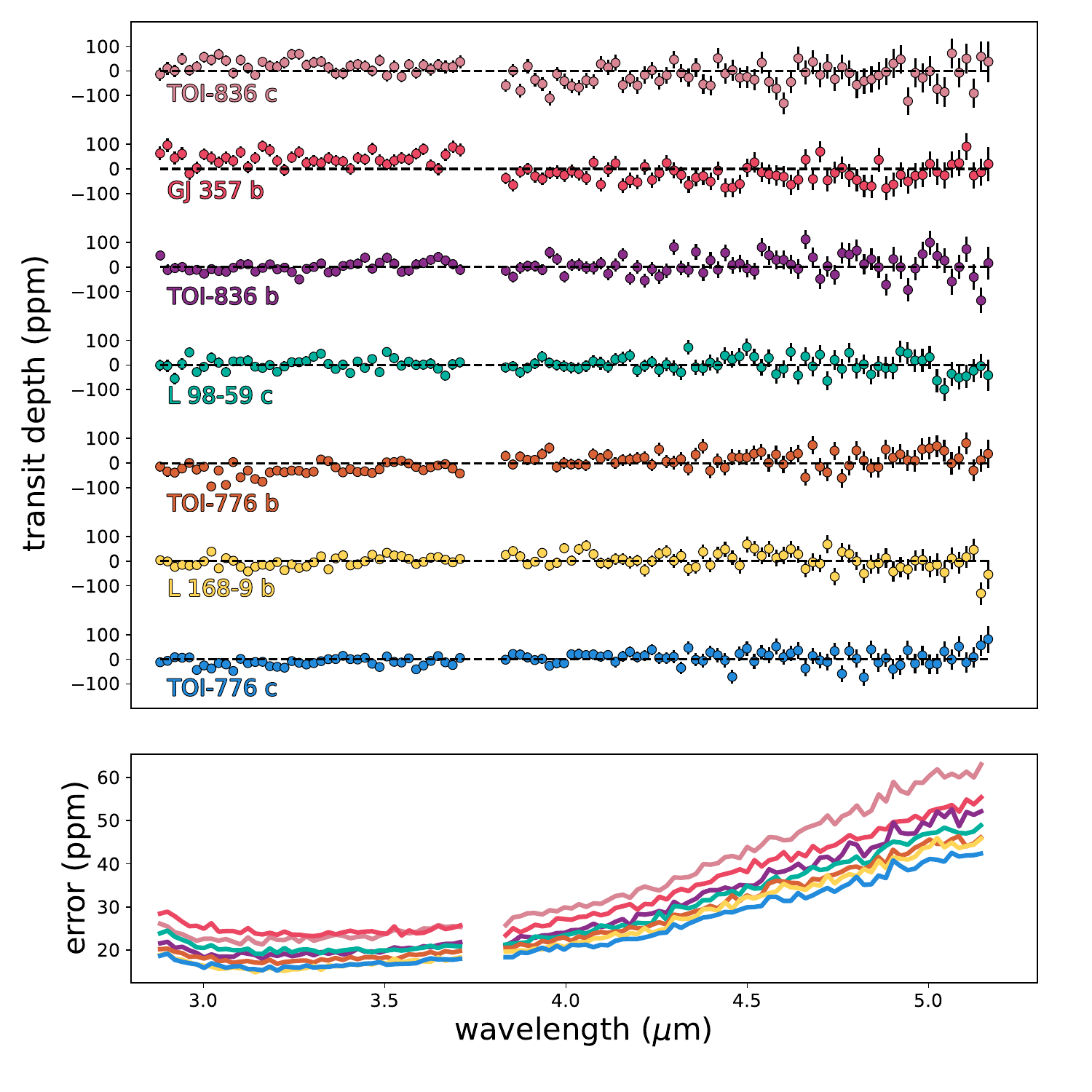}
            \caption{\textbf{Top:} Spectra for each planet in the sample after subtracting off the mean to make it easier to visualize the amplitude of the scatter. No attempt has been made to correct for offsets between detectors. \textbf{Bottom:} Error on transit depth measurements in parts-per-million. For GJ 357 b, TOI-836 b, and TOI-836 c these are the spectra obtained using the six-vector systematics model, while for TOI-776\,b, TOI-776\,c, L\,98-59\,c, and L\,168-9\,b these are the spectra obtained with the trend-only systematics model corresponding to the red points in Figure \ref{fig:pca_comparison}.}
            \label{fig:all_spectra_separate}
        \end{figure*}

         \begin{figure*}[htb!]
            \centering
            \includegraphics[width=\textwidth]{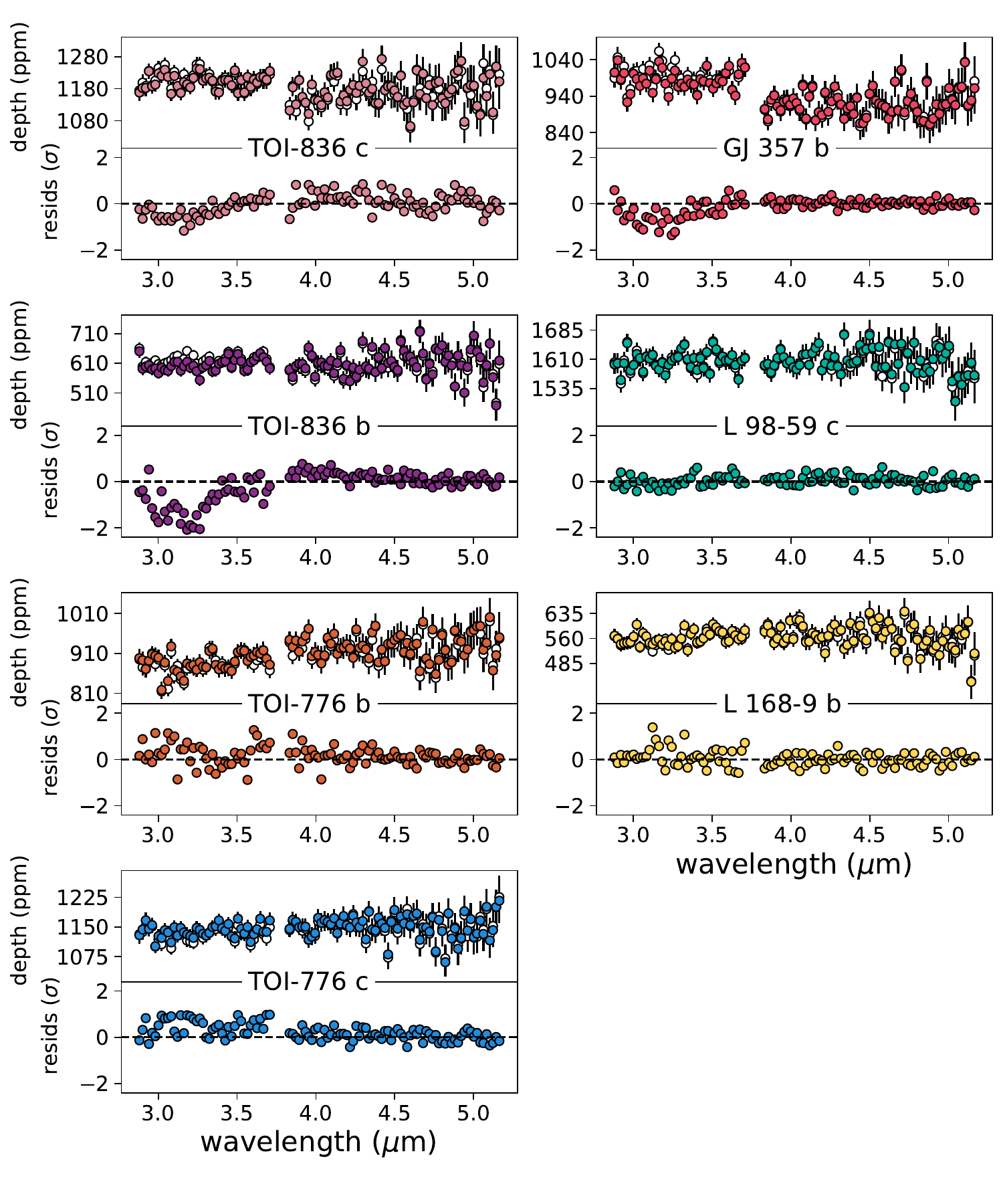}
            \caption{Comparison between spectra produced with the 0-vector systematics model (white points) and the 6-vector systematics model described above (colored points). The upper panel shows the two spectra, and the lower panel shows the residuals between them. For subsequent modeling and analysis we have elected to use the spectra corresponding to the 6-vector systematics model for GJ\,357\,b and TOI-836\,b \& c, all of which were observed with three groups and all of which show a systematic offset with respect to the other spectra between 3.0 and 3.5 $\mu$m. For all other targets we use the spectra corresponding to the 0-vector systematics model.}
            \label{fig:pca_comparison}
        \end{figure*}

        \begin{figure*}[htb!]
            \centering
            \includegraphics[width=1
            \textwidth]{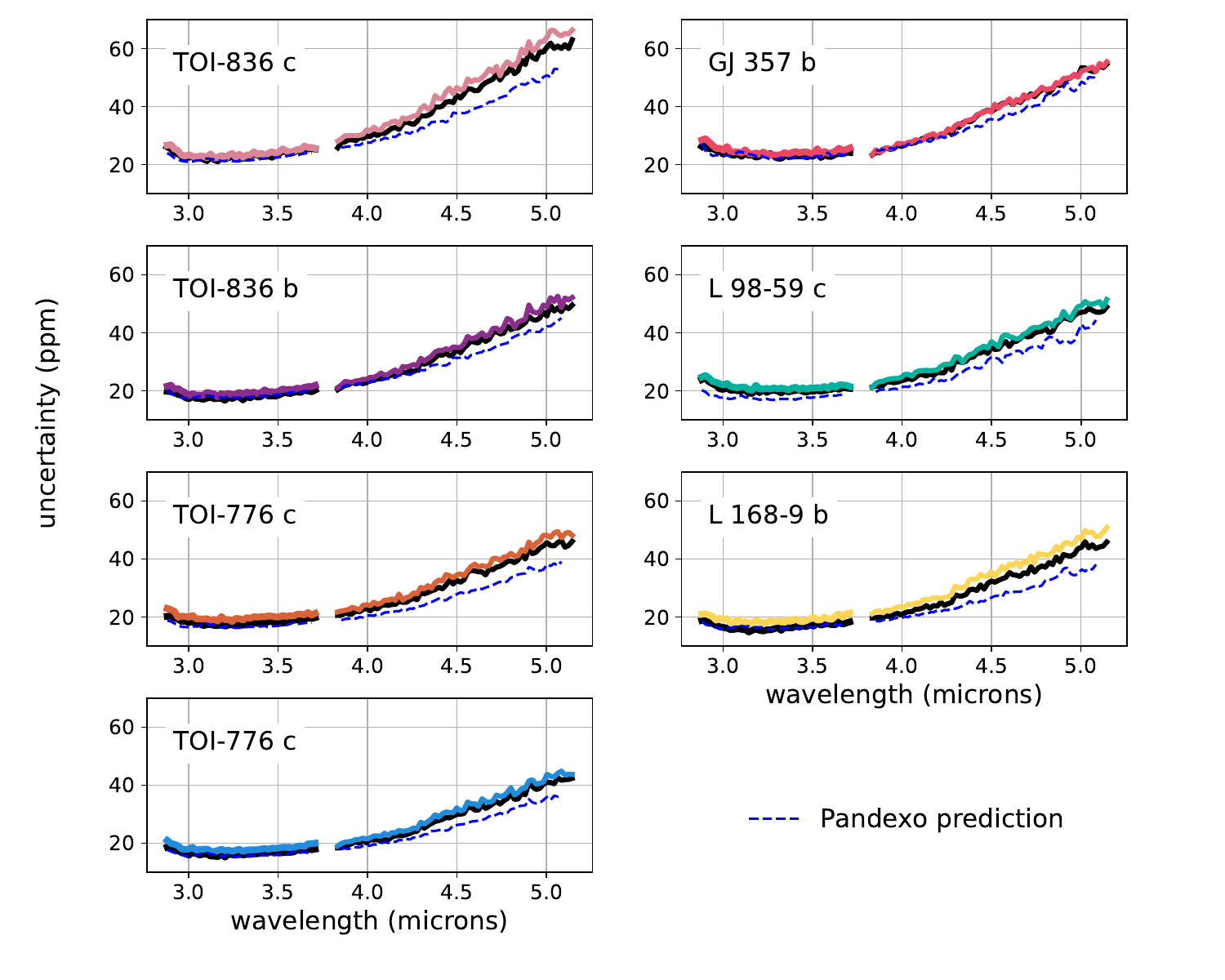}
            \caption{Comparison between precision of the spectra produced with the 0-vector systematics model (black line) and the 6-vector systematics model described above (colored lines). The blue dashed lines show the \texttt{pandexo} prediction, which is computed using the planetary parameters derived from our white lightcurve fits.}
            \label{fig:error_comparison}
        \end{figure*}
         
         We produce two transmission spectra for each planet: one with the same 6-vector systematics model used for the white light curves and one without the basis vectors that includes only the out-of-transit flux and the slope of the linear trend (the 0-vector model). Figure \ref{fig:pca_comparison} shows both versions of the transmission spectra along with the residuals between the two, and Figure \ref{fig:error_comparison} overplots the precision for both versions of the transmission spectra along with the \texttt{pandexo} predicted precision. In Table \ref{tbl:pandexo_comparison} we give approximate inflation factors to recover the observed precision for the transit depth given the \texttt{pandexo} prediction for the transit depth uncertainty for each version of the spectra. 
         
         We also experimented with optimizing the number of basis vectors in the systematics model for each spectral bandpass by testing every possible subset of the six basis vectors using a generalized least-squares fit and finding the subset that minimized the reduced $\chi^2$. We found that there was no subset of the six basis vectors for which the fitted model had a substantially improved reduced $\chi^2$ over the full 6-vector model, and we further found that the optimized systematics models resulted in spectra that were nearly identical to the 6-vector systematics model for every planet. 

         \begin{deluxetable*}{ccccccccc}[htb!]
        \tablehead{\colhead{target} & \colhead{NRS1 (0-vector)} & \colhead{NRS2 (0-vector)} & \colhead{NRS1 (6-vector)} & \colhead{NRS2 (6-vector)}}
        \startdata
            TOI-836\,c & 1.06  &  1.13  &  1.08  &  1.21  \\ 
            GJ\,357\,b & 1.02  &  1.06  &  1.07  &  1.07  \\ 
            TOI-836\,b & 1.00  &  1.07  &  1.07  &  1.13  \\ 
            L\,98-59\,c & 1.15  &  1.13  &  1.21  &  1.19  \\ 
            TOI-776\,b & 1.07  &  1.16  &  1.18  &  1.23  \\ 
            L168-9\,b & 1.00  &  1.15  &  1.15  &  1.27  \\ 
            TOI-776\,c & 1.04  &  1.12  &  1.13  &  1.18  \\ 
            \hline
            average & 1.05  &  1.12  &  1.13  &  1.18  \\ 
        \enddata
        \caption{Table of factors for converting between the average spectral precision predicted by \texttt{pandexo} and the average observed spectral precision for each detector for the spectrum obtained using each systematics model. The last row contains the average value of the inflation factor over all observations.}
        \label{tbl:pandexo_comparison}
    \end{deluxetable*}

         We found that the difference between the spectra obtained with each systematics model was only significant for three planets: GJ 357 b, TOI-836 b, and TOI-836 c. These targets are notable in that they were observed with three groups, while the remaining four targets were all observed with four or more groups. In Section \ref{sec:detector_systematics} we investigate the appearance of the systematics as a function of group number and demonstrate that the systematics are stronger for lower group numbers. For the other four planets in the sample, we determined that the use of the more complex 6-vector model for the spectral light curve fitting was not justified since it did not alter the final spectrum. We therefore elected to use the transmission spectra for the 6-vector systematics model for these three planets and the 0-vector systematics model for the remaining planets. The final versions of all spectra are shown in Figure \ref{fig:all_spectra_separate}. 

\section{NIRSpec Detector Systematics}
    \label{sec:detector_systematics}

\begin{figure*}[htb!]
        \centering
        \includegraphics[width=0.48\textwidth]{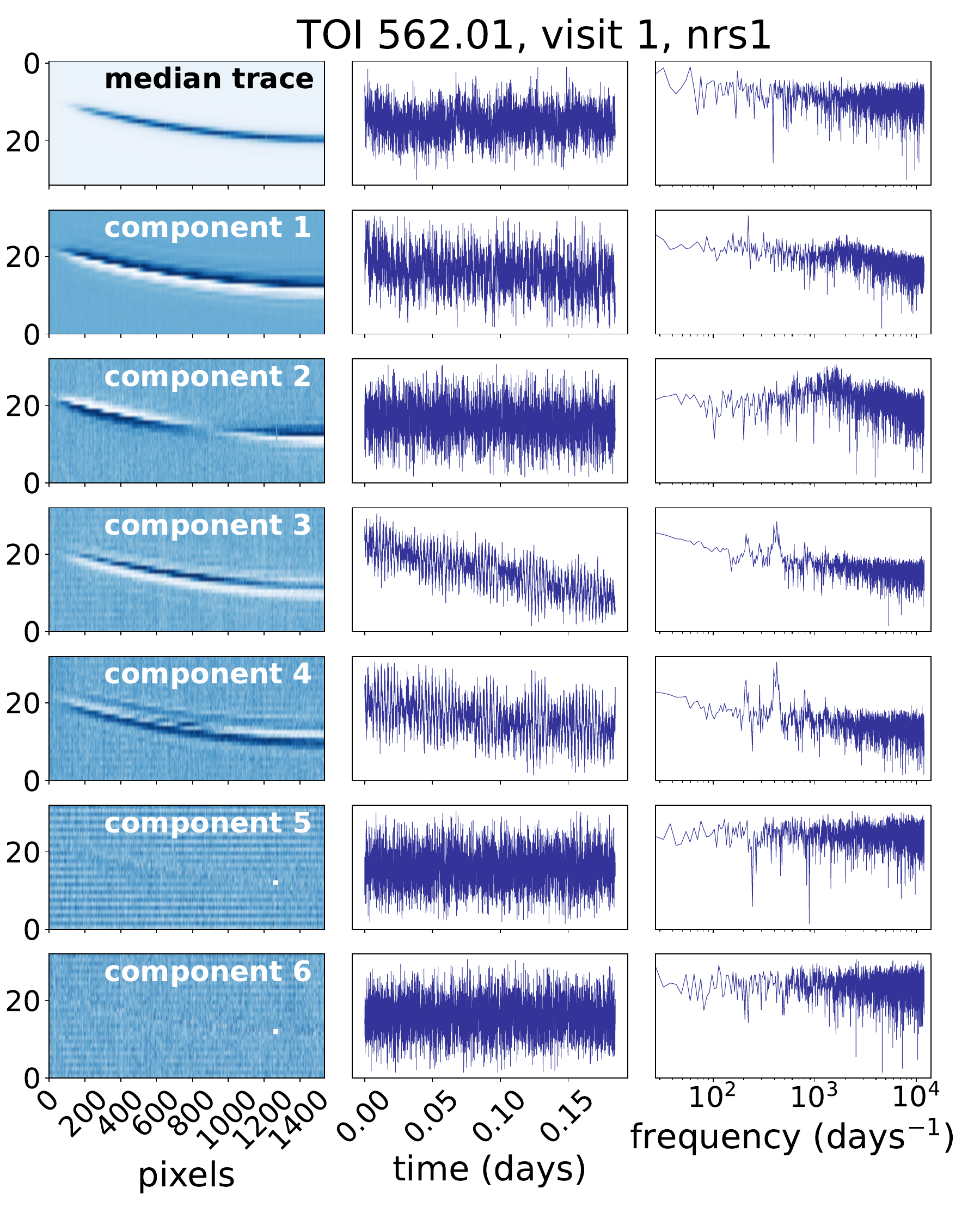}
        \includegraphics[width=0.48\textwidth]{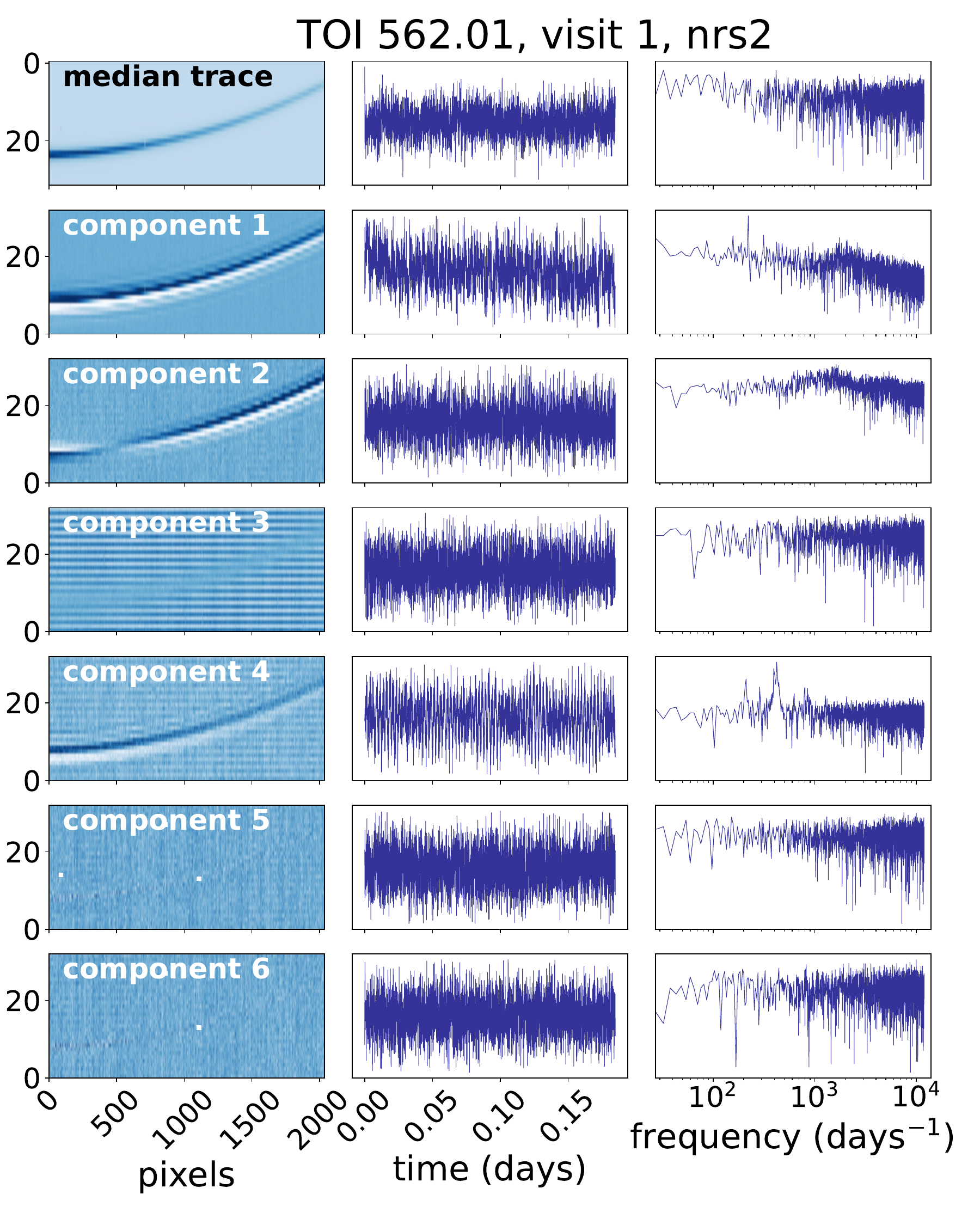}
        \caption{principal components of the relative pixel fluxes for the first visit of GJ\,357\,b for NRS1 (\textbf{left panel}) and NRS2 (\textbf{right panel}). \textbf{Left column of each panel}: Map of correlation coefficients between individual pixel fluxes and each principal component in the detector plane. Lighter colors indicate pixels which have a high positive correlation with the principal component and dark colored pixels have a large negative correlation. \textbf{Center column of each panel}: principal components of the relative pixel fluxes. \textbf{Right column of each panel}: Power spectral density of the principal component. The full figure set (27 images) is available in the online journal.}
        \label{fig:pca_single}
    \end{figure*}

    We now investigate the origin of the principal component vectors that form the basis for our systematics model. Throughout this discussion we use the single transit of GJ\,357\,b as an illustrative example. The first six principal components for this observation/detector are shown in the right column of both panels in Figure \ref{fig:pca_single}. 
    We compute the power spectral densities (PSDs) for each component, which we use to identify common frequencies at which the power spectrum peaks. These are plotted in the center column of the same figure. The PSD for the second component in both detectors has a wide peak corresponding to a one-minute timescale, while the third and fourth components in NRS1 and the fifth component in NRS2 have sharp peaks corresponding to 3.66 minutes, a second peak at a slightly shorter frequency, and a hint of additional peaks at harmonics of these frequencies.

    For each principal component, we compute the Pearson correlation coefficient between the component and the relative flux in each pixel. We map the correlation coefficients across the detector plane in the leftmost column of Figure \ref{fig:pca_single}. These correlation maps demonstrate that the components originate from changes in the shape and position of the trace on the detector. To see this, we examine the first component for the example of GJ 357 b. The lower edge of the trace has a lighter color, indicating that the pixels in this region are positively correlated with the principal component vector. In other words, the first principal component is tracking the illumination of these pixels. Meanwhile, the upper edge of the trace is darker in color indicating a negative correlation: the principal component vector increases when these pixels are less illuminated. From this we conclude that the first principal component arises from the shifting of the trace in the y-direction, possibly with some contribution from shifts in the x-direction due to the curvature of the trace. 

    Following this same logic, the second principal component must be associated with the rotation of the trace. To the left of the $\sim800^\mathrm{th}$ pixel column this component is tracking an upward motion in the trace, and to the right of the $\sim800^\mathrm{th}$ pixel column this component tracks a downward shift. The third and fourth PCA vector both appear to originate from a periodic de-focusing of the trace, as evidenced by the switch in the sign of the correlation as we move from the edge of the trace to its center. The correlation map for the fourth PCA vector shows a similar structure, except that the sign of the correlation changes twice as we move from the edge of the trace to the center between the 500th and 1000th pixel columns, suggesting that this is related to a higher-order change in the shape of the trace's cross section. The map for the fifth component shows faint vertical striping, pointing towards residual 1/f noise (this striping appears in the fourth and 6th components as well, indicating that the correspondence between the source of some variability and the principal components is not one-to-one). Both the fifth and sixth components show pattern of alternating positively and negatively correlated pixel rows, suggesting that this signal is related to alternating column noise (ACN) \citep{Rauscher2015}. Note that the engineering and science reference frames are rotated ninety degrees relative to each other, so that the ``columns'' referenced in ``alternating column noise'' are actually seen as rows in the science data. 

    Based on our PCA analysis, we have identified the following as potential sources of systematics in NIRSpec/G395H observations: 
    \begin{enumerate}
        \item \textbf{Shifting} in the position of the trace (component 1 in Figure \ref{fig:pca_single}).
        \item \textbf{De-focusing} of the trace, including both the first-order widening of the trace and a higher-order change in the trace profile seen in some of the COMPASS observations (e.g. components 3 and 4 in Figure \ref{fig:pca_single}). 
        \item \textbf{Rotating} of the trace (component 2 in Figure \ref{fig:pca_single}).
        \item \textbf{ACN} or alternative column noise, which is not related to the trace shape (components 5 and 6 in Figure \ref{fig:pca_single}). 
        \item Residual \textbf{1/f noise} (seen in component 5 of Figure \ref{fig:pca_single} alongside the alternating column noise).  
    \end{enumerate}

    The full set of components identified above does not appear in every single visit and detector. The most common components are the shifting component, which appears universally, and the de-focusing, rotation, and ACN components, which appear in the majority of observations. In addition to these common components, some visits and detectors display additional variability that typically arises from a small subset of pixels that do not show oscillatory behavior. Examples of these components are the third and sixth components from the NRS2 detector in Figure \ref{fig:pca_single}. The origin of these systematics is highly uncertain, but the discontinuities seen in the sixth component of NRS2 may be related to the ``popcorn mesa'' noise studied by \cite{Rauscher2004}, in which individual pixels appeared to switch between high and low voltage states during testing of a prototype of the NIRSpec detectors. In Figure \ref{fig:pca_single} the white squares mark the locations of pixels highly correlated with the corresponding principal component. In most cases these pixels do not coincide with the trace and therefore are likely to have a negligible effect on the extracted light curves. It may be possible to identify and remove these pixels prior to spectral extraction, however this is a topic that we leave to future work. 

\begin{figure*}
        \centering
        \includegraphics[width=1\textwidth]{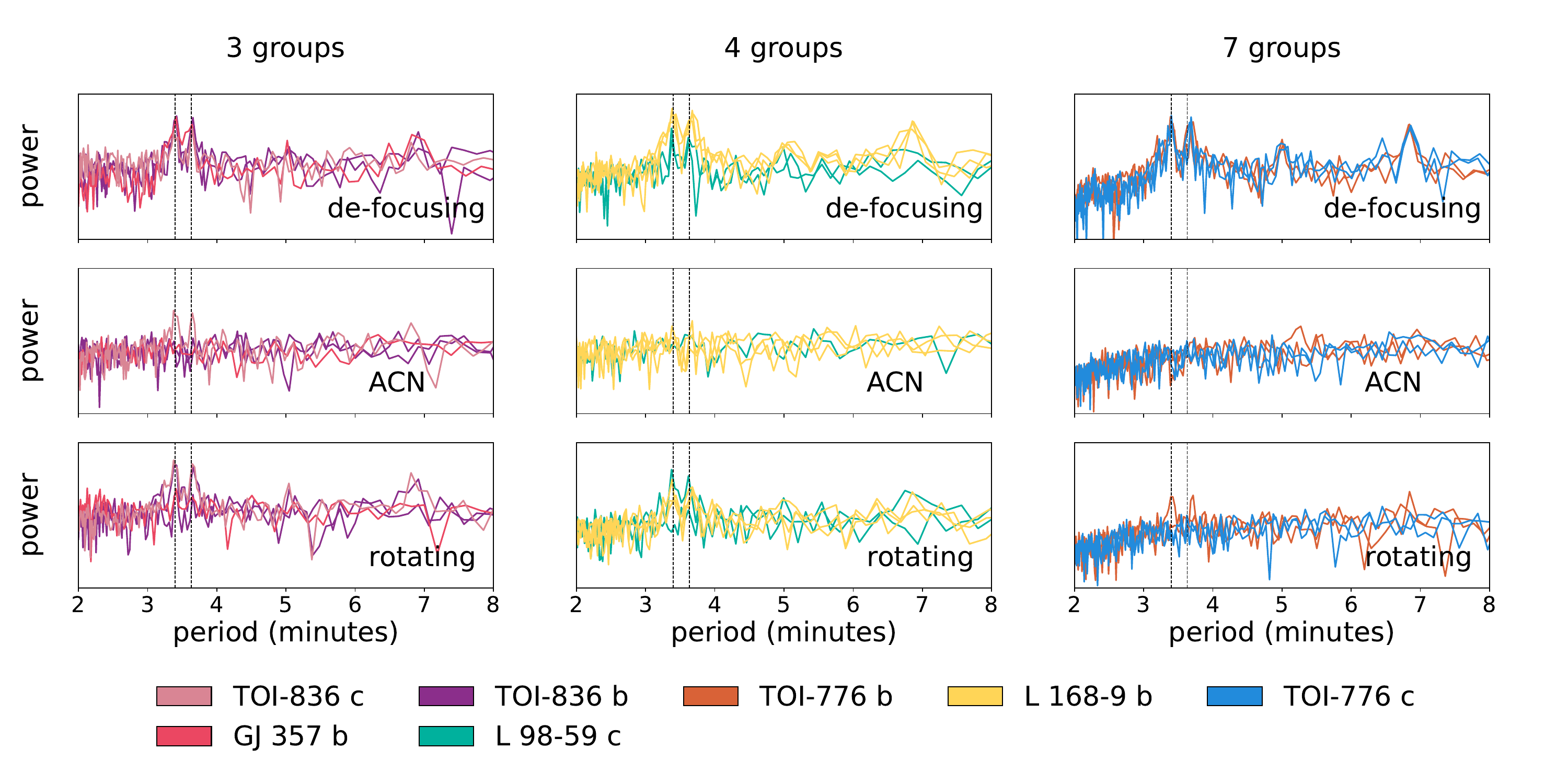}
        \caption{Superimposed power spectra for the principal components of the NRS1 RPF timeseries for all visits separated by three group observations (\textbf{left}), four group observations (\textbf{middle}), and seven group observations (\textbf{right}). The most prominent peaks in the power spectra are identified with vertical dashed lines at 3.4 and 3.6 minutes. These peaks do not appear in observation. The colors map to the colors for each planet's spectrum shown in Figure \ref{fig:all_spectra_separate}.}
        \label{fig:all_pca_psds_nrs1}
    \end{figure*}

\begin{figure*}
        \centering
        \includegraphics[width=1\textwidth]{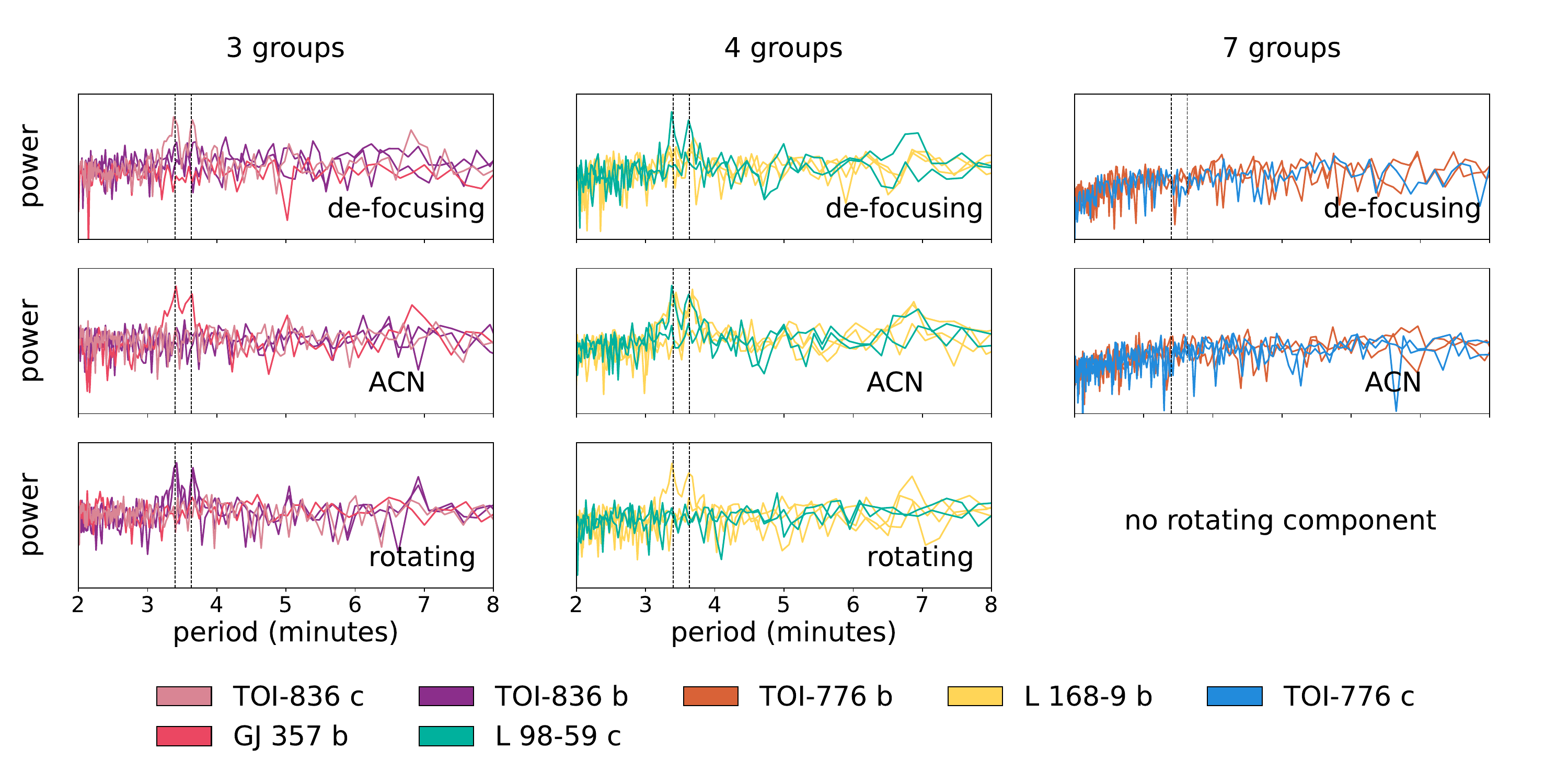}
        \caption{Superimposed power spectra for the principal components of the NRS2 RPF timeseries for all visits separated by three group observations (\textbf{left}), four group observations (\textbf{middle}), and seven group observations (\textbf{right}). The most prominent peaks in the power spectra are identified with vertical dashed lines at 3.4 and 3.6 minutes. These peaks do not appear in every observation. The colors map to the colors for each planet's spectrum shown in Figure \ref{fig:all_spectra_separate}.}
        \label{fig:all_pca_psds_nrs2}
    \end{figure*}

    In Figures \ref{fig:all_pca_psds_nrs1} and \ref{fig:all_pca_psds_nrs2} we show superimposed power spectra of the principal components associated with de-focusing, ACN, and rotation, for all visits separated by number of groups. The colors of the power spectra are unique for each target and map to the colors used in Figures \ref{fig:all_spectra_separate} and \ref{fig:pca_comparison}. By eye, we identify oscillations on approximate timescales of 3.4 and 3.63 minutes that are strongest for the de-focusing and rotation components, but which also appear in the ACN component in several cases. In general, it is less common to see prominent peaks in the ACN component than the other components, suggesting that this effect mostly contributes to the white noise. 
    
    A strong candidate for the oscillations on 3.4 and 3.63-minute timescales is thermal cycling of the heaters in the ISIM Electronics Compartment, which according to \cite{Rigby2024} results in ``a semi-periodic variation primarily in astigmatism'' on timescales of 2-4 minutes. This corresponds well to the observed timescales of the oscillations and explains their prominence in the de-focusing component, which would correspond to the astigmatism discussed in \cite{Rigby2024}. Why these oscillation timescales also appear in the rotation component is unclear, as is the origin of the rotation itself. 

    All of the oscillation timescales that we observe are much shorter than a typical transit duration, with the result that they have only a small impact on the measurement of the transit depth. However, they can have a significant impact on transit timing and duration measurements, because they occur on timescales close to those of the transit ingress and egress. 

    \subsection{Dependence on Group Number}

\begin{figure*}
            \centering
            \includegraphics[width=1\textwidth]{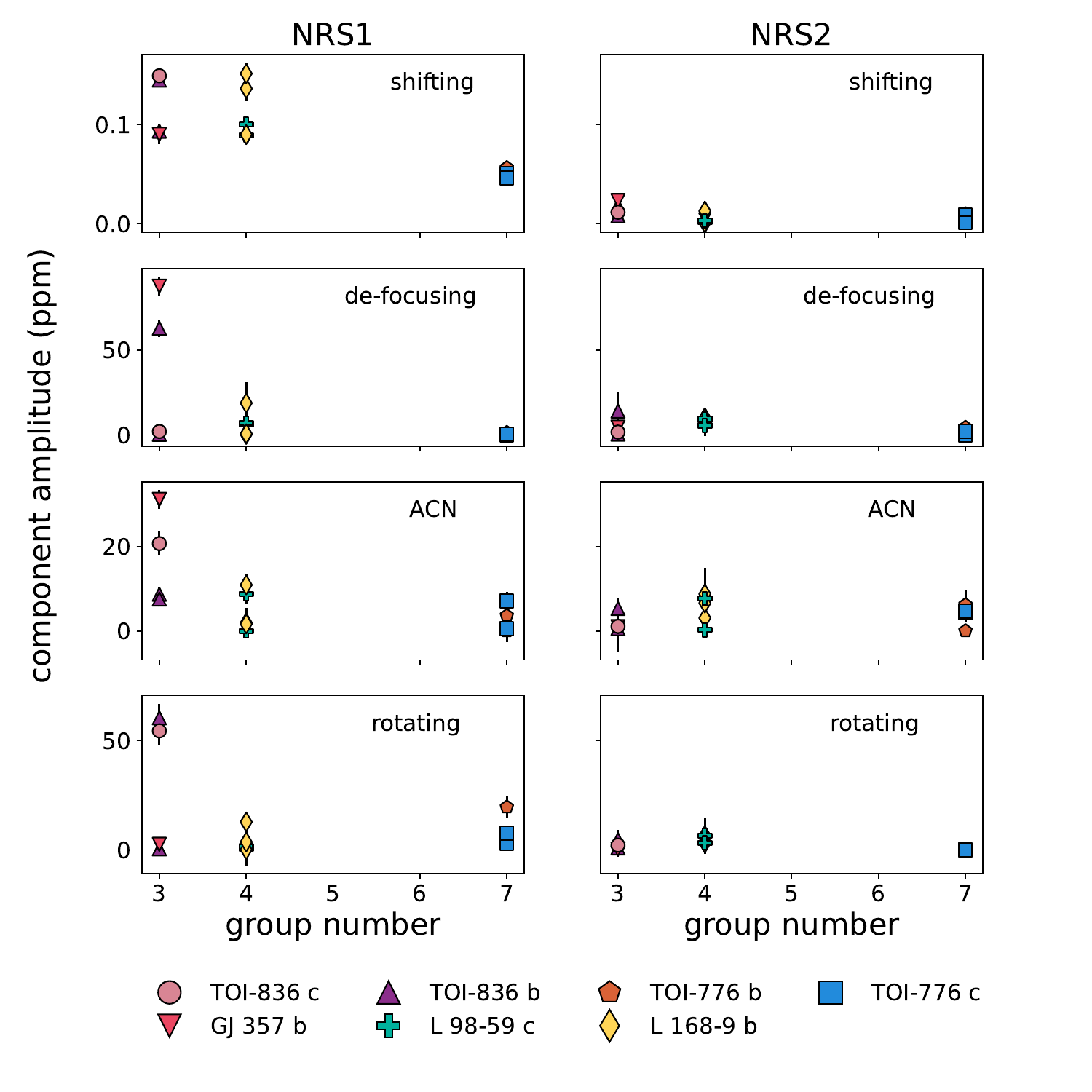}
            \caption{Maximum-likelihood amplitudes and 1-$\sigma$ error bars for the PCA components associated with each mode as a function of the number of groups for the observation. While there is significant scatter in the component amplitudes at low group numbers, we observe that most tend to decrease for larger numbers of groups and are generally smaller in NRS2 than NRS1. For some targets and visits we were not able to identify all modes, so some data points are missing.}
            \label{fig:pca_amplitudes_all}
        \end{figure*}

        By examining the white light curve posterior samples for the coefficients of each PCA basis vector we can examine the strength of each source of systematics as a function of the number of groups used for each observation. We examine the components associated with the first four sources of systematics above: shifting, de-focusing, rotating, and ACN. We neglect the term associated with residual 1/f noise because this component is not prominent in most observations. Examining Figure \ref{fig:pca_amplitudes_all} we first note that the contributions from all components tends to be stronger in NRS1 than NRS2. This is consistent with the observation that white light curves for NRS1 display more red noise than NRS2, as supported by Figure \ref{fig:allan_var_all}. The amplitude of the shifting component is small compared to the others for both detectors. 

        For NRS1, both the average amplitude of each component and the scatter in the amplitudes between targets and visits decreases as we move from three groups to four groups. The amplitude of the shifting and de-focusing components decreases further as we move to seven groups, while it appears unchanged for the ACN and rotating components between four and seven groups. These results confirm that low group number observations should be treated with caution, as has been noted in previous studies of datasets with small group numbers \citep{Alderson2024, Sikora2025}. Given that the de-focusing component reflects systematics on 2-4 minute timescales, this may be especially important for analyses that depend on the shape of the transit ingress and egress, such as transit timing/duration variations, limb-darkening measurements, investigations into planetary and/or stellar oblateness. 

        Recently, \cite{Luque2024} applied a similar ICA-based systematics model to phase curve observations of TOI-1685\,b taken with 16 groups and found that the model was unsuccessful at removing correlated noise. Their Figures 7 and 8 show their ICA basis vectors and associated Lomb-Scargle periodograms, which do not exhibit the same periodic behavior we observe in our PCA basis vectors. This additional datapoint supports the notion that systematics related to time-variable trace morphologies are less of any issue for observations using high numbers of groups per integration. 

    \subsection{Wavelength Dependence}

\begin{figure*}[htb!]
            \centering
            \begin{subfigure}[t]{0.45\textwidth}
                \centering
                \includegraphics[width=1.0\linewidth]{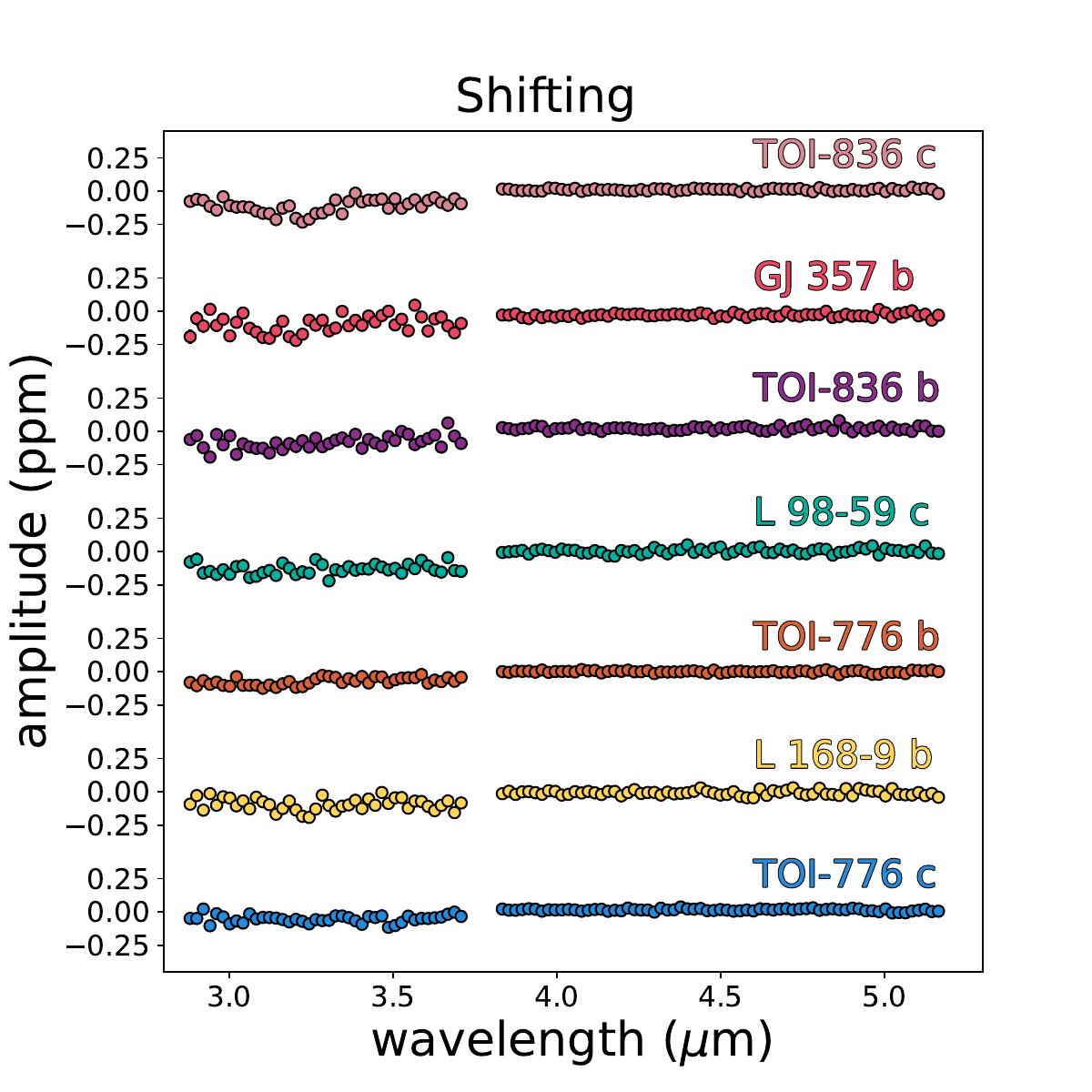} 
            \end{subfigure}
            \begin{subfigure}[t]{0.45\textwidth}
                \centering
                \includegraphics[width=1.0\linewidth]{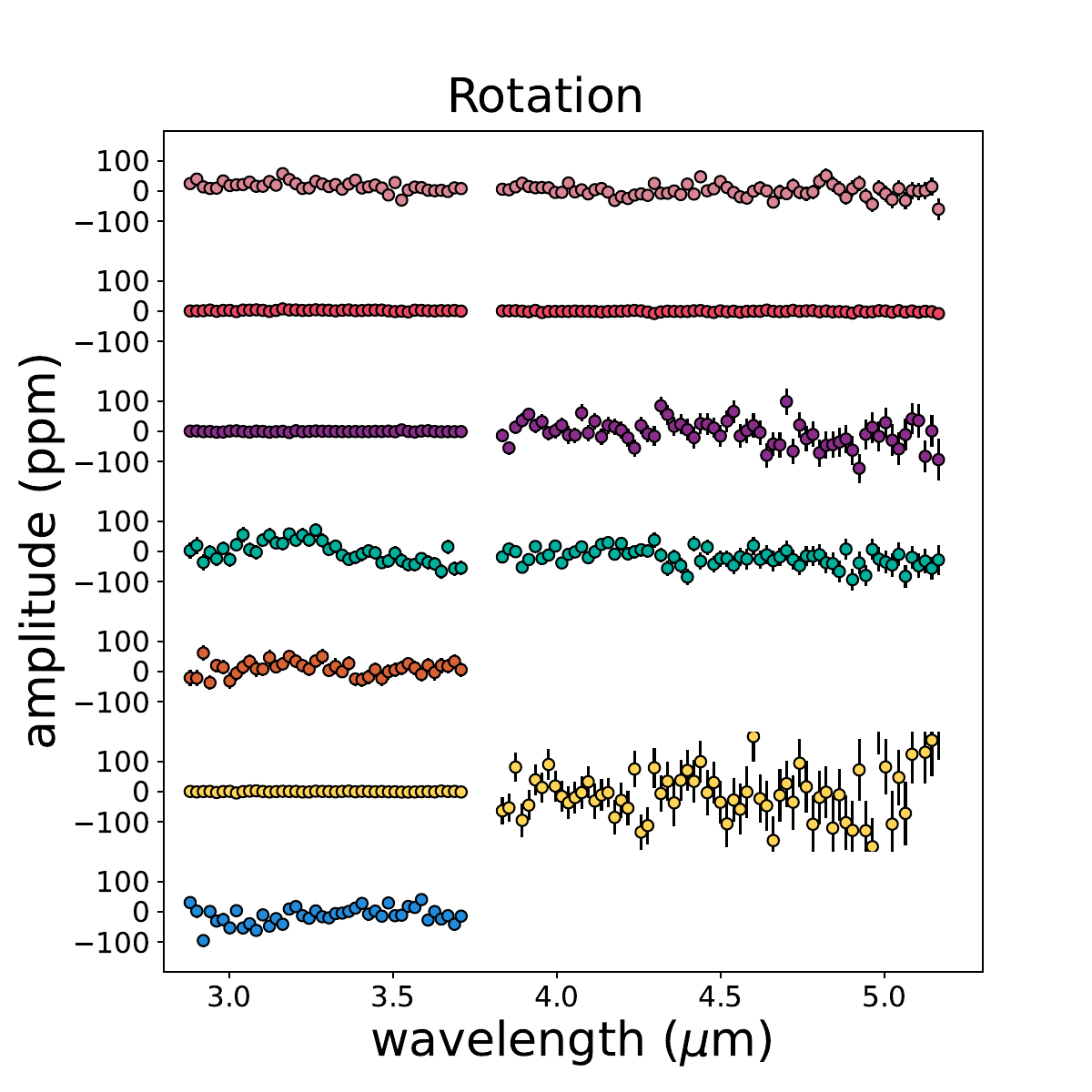} 
            \end{subfigure}
            \begin{subfigure}[t]{0.45\textwidth}
                \centering
                \includegraphics[width=1.0\linewidth]{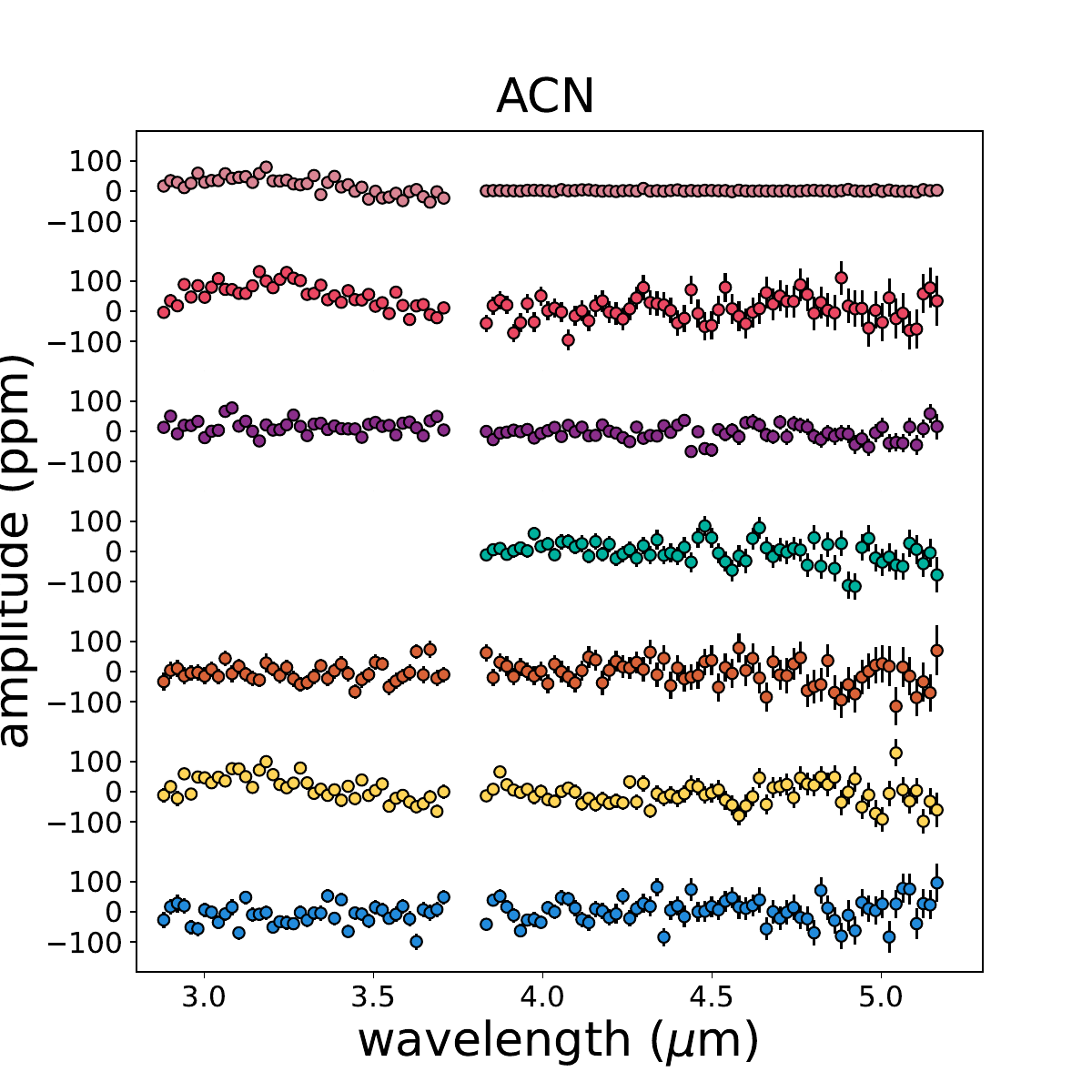} 
            \end{subfigure}
            \begin{subfigure}[t]{0.45\textwidth}
                \centering
                \includegraphics[width=1.0\linewidth]{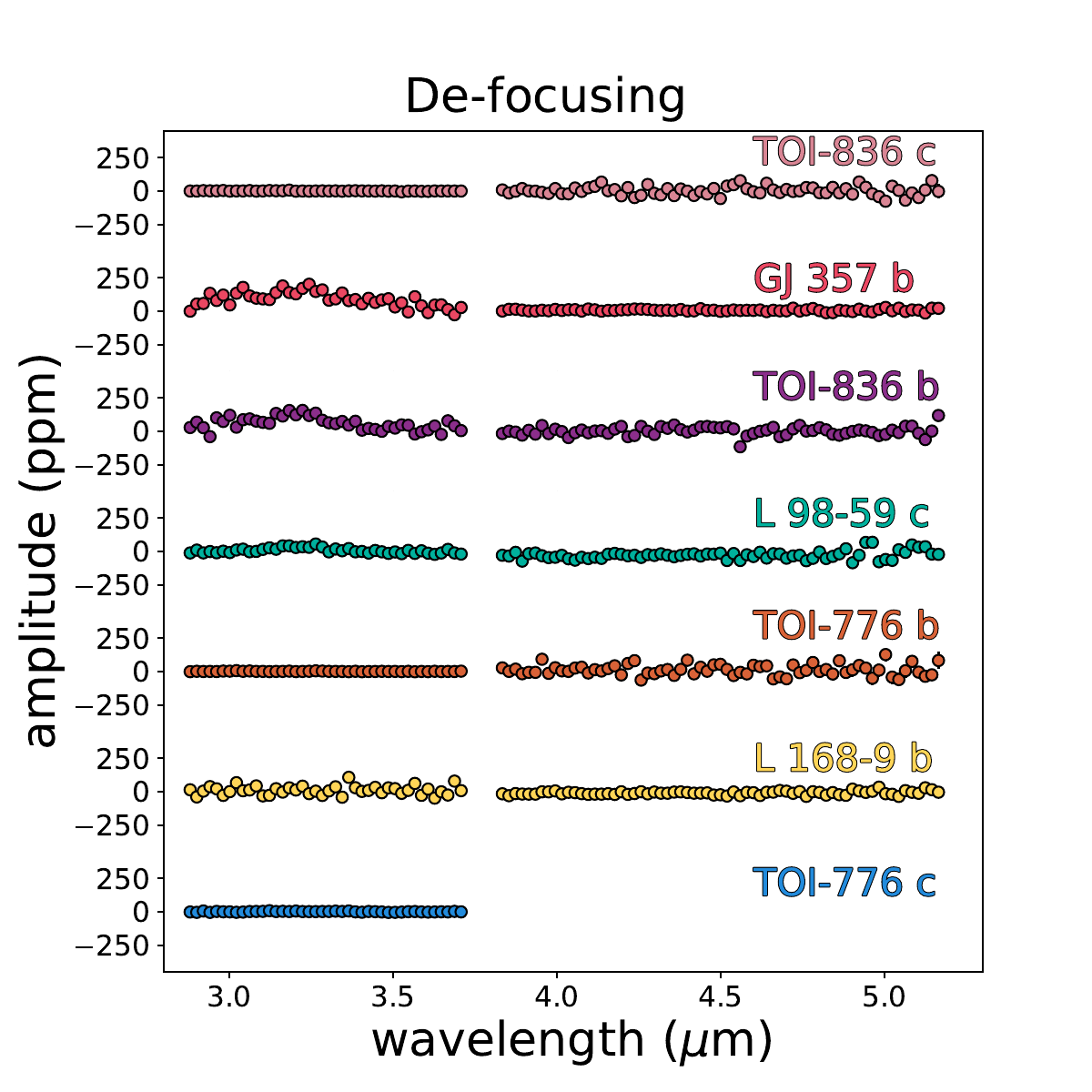} 
            \end{subfigure}
            \caption{Median estimates of the amplitudes of the PCA basis vectors as a function of wavelength for the first visit of each target. Missing data indicates that we were not able to attribute a basis vector to the corresponding mode for that dataset. \textbf{Upper left:} Amplitudes for the shifting mode. Because the trace position is variable during the spectral extraction, this component contributes very little to the red noise in the final light curve, as evidenced by the small amplitudes for all targets. \textbf{Upper right:} The rotation mode. \textbf{Lower left:} Alternating column noise (ACN). \textbf{Lower right}: The defocusing mode.}
            \label{fig:pca_wav_specs}
        \end{figure*}

        We now explore the wavelength dependence of the NIRSpec systematics using the posterior samples from our MCMC analysis of the spectral light curves. For each wavelength bin we take the median of the coefficient for each principal component in the systematics model. We then multiply this by the RMS amplitude of the principal component vector itself to obtain the amplitude of that component. In Figure \ref{fig:pca_wav_specs} we show the wavelength dependence of the amplitudes for the shifting, de-focusing, rotating, and ACN-associate components for each target. 

        The components that do show wavelength dependence have a common profile, in which the absolute value of the amplitude of the component increases from 2.8 to 3.2\,$\mu$m where it peaks before falling back to zero around 3.5\,$\mu$m. This behavior is most prominent in the ACN component, but also occurs in the rotation and de-focusing components. 

        The fact that the systematics are strongest in NRS1 aligns with the observation that the white light curves from NRS1 display more systematics than NRS2. It appears that the systematics seen in the white light curve are due to the behavior of the detector between 2.8 and 3.5 $\mu$m, which is exactly the region for which a discrepancy is seen between the spectra extracted with the 6-vector systematics model and those extracted with the 0-vector systematics model in Figure \ref{fig:pca_comparison}. We therefore caution that G395H transmission spectra may display spurious features in this wavelength range when higher-order changes in trace morphology, beyond positional shifts, are not accounted for. As previously discussed, this is particularly true for observations with smaller number of groups per integration, but Figure \ref{fig:pca_wav_specs} also shows that observations with larger group numbers can show wavelength-dependent systematics (see the rotation component for TOI-776 c and the ACN component for L 168-9 b).

    \subsection{Nonlinearity and the Brighter-Fatter Effect}

\begin{figure}
            \centering
            \includegraphics[width=0.5\textwidth]{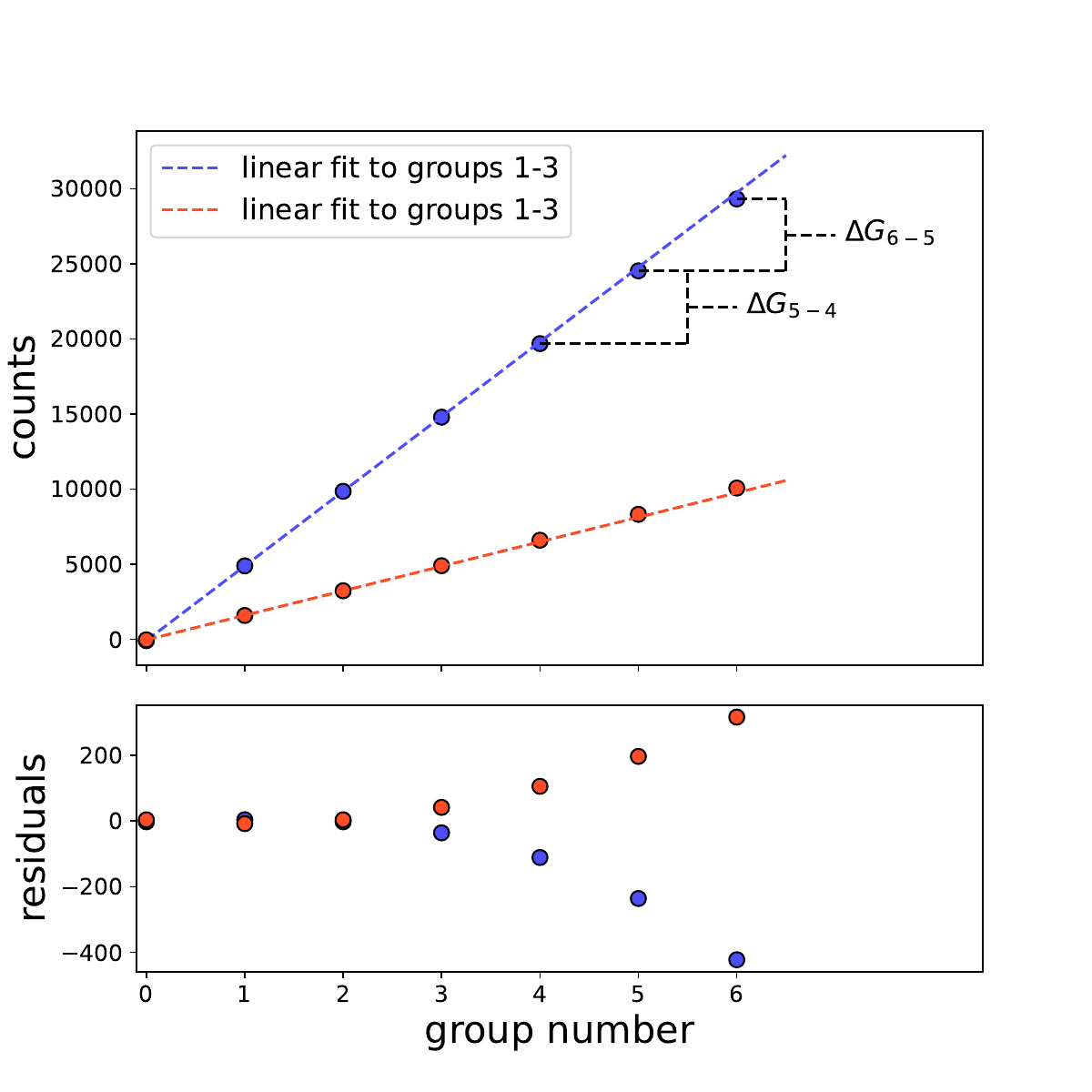}
            \caption{\textbf{Top:} Counts as a function of group number for a pixel near the center of the trace (blue) and an adjacent pixel (orange). The dashed line is a linear fit to the first three groups. \textbf{Bottom:} The residuals from the linear fit to the first three groups for each pixel. Note the nonlinearity at higher group numbers, with the pixel near the center of the trace showing a deficit of counts compared to the expectation for a linear accumulation of photoelectrons and the adjacent pixel showing an excess.}
            \label{fig:bfe_pixels}
        \end{figure}

        While the systematics model we have implemented is effective at mitigating correlated noise on short timescales, Figure \ref{fig:allan_var_all} shows that the model does not account for all of the observed red noise. Residual red noise can also be seen in Figure \ref{fig:wlc_single} and the accompanying figure set. One possible culprit is the brighter-fatter effect (BFE). BFE is a phenomenon affecting infrared detectors wherein the accumulation of photoelectrons in a pixel near the center of an under-sampled PSF causes changes in the shape of the electric potential and repulses new incoming photoelectrons to adjacent pixels. In addition to causing the PSF to become fatter in brighter regions of the detector, BFE causes the charge as a function of groups to become nonlinear as the pixel gets closer to saturation (see Figure \ref{fig:bfe_pixels}). The brighter-fatter effect has been studied for the MIRI Si:As IBC detectors \citep{Argyriou2023, Gasman2024}, for WFIRST's H4RG-10 detectors \citep{Hirata2020}, and for Teledyne H4RG detectors \citep{Plazas2018} and the Teledyne H2RG detectors used by JWST's NIRISS and NIRSpec instruments \citep{Goudfrooij2024, Alam2026}. \cite{Alam2026} demonstrates that the brighter-fatter effect can acquire time-dependence in timeseries observations due to its dependence on the illumination of the detector. However the degree to which this effect impacts transit depth measurements is still understudied.

\begin{figure*}
            \centering
            \includegraphics[width=1\textwidth]{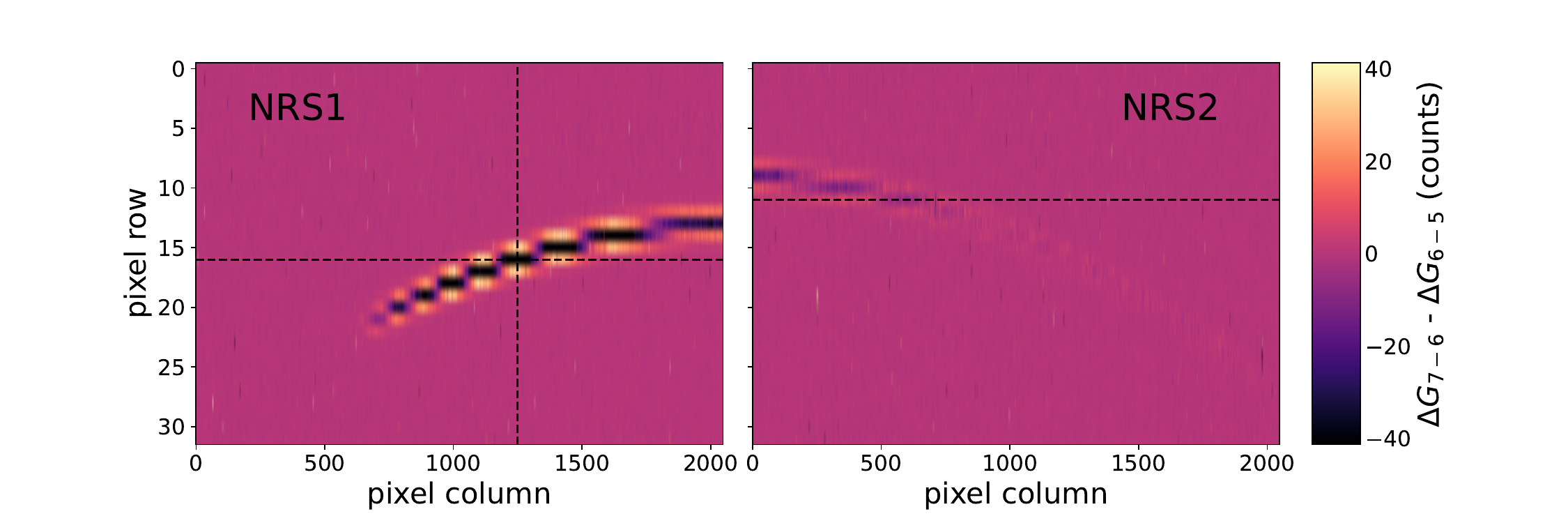}
            \caption{Approximate second derivative at the sixth group of the median integration for each pixel on the two NIRSpec detectors for the first visit of TOI-776\,c. The dashed horizontal lines indicate the rows used to compute Figure \ref{fig:bfe_groups} and the dashed vertical line in NRS1 indicates the column from which the pixels highlighted in Figure \ref{fig:bfe_pixels} were taken.}
            \label{fig:bfe_map}
        \end{figure*}

\begin{figure}
            \centering
            \includegraphics[width=0.5\textwidth]{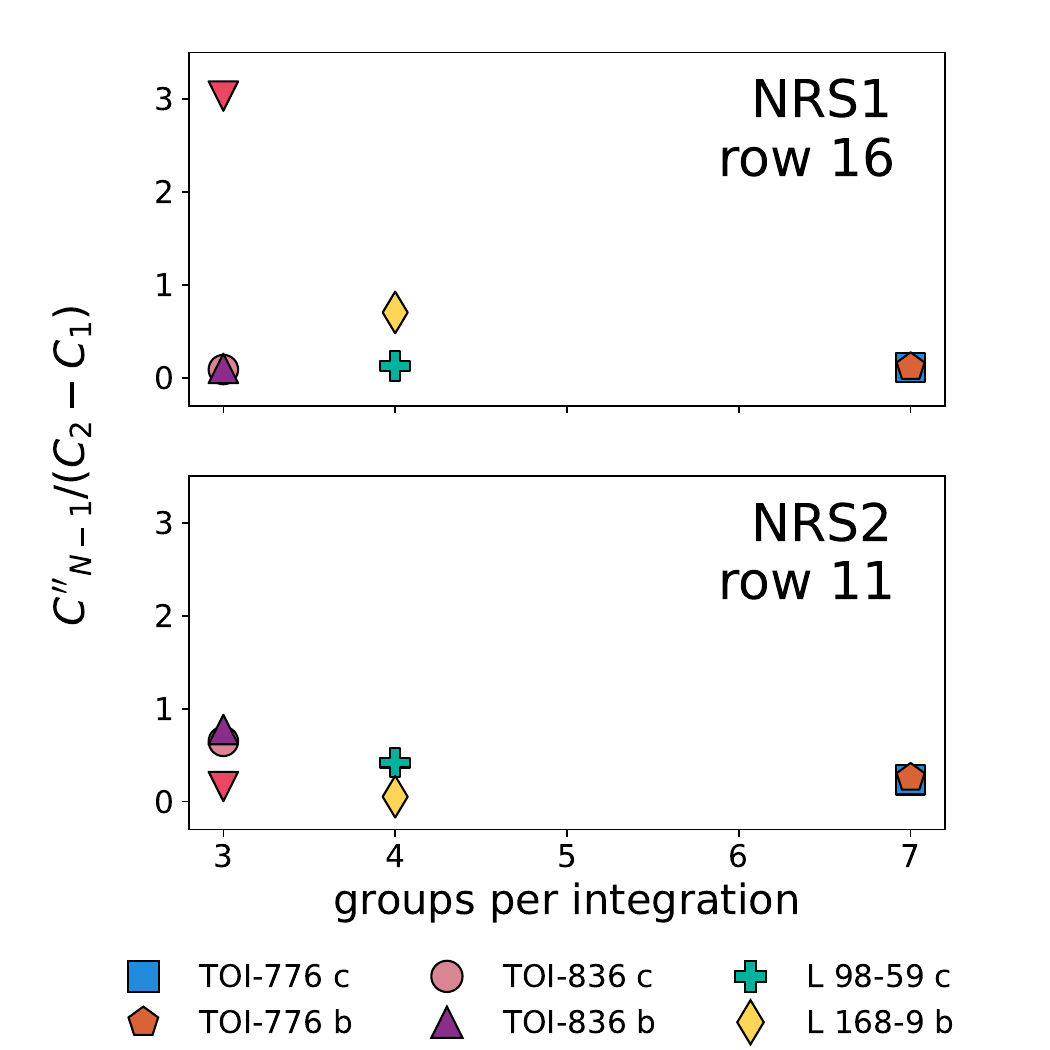}
            \caption{Approximated second derivative at the end of the ramp normalized by the slope at the beginning of the ramp for the median integration as a function of the number of groups.}
            \label{fig:bfe_groups}
        \end{figure}

        We investigate the presence of the brighter-fatter effect in our data using the observations from the second visit of TOI-776 b. Because these observations used seven groups, we are able to use the first several groups for which the accumulation of counts is relatively linear as a baseline to explore the nonlinearity of the last several groups. Starting with detector images that that have been destriped at the group level, we construct a median integration by taking the median detector image for each group across all integrations in the observation. Figure \ref{fig:bfe_groups} shows the median accumulated counts as a function of group for two pixels on the NRS1 detector: one near the center of the trace (blue) and its neighboring pixel (red). The location of this central pixel is identified on the NRS1 detector in Figure \ref{fig:bfe_map} by the intersection of the horizontal and vertical dashed lines. To highlight the nonlinearity caused by the brighter-fatter effect we have fit a linear model to the first three groups and the examined the residuals between this model and all seven groups. These residuals are shown in the bottom panel of the figure. For the central pixel we see that the counts start out linear, but fall increasingly below the linear model as the integration proceeds. For its neighboring pixel the counts begin linear and then fall increasingly above the linear model. This is the expected behavior for the brighter-fatter effect: As the charge accumulates in the central pixel, photoelectrons that would otherwise fall on that pixel are diverted to neighboring pixels resulting in an excess of flux for the neighbors and a deficit for the central pixel which only gets worse as the integration proceeds.

        We construct a diagnostic of the BFE for each pixel, denoted $C''_{N-1}$ which approximates the second derivative of the counts ($C_i$) as a function of group number ($i$) at the location of the second to last group ($C_{N-1}$ where $N$ is the number of groups for the observation). In Figure \ref{fig:bfe_map} we show how this value varies across the NRS1 and NRS2 detectors. We see that the approximated second-derivative is negative near the center of the trace in both NRS1 and NRS2, and then becomes positive at the edges of the trace, again as expected given our understanding of the brighter-fatter effect. Notably, the effect appears much stronger in NRS1 than NRS2.

        We compute the difference between the maximum and minimum values of $C''_{N-1}$ for detector row 16 in NRS1 and row 11 in NRS2 for the first visit of each target. We normalize this value by the median of $C_{2} - C_{1}$, the average slope of counts between the first and second groups, taken over all pixels. This normalization is necessary if we wish to compare charge migration between different targets, because higher count rates from brighter stars will always result in more charge migration between groups. We choose to normalize by the slope at the beginning of the integration because the ramp is expected to be linear at this point, although the expectation of linearity at the beginning of the count accumulation ramp may break down for observations with very small group numbers. These values, which are plotted in Figure \ref{fig:bfe_groups}, indicates the extent to which the brighter-fatter effect impacts the linearity of the up-the-ramp samples for each observation. The first thing we note is that the nonlinearity varies more from observation to observation for NRS1 than NRS2. It also appears to be the case that the observations with seven groups experience less, and more consistent, charge migration than those observations with three or four groups. However, it is difficult to make precise statements about trends in charge migration with group number for such a small sample of observations.
        

        While a correction step to mitigate the brighter-fatter effect has been implemented for NIRISS in the JWST calibration pipeline \citep{Goudfrooij2024}, there is as of yet no correction for NIRSpec. While a deeper exploration of the impacts of the brighter-fatter effect on transit depth measurements and ultimately the transmission spectrum itself is beyond the scope of this work, it is clear that the brighter-fatter effect impacts merits dedicated investigation in future work.

\section{Analysis of Spectra}
    \label{sec:spec_analysis}

    To arrive at a final set of spectra for our analysis, we select the spectra corresponding to the 6-vector model for the three-group observations (GJ\,357\,b, TOI-836\,b, and TOI-836\,c) for which correlated residuals were observed between the 0-vector model and the shifts-only model in Figure \ref{fig:all_spectra_separate} and for which the Bayesian Information Criterion demonstrated a preference for the 6-vector model in the white light curve analysis. For the remaining four planets (TOI-776\,b, TOI-776\,c, L\,98-59\,c and L\,168-9\,b) we select the spectra obtained with the 0-vector model, since the BIC did not justify the use of the more complex model for their white light curves and the residuals in Figure \ref{fig:all_spectra_separate} appear to be Gaussian distributed. 

    In this section we first carry out least squares fits of flat line models with and without offsets between NRS1 and NRS2. We then use grids of thermochemical equilibrium models to determine limits on the composition of these atmospheres. Last, we analyze combined (or composite) spectra of the targets in order to search for emergent transmission features. 

    \subsection{Flat Spectral Models}
    
        We begin our analysis by considering a model consisting of a flat line (i.e. a featureless spectrum) with an optional offset between the NRS1 and NRS2 portions of the spectra. We carry out a two generalized least squares fits for each spectrum: one with an offset and one with the offset fixed to zero. Table \ref{tbl:flat_spectrum_fits} contains the best-fit offset parameter, reduced chi-square for the fit with and without an offset, and the difference in the Bayesian information criteria between the two cases. Following \cite{Raftery1995}, we find that the $\Delta_\mathrm{BIC}$ values indicate there is strong evidence for an offset for TOI-836\,c, GJ\,357\,b, TOI-776\,b, and TOI-776\,c, while there is moderate evidence for the model without an offset for the remaining targets. The reduced chi-square values indicate that a flat spectrum model, either with or without an offset, is an adequate model for every spectrum. This result is consistent with previous analyses of each target that found that no putative features in the spectra rose to the level of a detection. We therefore proceed with our analysis under the assumption that there are no detectable molecular features in any of the spectra we present in this work. 

    \begin{deluxetable*}{ccccccccc}[htb!]
        \tablehead{\colhead{target} & \colhead{offset (ppm)} & \colhead{$\chi^2/N$ no offset} & \colhead{$\chi^2/N$ offset} & \colhead{$\Delta_\mathrm{BIC}$ (no offset - offset)}}
        \startdata
                TOI-836 c  &  -46.33$\pm$5.99  &  1.87  &  1.10  &  21.96 \\
                GJ 357 b  &  -66.27$\pm$5.75  &  2.63  &  0.96  &  72.26 \\
                TOI-836 b  &  4.18$\pm$4.98  &  1.31  &  1.26  &  -3.94 \\
                L 98-59 c  &  -0.54$\pm$4.93  &  1.03  &  0.99  &  -2.70 \\
                TOI-776 b  &  43.07$\pm$4.58  &  3.08  &  1.23  &  57.63 \\
                L 168-9 b  &  14.16$\pm$4.32  &  1.38  &  1.27  &  -2.70 \\
                TOI-776 c  &  19.26$\pm$4.19  &  1.31  &  0.86  &  15.67 \\
        \enddata
        \caption{Results of a generalized least squares fit to the spectra presented in Figure \ref{fig:all_spectra_separate} for featureless models with and without an offset between NRS1 and NRS2. A larger positive $\Delta_{\mathrm{BIC}}$ values in the fifth column indicates a preference for the model with an offset, while a larger negative value would indicate a preference for the model without an offset. Applying the guidance of \cite{Raftery1995} we find strong evidence for an offset in cases where $\Delta_\mathrm{BIC}>10$, and moderate evidence for a model without an offset when $-10<\Delta_\mathrm{BIC}<-3.2$.  \label{tbl:flat_spectrum_fits}}
    \end{deluxetable*}

    \subsection{Thermochemical Equilibrium Modeling}

        Using our uniformly reduced multi-visit spectra, we carry out thermochemical equilibrium modeling following a methodology similar to that carried out previously for each target. The purpose of repeating this exercise is twofold: First, we wish to investigate the robustness of these lower limits to different choices made during the reduction and light curve fitting steps, and second, we wish to provide a set of lower limits derived from a uniform reduction and analysis. We begin by computing a grid of model atmospheres for each planet in terms of bulk metallicity and ``opaque pressure level'' \citep[e.g.,][]{Wallack2024}. The opaque pressure level is the pressure above which the atmosphere is clear, and below which all wavelengths of light are uniformly absorbed. In reality this could represent the top of a cloud deck, a haze layer, or a planetary surface. For our forward model we use \texttt{Picaso} \citep{Batalha2019} with equilibrium chemical abundances computed by \texttt{easyCHEM} \citep{Lei2024}. While we use the full set of molecules output by \texttt{easyCHEM} to calculate the mean molecular weight of the atmosphere, we include only CO, H$_2$O, CH$_4$, PH$_3$, CO$_2$, NH$_3$, H$_2$S, VO, TiO, Na, and K as absorbing species, with opacities from the Zenodo v2 database of Resampled Opacities \citep{Batalha2025}. We compute the equilibrium chemistry for C/O ratios of 0.5$\times$ Solar, 1.0$\times$ Solar, and 1.5$\times$ Solar \citep[where Solar C/O=0.55;][]{Asplund2009} in order to investigate the dependence of the resulting limits on C/O ratio. Table \ref{tbl:planetary_stellar_params} lists the planetary parameters that were used to compute our forward models. The forward models are computed at $R=150$ and then binned to the resolution of our data.

        Next, we interpolate over these model grids to produce smoothly varying spectra as a function of metallicity and opaque pressure level. We carry out an MCMC simulation for each planet with the metallicity relative to Solar $m$, logarithm of the opaque pressure level $\mathrm{log}(P_\mathrm{op})$, and an offset between the model atmosphere and the NRS1 and NRS2 segments of the spectrum, $A_\mathrm{NRS1}$ and $A_\mathrm{NRS2}$, as input parameters. For each set of input parameters we use the interpolated grid to produce a model spectrum, and then we compute the log-likelihood of the observed spectrum given the set of model parameters as
        \begin{equation}
            \label{eqn:likelihood}
            \mathrm{log}\ \mathcal{L} = -\frac{1}{2}\sum_i\left[(S_i-M_i)^2/\sigma_i^2 - \mathrm{log}(2\pi\sigma_i^2)\right],
        \end{equation} where $S_i$ takes on the value of the transit depth in the $i^\mathrm{th}$ wavelength bin, $\sigma_i$ is the 1-$\sigma$ uncertainty on the transit depth, and the model values $M_i$ are given by 
        \begin{equation}
            M_i = F_i(m, \log P_\mathrm{op}) + \begin{cases}
                A_\mathrm{NRS1}\ \mathrm{if}\ \lambda_i< 3.8\ \mu m \\
                A_\mathrm{NRS2}\ \mathrm{if}\ \lambda_i > 3.8\ \mu m
            \end{cases}
        \end{equation} where $F_i$ is the value of the model in the $i^\mathrm{th}$ wavelength bin and 3.8 $\mu m$ is an arbitrary cutoff chosen to fall in the detector gap between NRS1 and NRS2. The prior on metallicity is uniform between 10 and 1000$\times$ Solar, and the prior on $\mathrm{Log}(P_\mathrm{op})$ is uniform between -6 and -1. We do not place a prior on the offset parameters. 
    
        For each target we simulate 8 MCMC chains for 1,000,000 steps each, discarding the first 20,000 steps as burn-in (2\% of the total simulation). This results in an effective sample size of at least 80,000 for each parameter based on an analysis of the autocorrelation length and $\hat{R}$ parameters equal to 1.0 to three decimal places indicating that the simulations are well-converged, with both quantities computed using \texttt{arviz} \citep{arviz}. We see no evidence of multimodality in the posteriors.
    
        In addition to exploring the two dimensional metallicity/opaque pressure level posterior space, we also carry out a set of MCMC simulations for clear atmospheres in which we fix the opaque pressure level to 100 bar. Other than fixing the opaque pressure level, we follow the same procedure as for the previous simulations. 
        
        In order to control for the extent to which our methodology contributes to any disagreement between our results and those of previous COMPASS analyses, we also compute the same limits for metallicity and opaque pressure level using identical pipelines from the original works. As described in other works \citep[e.g.][]{Wallack2024, Teske2025, Alderson2025} we compute a similar model grid using \texttt{PICASO} and \texttt{photochem} \citep{Wogan2025} to compute spectra at chemical equilibrium abundances. We choose two values of C/O (0.5$\times$Solar and 1$\times$Solar) and 20 log-even spaced metallicities between $\log$M/H=[0,3]. We fit for an offset between NRS1 and NRS2 using \texttt{ultranest} \citep{ultranest}. Then, we compute a reduced $\chi^2$ per data point of the NRS1/NRS2 offset corrected spectrum, which we can then be converted to a p-value and associated $\sigma$-confidence. The effect of clouds is added to the spectrum by including a thick ($\tau=10$) gray opacity source at a specified pressure level. Our grid includes 5 opaque pressure levels spanning 1-10$^{-4}$~bars. 

    \subsection{Composite spectra}

        We now produce and analyze two composite spectra by summing over the transit depths in each wavelength bin for a collection of spectra. The error bars for each composite spectrum are similarly obtained by adding the individual errors in quadrature for each wavelength bin. The first composite spectra we produce is for the collection of super-Earths in our sample (GJ\,357\,b, L\,98-59\,c, L\,168-9\,b, TOI-836\,b and TOI-776\,b), and the second is for the two sub-Neptunes (TOI-776\,c and TOI-836\,c). As we did for the individual spectra, we compute a generalized least squares fit for a flat spectrum. Unlike for the individual spectra, however, we only consider fits that include an offset between NRS1 and NRS2. This is because the preference for an offset in at least some of the individual spectra implies that, since the offsets are not removed prior to computing the composite spectra, they should be included in our modeling. 

        The purpose of producing these composite spectra is to test the hypothesis that the individual spectra do contain features, but that these features are of such a low amplitude that they're entirely obscured by the white noise. If multiple of the spectra contained the same low-level features, however, we might expect that upon co-adding the spectra the white noise would bin down while the features would reinforce each other, resulting in a feature associated with one or more species that these atmospheres have in common. The division of the sample between super-Earths and sub-Neptunes is meant to reflect the expectation that sub-Neptunes and super-Earths may have very different atmospheric chemistries resulting such that their features would destructively interfere upon summing over the spectra.
        
\section{Atmospheric Composition Results}
    \label{sec:results}

\begin{figure*}[htb!]
        \centering
        \includegraphics[width=1\linewidth]{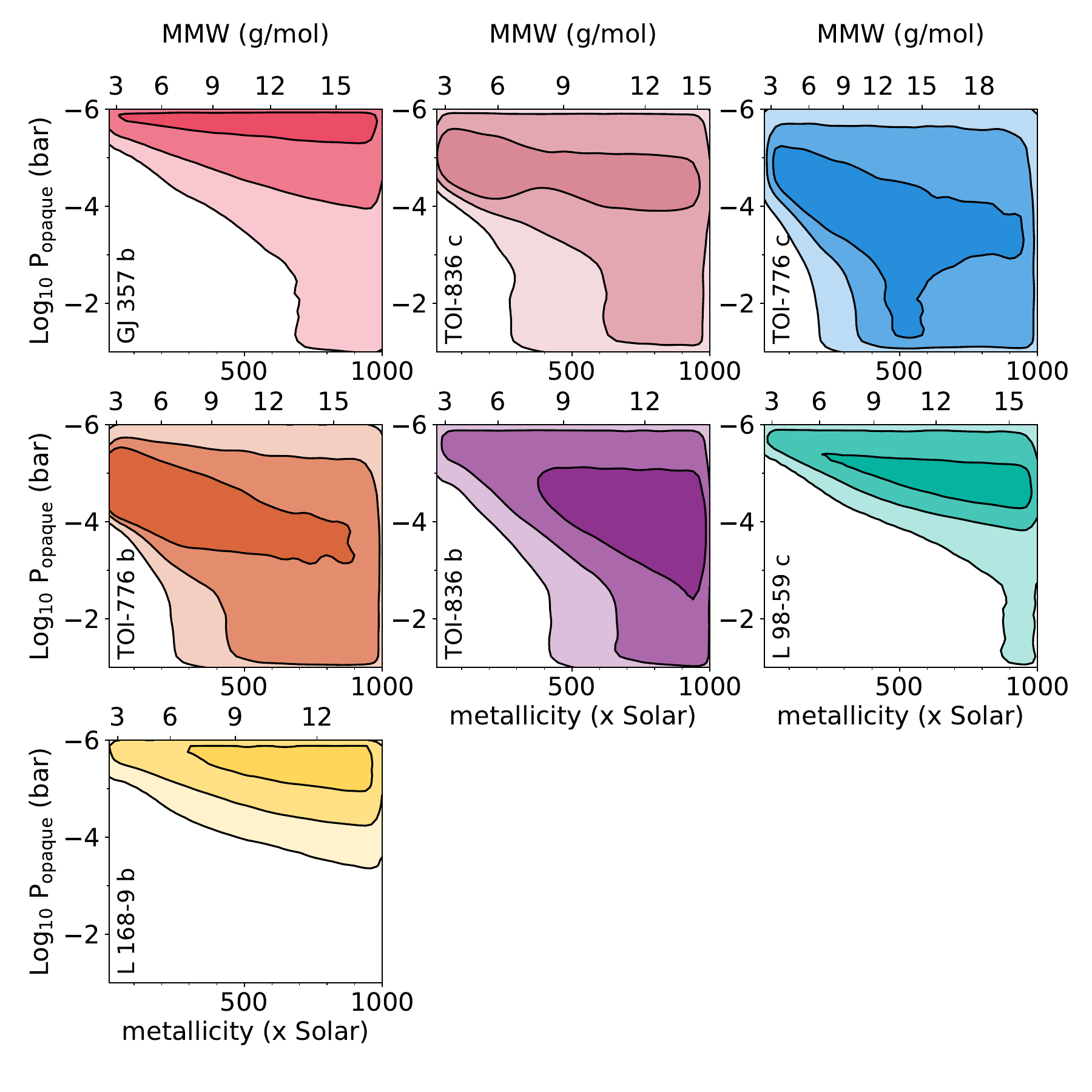} 
        \caption{Constraints on metallicity and opaque pressure levels for the seven COMPASS planets. The regions plotted are, from light to dark, the 98.9\% (3-$\sigma$), 86.4\% (2-$\sigma$), and 39.3\% (1-$\sigma$) highest density intervals. The white region represents models that are excluded at greater than 3-$\sigma$. The model grids used to derive these results are computed for thermochemical equilibrium with C/O=1$\times$ Solar. Note that the two-dimensional credible interval contours do not match the one-dimensional constraints from Figure \ref{fig:individual_constraints_clear} in the clear atmosphere limit. This is expected: In both the 1D and 2D cases the limits are chosen such that 99.5\% (corresponding to 3-$\sigma$ for a 1D Gaussian distribution) and 98.9\% (corresponding to 3-$\sigma$ for a 2D Gaussian distribution) of the probability mass is inside the constraint. In the 2D case that probability mass occurs largely at high opaque pressure levels, which tends to draw the constraint to higher metallicities at low opaque pressure levels. However, a slice of the 2D posteriors taken at 100 bars of pressure would recover the metallicity constraints shown in Figure \ref{fig:individual_constraints_clear}}.
        \label{fig:mcmc_individual}
    \end{figure*}

    Figure \ref{fig:mcmc_individual} shows the 99.5\% credible intervals for each individual spectrum in terms of the metallicity and opaque pressure level marginalized over the detector offsets assuming C/O=1$\times$ Solar. We find that clear atmospheres with metallicities below several hundred times Solar are ruled out for all targets, in broad agreement with the results of similar equilibrium modeling undertaken in earlier studies on these planets. In the low metallicity limit, opaque pressure levels of $10^{-6}$ to $10^{-4}$ bars are required to explain these flat spectra. 

    \subsection{Influence of C/O Ratio}

\begin{figure}
            \centering
            \includegraphics[width=0.5\textwidth]{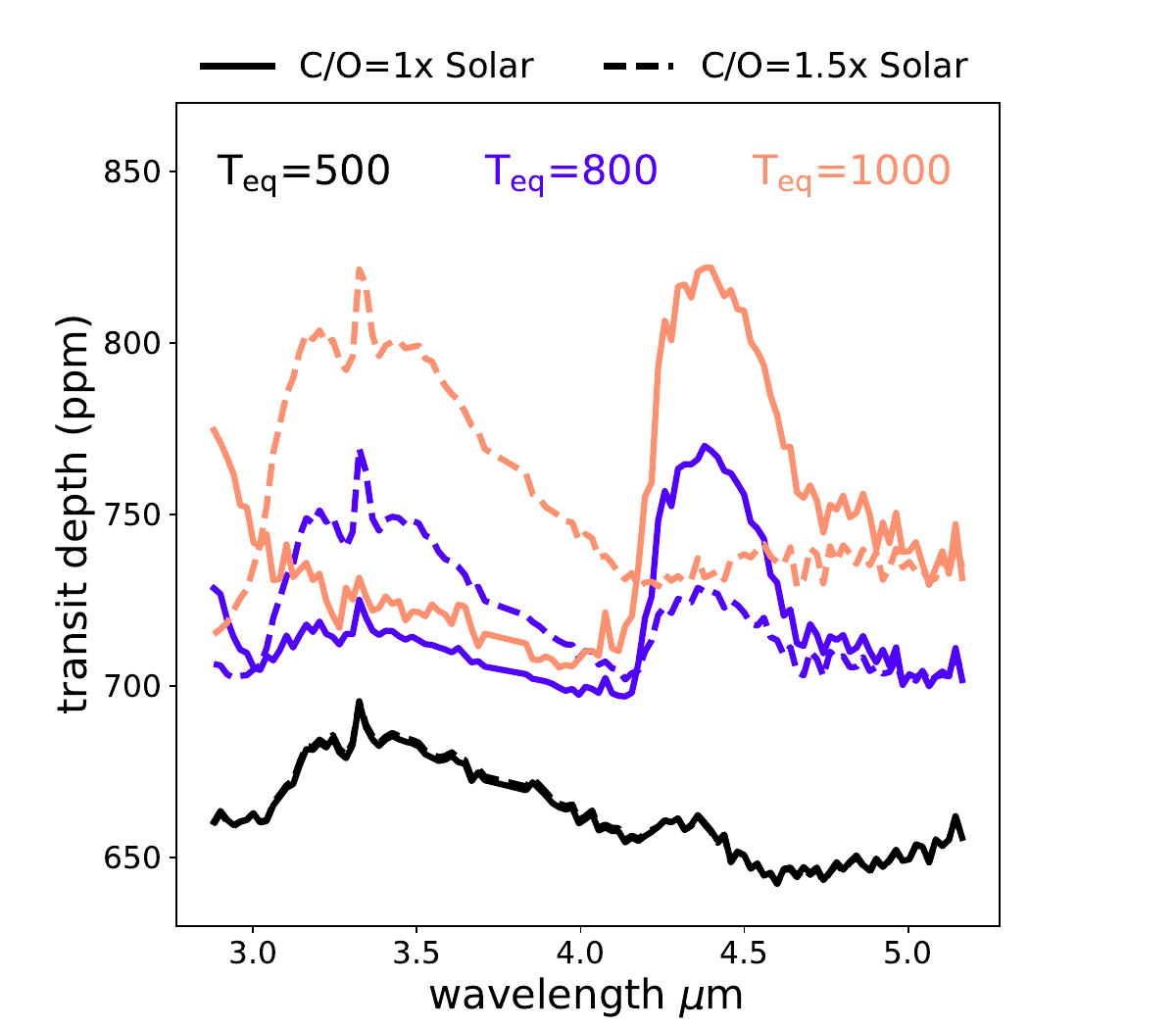}
            \caption{Model spectra for a planet with the mass and radius of TOI-836\,b for 100$\times$ Solar metallicity demonstrating that equilibrium atmospheres with C/O=1$\times$ Solar and C/O=1.5$\times$ Solar are nearly indistinguishable for $T_\mathrm{eq}=500$ K, begin to differ by $T_\mathrm{eq}=800$ K, and appear quite distinct at $T_\mathrm{eq}=1000$ K due to the differing relative abundances of CH$_4$ and CO$_2$.}
            \label{fig:example_spectra_CO}
        \end{figure}

        For planets with equilibrium temperature below about 1000 K, changes in the C/O ratio have minimal impacts on the transmission spectrum \citep{Madhusudhan2012}. As most of the planets in our sample are in this regime, the results of our thermochemical equilibrium modeling are robust to changes in C/O, justifying the choice to assume C/O=1$\times$ Solar. However, TOI-836\,b and L\,168-9\,b, with equilibrium temperatures of 870 K and 1000 K respectively, are warm enough to be influenced by the choice of C/O ratio. At these temperatures the mixing ratios of H$_2$O and CO$_2$ can change substantially between C/O=1$\times$ Solar and C/O=1.5$\times$ Solar, as illustrated in Figure \ref{fig:example_spectra_CO}, which in turn drives changes in the posteriors for opaque pressure level versus metallicity. 
        
\begin{figure*}[htb!]
            \centering
            \includegraphics[width=1\textwidth]{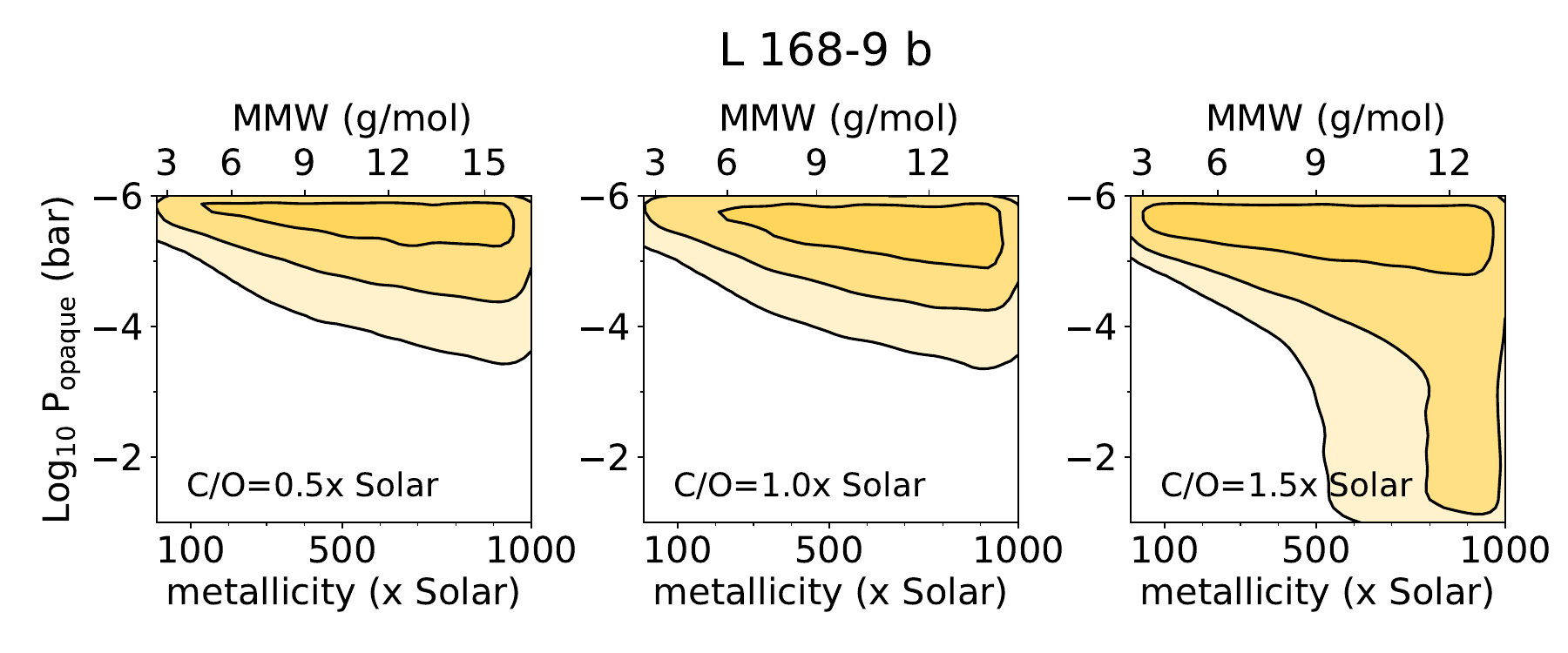}
            \includegraphics[width=1\textwidth]{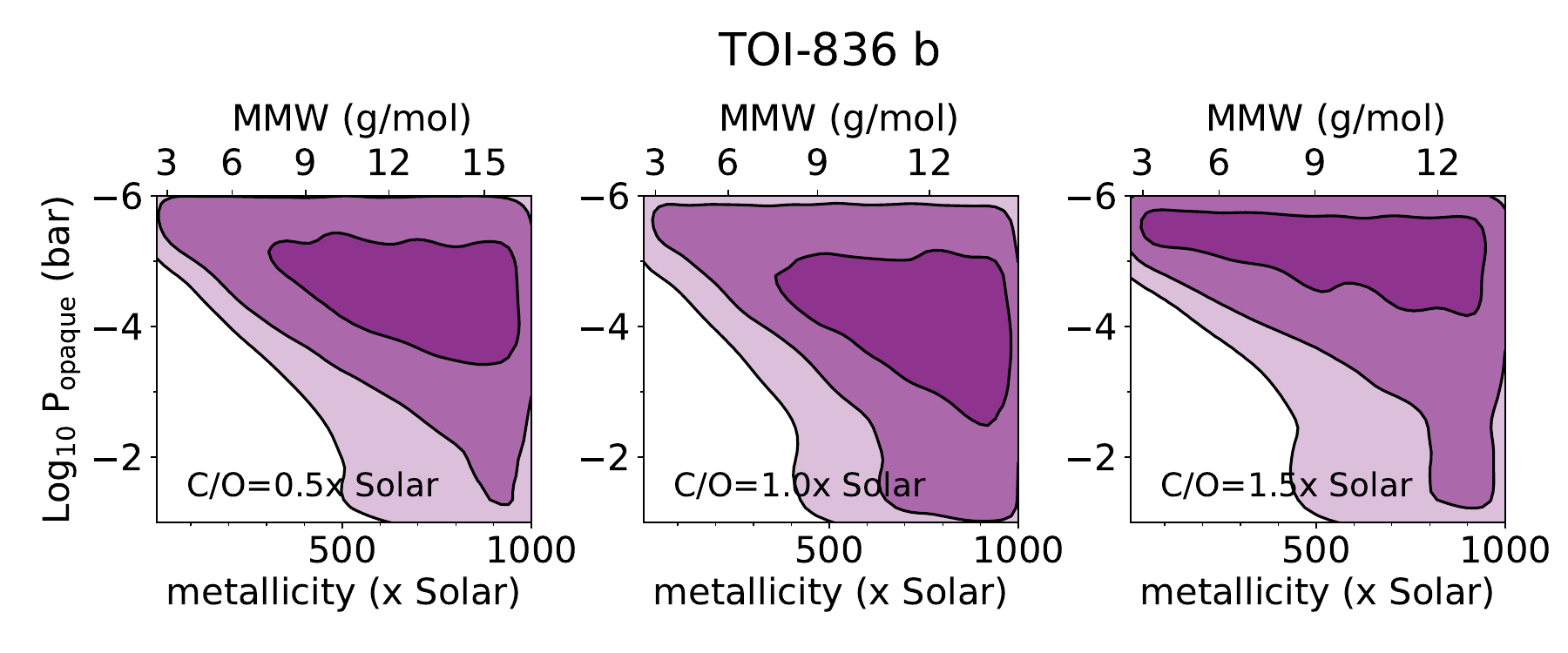} 
            \caption{Posterior distributions for opaque pressure level and metallicity at three different C/O ratios (\textbf{left:} 0.5$\times$ Solar, \textbf{middle:} 1$\times$ Solar, and \textbf{right}: 1.5$\times$ Solar) for the two warmest planets in the sample: TOI-836\,b ($T_\mathrm{eq}=870$ K) and L\,168-9\,b ($T_\mathrm{eq}=1000$ K).}
            \label{fig:constraints_13401_83602}
        \end{figure*}

        In Figure \ref{fig:constraints_13401_83602} we show these posterior distributions for these two warmer planets at three C/O ratios ranging from 0.5$\times$ Solar to 1.5$\times$ Solar. We find that the posteriors do not differ significantly between C/O=0.5$\times$ Solar and 1$\times$ Solar. However, between C/O=1$\times$ Solar and 1.5$\times$ Solar the picture changes substantially for L\,168-9\,b. For this planet the assumption that C/O=1.5$\times$ Solar atmosphere allows for the possibility of atmospheres with metallicity above 500$\times$ Solar, whereas assuming a smaller C/O ratio would result in ruling out atmospheres with metallicity up to 1000$\times$ Solar. For TOI-836\,b the change is less profound, but we can begin to see the shape of the posterior shift towards favoring high opaque pressure levels over high metallicities as the C/O ratio increases. The example of the two warm planets TOI-836 b and L 168-9 b highlights the need to explore a range of C/O ratios when constraining metallicity for planets with equilibrium temperatures above about $\sim800$ K.

\begin{figure}
            \centering
            \includegraphics[width=0.5\textwidth]{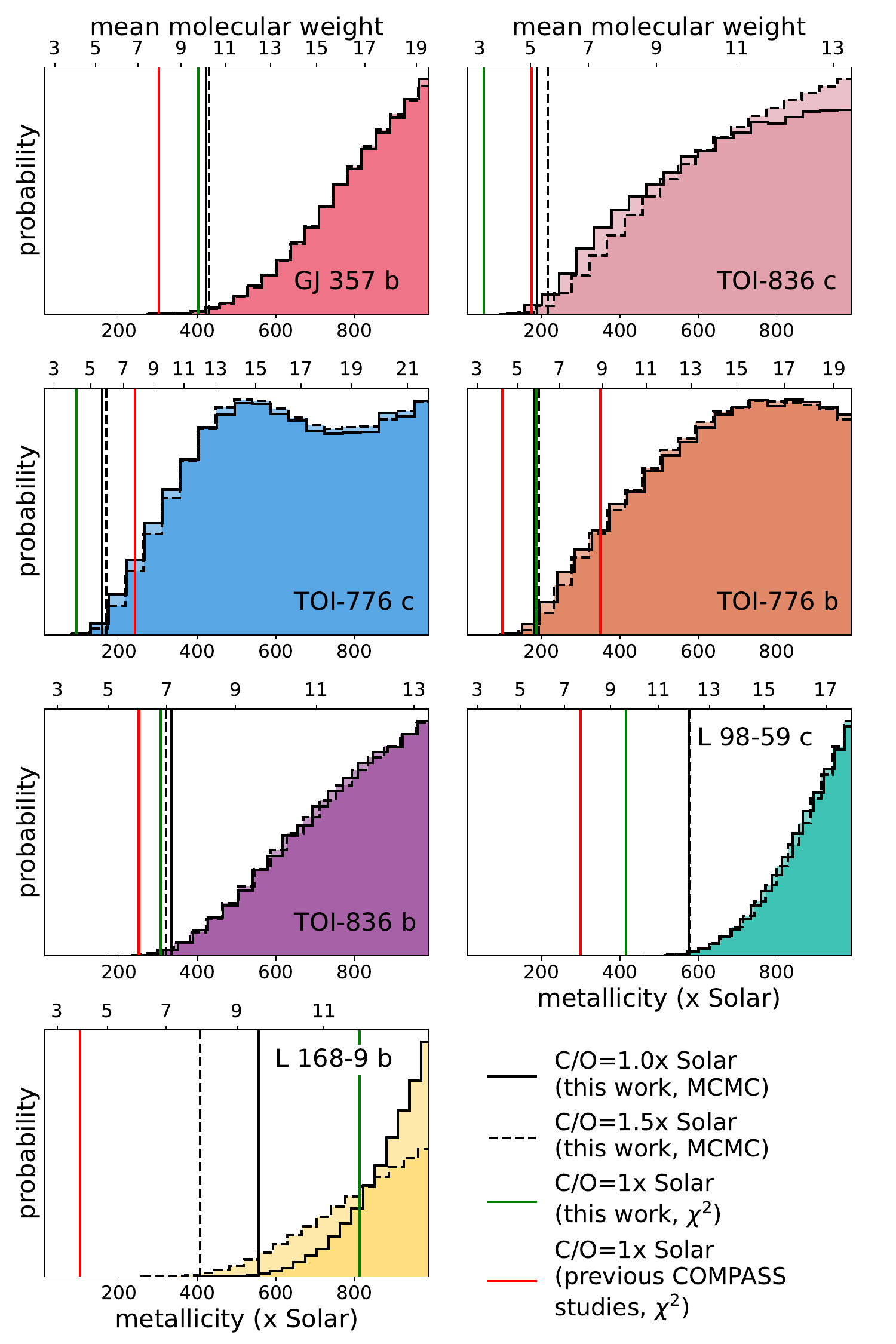}
            \caption{Posterior distributions and 99.5\% confidence lower limits} for the metallicity of each COMPASS target for clear atmospheres. The solid red lines represent metallicity lower limits from previous studies of each planet. For TOI-776\,b the two red lines represent the reported lower limits for the first and second visits individually, as a lower limit for the combined multi-visit spectrum was not reported in \cite{Alderson2025} The solid black lines correspond to the posterior distribution and lower limit for C/O=1$\times$ Solar and the dashed lines are for C/O=1.5$\times$ Solar.
            \label{fig:individual_constraints_clear}
        \end{figure}

        For the clear atmosphere case, shown in Figure \ref{fig:individual_constraints_clear}, we find that our lower limits are broadly consistent with earlier studies (red vertical lines in Figure \ref{fig:individual_constraints_clear}), with the exception of L\,168-9\,b and L\,98-59\,c. We also find that the results of the MCMC simulations match closely the results obtained by computing the $\chi^2$ between our spectra and the model grids (green vertical lines in the Figure \ref{fig:individual_constraints_clear}), again with the exception of L\,168-9\,b and to a lesser extent L\,98-59\,c.
        
        In the case of L\,168-9\,b, for which we find a metallicity lower limit of $\sim580\times$ Solar for the C/O=1$\times$ Solar case, the discrepancy is at least in part driven by the planet parameters used in our forward models. \cite{Alam2025} used the mass, radius, and equilibrium temperature from \cite{AstudilloDefru2020}, while we use \cite{Hobson2024}. When we substitute the \cite{AstudilloDefru2020} for our forward model grid computation, we find a metallicity lower limit of 335$\times$ Solar for the C/O=1$\times$ Solar case. This is significantly closer, but still much higher, than the limit of $\sim100\times$ Solar found by \cite{Alam2025}.
        
        Previous COMPASS studies obtained their metallicity lower limits by mapping the contour along which gridded models could be rejected at 3$\sigma$. While we expect our MCMC-based methodology to give results that are consistent with this procedure, it is possible that some discrepancies could arise from the sensitivity of the $\sigma$-mapping procedure to the size of the error bars for each depth measurement. This sensitivity arises because of the dependence of the $\sigma$-contours on the $\chi^2$ value. The additive log-variance term in Equation \ref{eqn:likelihood} reduces the sensitivity of the MCMC simulation to the error bars. 
        
        While this sensitivity may partially explain the difference between the lower limits derived from the MCMC method and those derived from the $\sigma$-mapping method for L\,98-59\,c and L\,168-9\,b (the black and green vertical lines in Figure \ref{fig:individual_constraints_clear}), it cannot explain the discrepancy between the lower limits derived in previous COMPASS studies and those found in this work since the two methods are in such close agreement for the other targets. A possible explanation is found in the fact that the change in the amplitude of transmission features with changes in metallicity grows smaller as metallicity increases. Our L\,168-9\,b and L\,98-59\,c spectra constrain the metallicity to be quite high, putting us in a regime where the feature amplitude, and therefore the $\chi^2$ of the models with respect to the data, are relatively insensitive to changes in metallicity. This means that large changes in the metallicity are necessary to effect small changes in the quality of the fit, making the metallicity lower limit sensitive to small changes in the spectrum between reductions as well as to the goodness of fit metric used (i.e. $\chi^2$ versus the log-likelihood). 

    \subsection{Results of Composite Spectra Analysis}

        We can formulate a hypothesis regarding the existence of low-amplitude transmission features as follows: We hypothesize that transmission features exist in at least some of our seven spectra, but they too small in comparison to the uncertainty to achieve statistical significance. Upon co-adding the spectra the uncertainty would bin down as approximately $1/\sqrt{N}$ where $N$ is the number of spectra being co-added. If it is the case that low-amplitude features exist in the spectra, if these features sit at the edge of statistical significance, and if enough planets in the sample share the same features (e.g. CO$_2$ or CH$_4$), then we would that upon co-adding the spectra to reduce the effective uncertainty these features may become visible. 

        To test this hypothesis we compute two composite spectra: One for the sub-Neptunes TOI-776\,c and TOI-836\,c, and the other for the five super-Earths (L\,98-59\,c, L\,168-9\,b, GJ\,357\,b, TOI-836\,b, and TOI-776\,b). Each composite spectra is produced by co-adding the group of spectra with error bars are obtained by summing the error bars for the individual spectra in quadrature. Offsets were not removed prior to co-adding the spectra, so the offset in the composite spectra are the sum of the individual offsets. 

\begin{figure*}
            \centering
            \includegraphics[width=\textwidth]{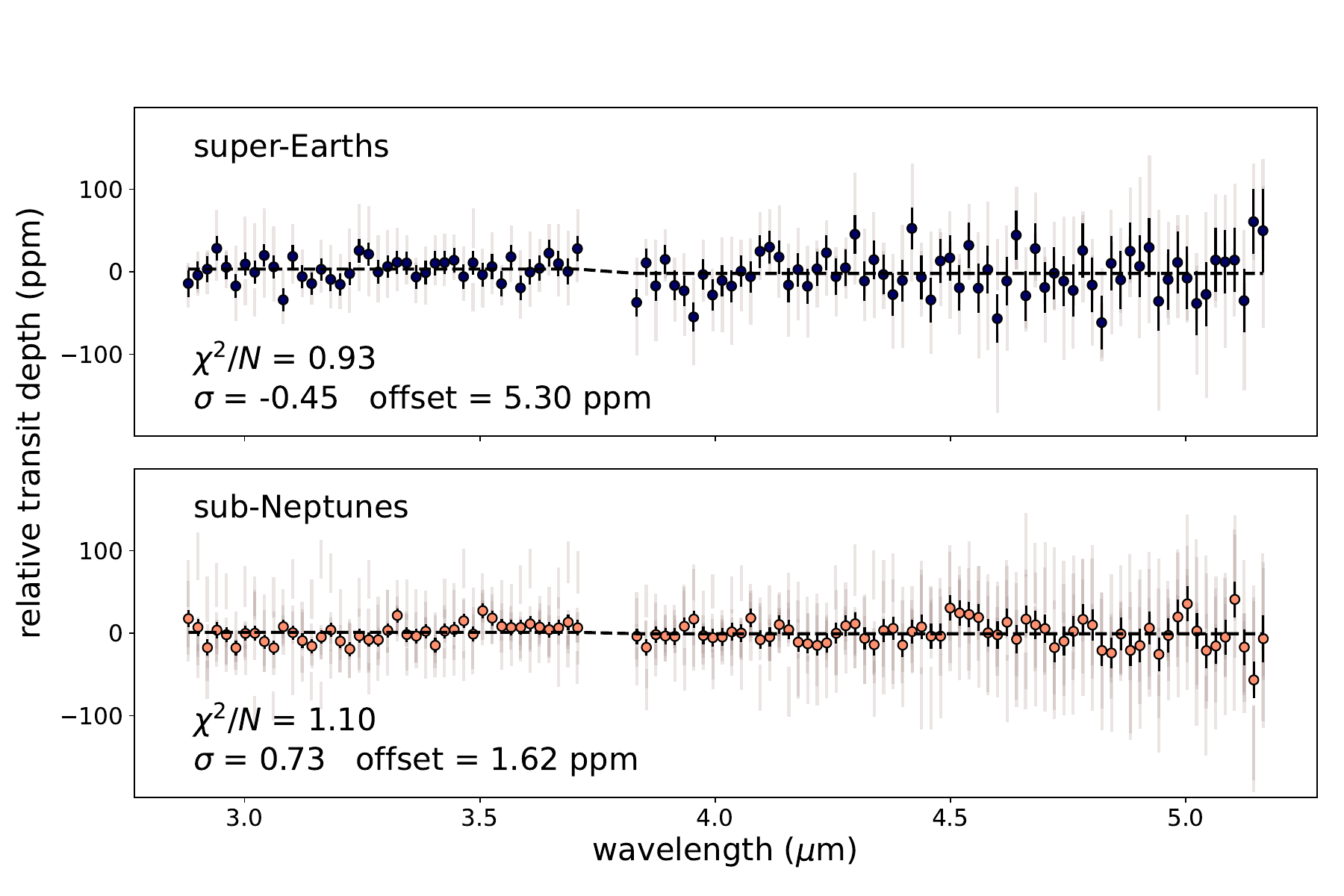}
            \caption{Composite spectra for the five super-Earths (\textbf{top}) and two sub-Neptunes (\textbf{bottom}). The dashed line represents the best-fit flat line model including an offset between NRS1 and NRS2, and the $\chi^2/N$ and $\sigma$ rejection threshold values refer to this best-fit model. Note that $\sigma<0$ indicates $\chi^2/N<1$.}
            \label{fig:composite_spectra}
        \end{figure*}

        Figure \ref{fig:composite_spectra} shows the two composite spectra. Upon visual inspection we see no indication of any spectral features in either spectrum, and both are statistically consistent with a flat line model including an offset between NRS1 and NRS2. This is an unsurprising result for this exploratory analysis: It need not be the case that the planets in either subsample share features in common even if they do possess clear atmospheres. Furthermore, the small number of samples results in a relatively small reduction in the error bars; the uncertainty on the transit depth measurement at 4 $\mu$m is 11.5 and 19 ppm for the super-Earths and sub-Neptunes respectively, compared to the 20-30 ppm uncertainties for the individual spectra. For reference, the approximate expected amplitude of the CH$_4$ feature for L\,98-59\,c at a metallicity of 1000$\times$ Solar is 25 ppm, and for TOI-836\,c the CH$_4$ feature at 500$\times$ Solar metallicity has an amplitude of about 20 ppm. 

\begin{figure*}
            \centering
            \includegraphics[width=\textwidth]{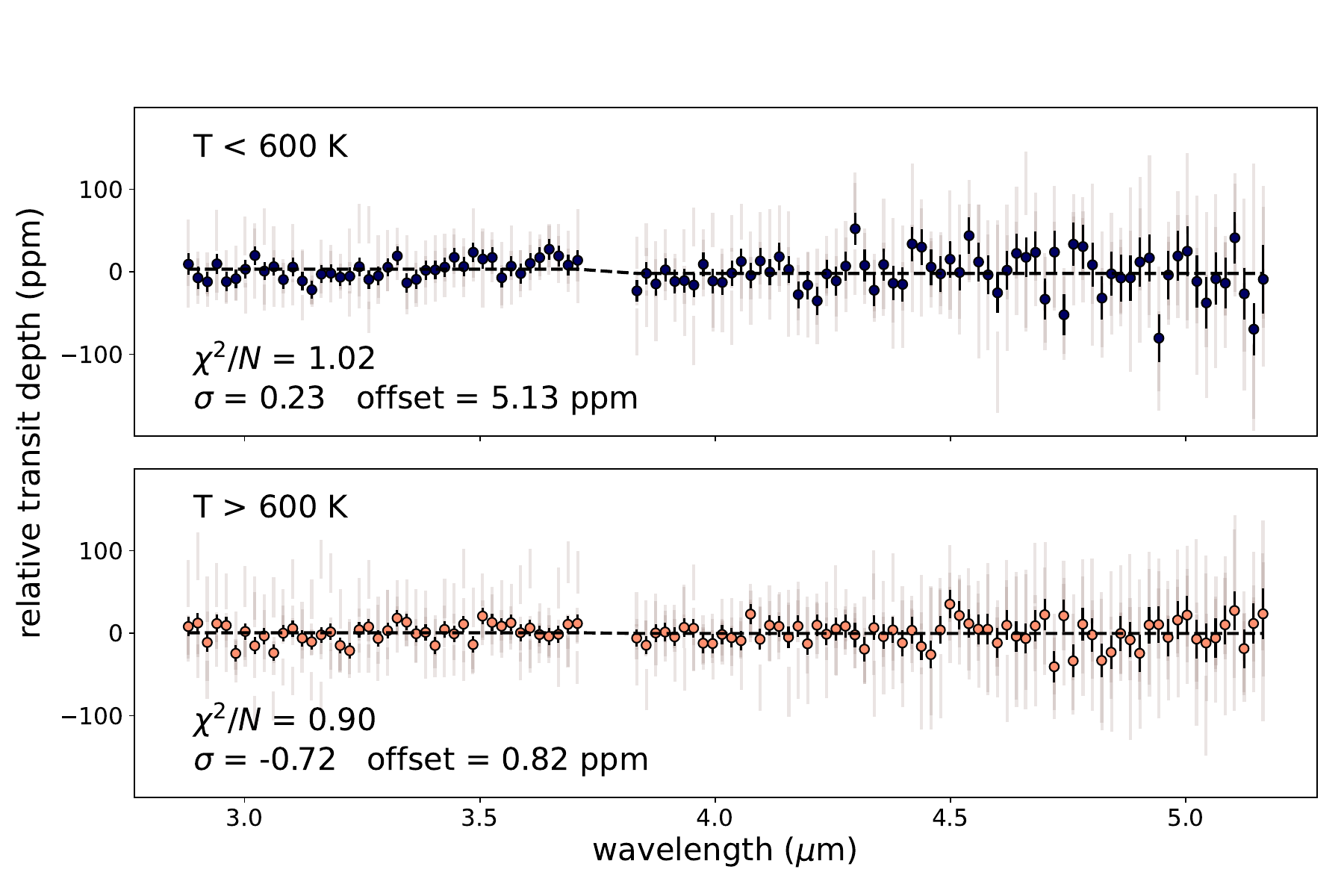}
            \caption{Composite spectra for the four planets with $<600K$ (\textbf{top}) and three planets with $T > 600K$ (\textbf{bottom}). The dashed line represents the best-fit flat line model including an offset between NRS1 and NRS2, and the $\chi^2/N$ and $\sigma$ rejection threshold values refer to this best-fit model.}
            \label{fig:composite_spectra_temperature}
        \end{figure*}

        Recent work by \cite{Kirk2025} shows that co-added spectra are unlikely to be informative when the planetary atmospheres in question are insufficiently self-similar. In particular, for equilibrium atmospheres the temperatures should not span too large a range and should not cross any important chemical transitions, such as the transition between atmospheres where CO$_2$ is the dominant Carbon molecule and those atmospheres where CH$_4$ is the dominant Carbon molecule. This occurs at around 600K for planets with Solar C/O ratio. We therefore compute a second set of composite spectra where we have grouped the planets by equilibrium temperature, separating them into a sample of planets with $T < 600K$ and another with $T > 600K$. In this grouping the composite spectrum for the cool planets might be expected to show evidence for a methane feature, while the warmer planets might be expected to show evidence for a CO$_2$ feature. These spectra are shown in Figure \ref{fig:composite_spectra_temperature}. We again find no evidence that the spectra are statistically non-flat.

\section{Discussion}
    \label{sec:discussion} 

    There are a range of possible explanations for the lack of transmission features for these seven COMPASS targets. The majority of the sample is made up of super-Earths, for which atmosphere detections have proven challenging \citep[see, e.g.][as well as previous COMPASS papers]{May2023, Cadieux2024, Moran2023}. Of the ten super-Earths observed by JWST in transmission in the near-infrared, only one claim of an atmospheric detection has been made, for L\,98-59\,b \citep{Bello-Arufe2025}. If these planets do possess atmospheres, they may be high mean molecular weight secondary atmospheres. Compositions with mean molecular weights greater than five to ten g/mol are permitted for all five super-Earths in our sample given the precision of the transmission spectra presented here. Future JWST investigations including the Rocky Worlds Director's Discretionary Time program and the Charting the Cosmic Shoreline Guest Observer program (\#7073, PI: Lustig-Yaeger) will help to answer lingering questions about whether these small worlds can harbor atmospheres, at least in the case of the M-dwarf host stars targeted by those programs. 

    The two sub-Neptunes in our sample, TOI-776\,c and TOI-836\,c, both have bulk densities that imply the presence of volatiles, and Helium escape has been detected for TOI-836\,c, and Hydrogen escape has been observed for TOI-776\, demonstrating that they must retain at least some of their initial H/He envelopes \citep{Loyd2025, Zhang2025}. TOI-836\,c is a particularly interesting case due to its proximity in mass-radius space to TOI-421\,b, and K2-18\,b, both of which have atmosphere detections. However, TOI-836\,c has an equilibrium temperature of 665K, which places it in a range where previous studies have suggested that hazes may play a role in attenuating spectral features \citep{Brande2024, Crossfield2017}. Indeed, \cite{Wallack2024} finds that haze production rates on par with those predicted for GJ\,1214\,b, which is proximate to TOI-836\,c in mass, radius, and also temperature, could produce a high-altitude haze for TOI-836\,c. 

    We begin this discussion with a series of comparisons between the COMPASS targets and eleven other small planets observed by JWST in the near-infrared. We then consider the potential for future observations of these seven COMPASS targets to clarify our understanding of their atmospheres. 

    \subsection{Comparison of COMPASS Targets to other Small Planets Observed with JWST}

        Here we compare the seven COMPASS targets in this study to eleven other super-Earth and sub-Neptune exoplanets with radii below 3$R_\oplus$ that have been observed with JWST in the near-infrared. The goal of this comparison is to place the COMPASS targets in a broader context and to try to understand why some small planets show spectral features while others do not. As a reminder, we use the term ``super-Earths'' to refer to all planets below the radius gap at approximately 1.8$R_\oplus$, which includes the COMPASS targets GJ\,357\,b, L\,98-59\,c, L\,168-9\,b, TOI-836\,c, and TOI-776\,b. The remaining two COMPASS targets in our sample, TOI-776\,c and TOI-836\,c are sub-Neptunes by our definition. 

        The seven COMPASS targets studied in this work span from one to three Earth-radii in size. In this same radius range there are a total of six planets observed by JWST in the near-infrared which are thought to show spectral features. These are L\,98-59\,b \citep{Bello-Arufe2025}, GJ\,9827\,d \citep{Piaulet2024}, TOI-270\,d \citep{Benneke2024}, K2-18\,b \citep{Madhusudhan2023}, TOI-421\,b \citep{Davenport2025}, and GJ\,1214\,b \citep{Schlawin2024, Ohno2025}. There are an additional five small planets observed by JWST in the near-infrared which do not have transmission features. These are GJ\,3090\,b \citep{Ahrer2025}, GJ\,1132\,b \citep{May2023, Bennett2025}, GJ\,486\,b \citep{Moran2023}, LHS\,475\,b \citep{LustigYaeger2023}, and LHS\,1140\,b \citep{Cadieux2024}. 
    
\begin{figure}
            \centering
            \includegraphics[width=0.5\textwidth]{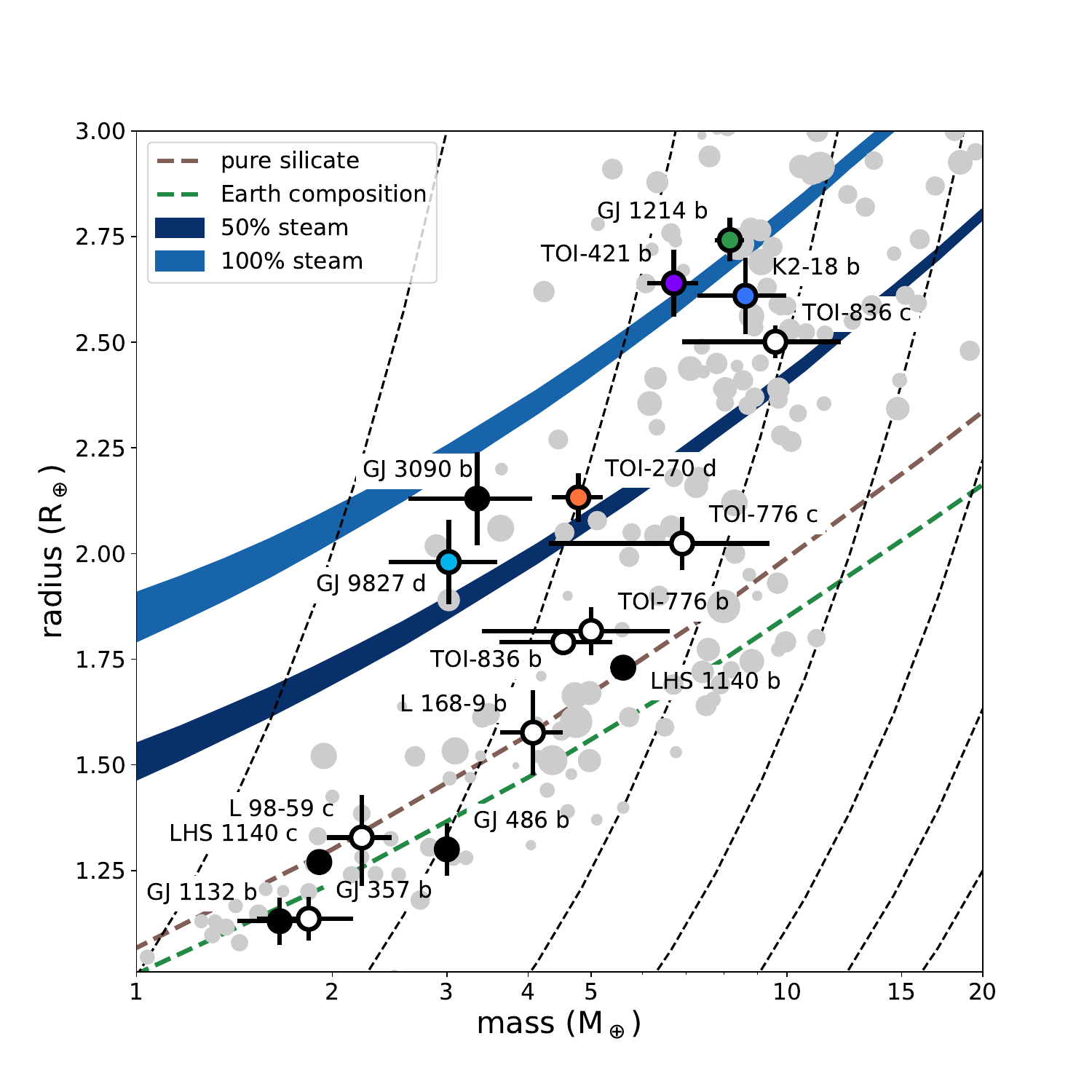}
            \caption{Mass-radius diagram highlighting the seven COMPASS targets analyzed in this study (white circles) compared to six other small exoplanets with detected transmission features (colored circles). The colors correspond to the model spectra in Figure \ref{fig:comparisons}. Gray circles represent other exoplanets meeting mass and radius precision cutoffs of 20\% or better. The dashed lines are models from \cite{Zeng2016} for Earth-like composition (green) and pure silicate bodies (brown). The blue shaded regions are for 50\% water/50\% silicate bodies (dark blue) and 100\% water worlds (light blue) from \cite{Aguichine2021}, and the shaded region represents the difference between a $T_\mathrm{eq}=400$K atmosphere at the bottom edge and a $T_\mathrm{eq}=1000$K atmosphere at the upper edge.} Note that L\,98-59\,b is not shown in this plot. With a mass and radius of 0.4$M_\oplus$ and $0.85R_\oplus$ \citep{Demangeon2021} it sits far below and to the left of the planets shown here.
            \label{fig:mass_radius}
        \end{figure}

\begin{figure}
            \centering
            \includegraphics[width=0.5\textwidth]{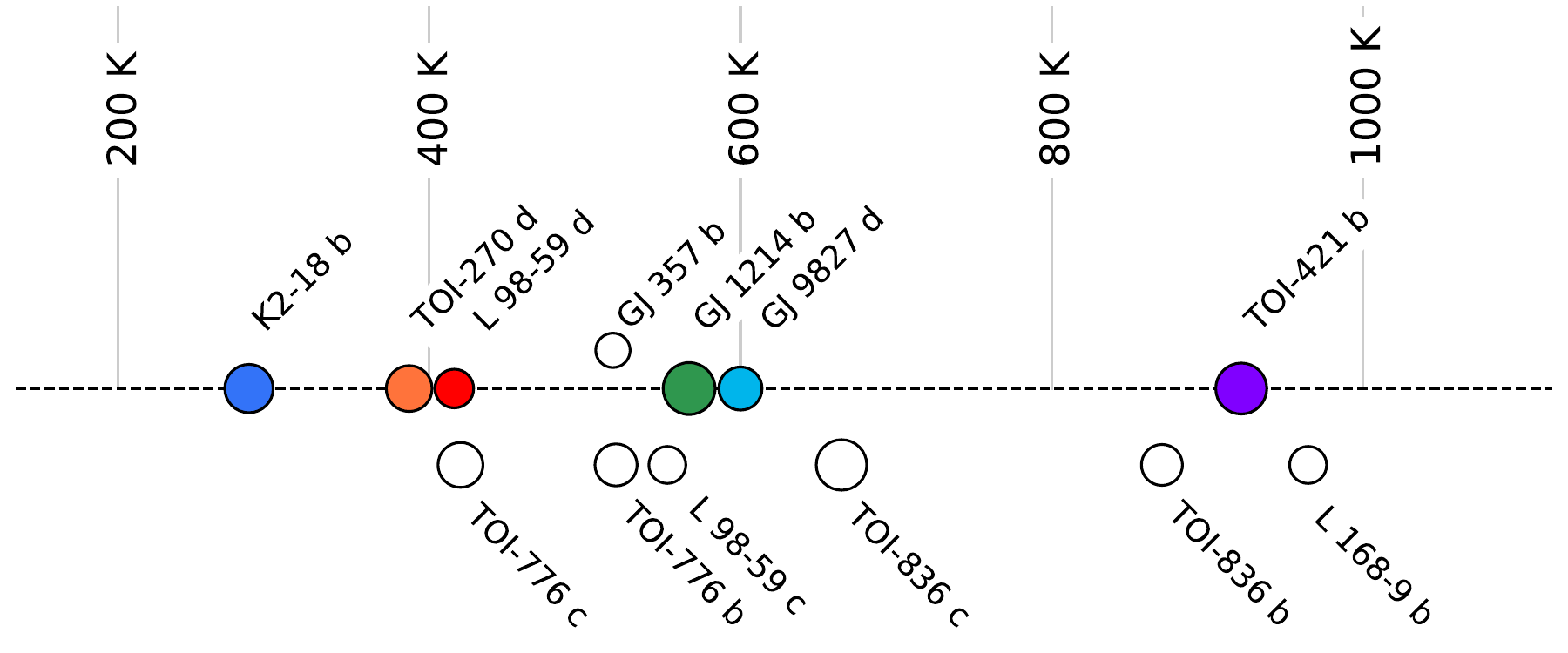}
            \caption{Equilibrium temperatures of the COMPASS planets studied in this work (open circles) compared to the six sub-Neptunes with transmission features (colored circles). The size of each planet is proportional to its radius. Offsets from the center line are for visibility.}
            \label{fig:temps}
        \end{figure}

        In addition to radius, these planets cover similar ranges in mass and equilibrium temperature to the COMPASS targets (see Figures \ref{fig:mass_radius} and \ref{fig:temps}). A possible trend emerges when we consider the distribution of planets with transmission features compared to those without: With a few exceptions (GJ\,3090\,b and TOI-836\,c), the planets with transmission features tend to sit near or above the radius expected for a 50\% steam atmosphere (see the dark blue line from \cite{Aguichine2021}) while the planets without transmission features tend to sit below this line, closer to the pure silicate model from \cite{Zeng2016}. A fuller understanding of the statistical significance of this trend and its implications for the exoplanet population would require an analysis that is beyond the scope of this work. However, this emerging trend suggests that atmospheres may be challenging to observe for denser worlds.
    
\begin{figure*}[t!]
            \centering
            \begin{subfigure}[t]{0.45\textwidth}
                \centering
                \includegraphics[width=1.0\linewidth]{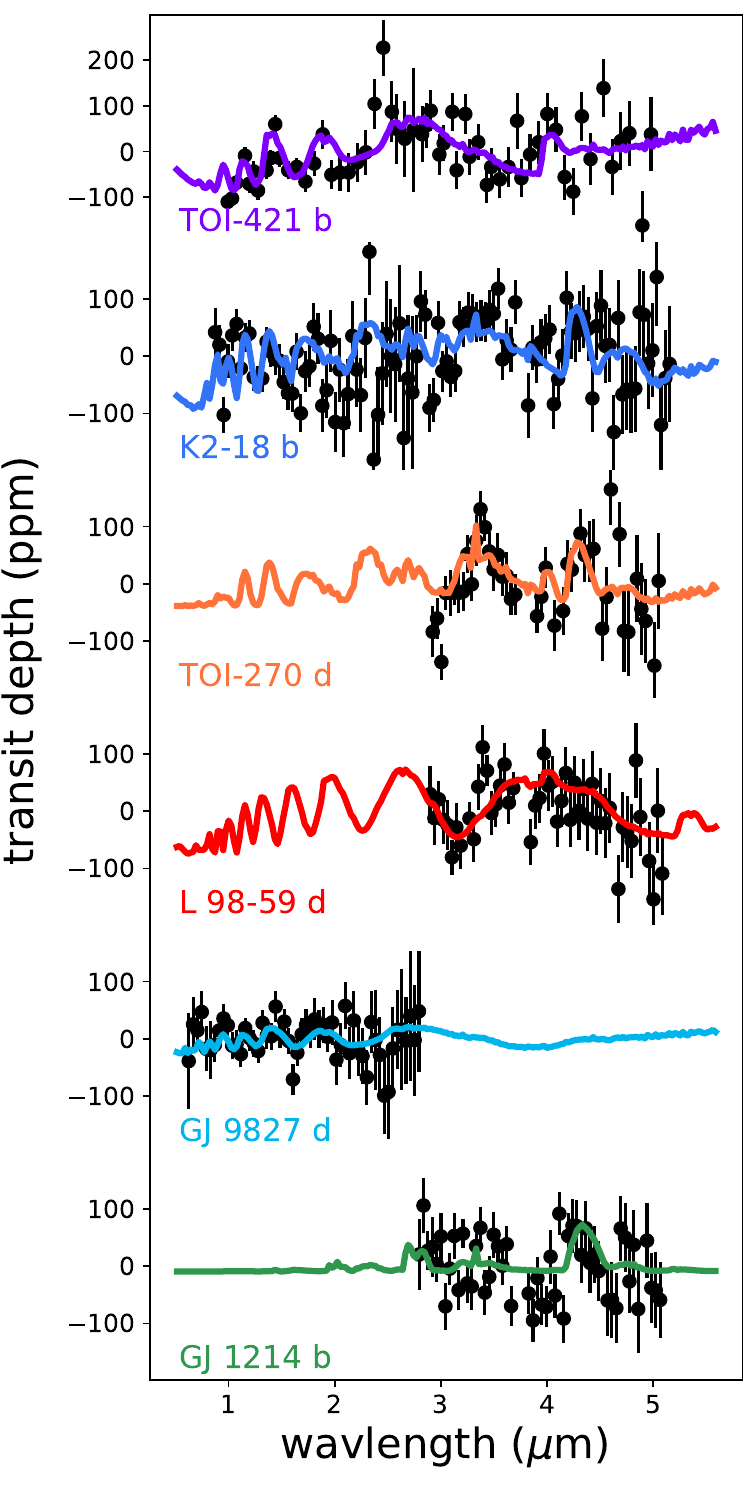} 
            \end{subfigure}
            \begin{subfigure}[t]{0.45\textwidth}
                \centering
                \includegraphics[width=1.0\linewidth]{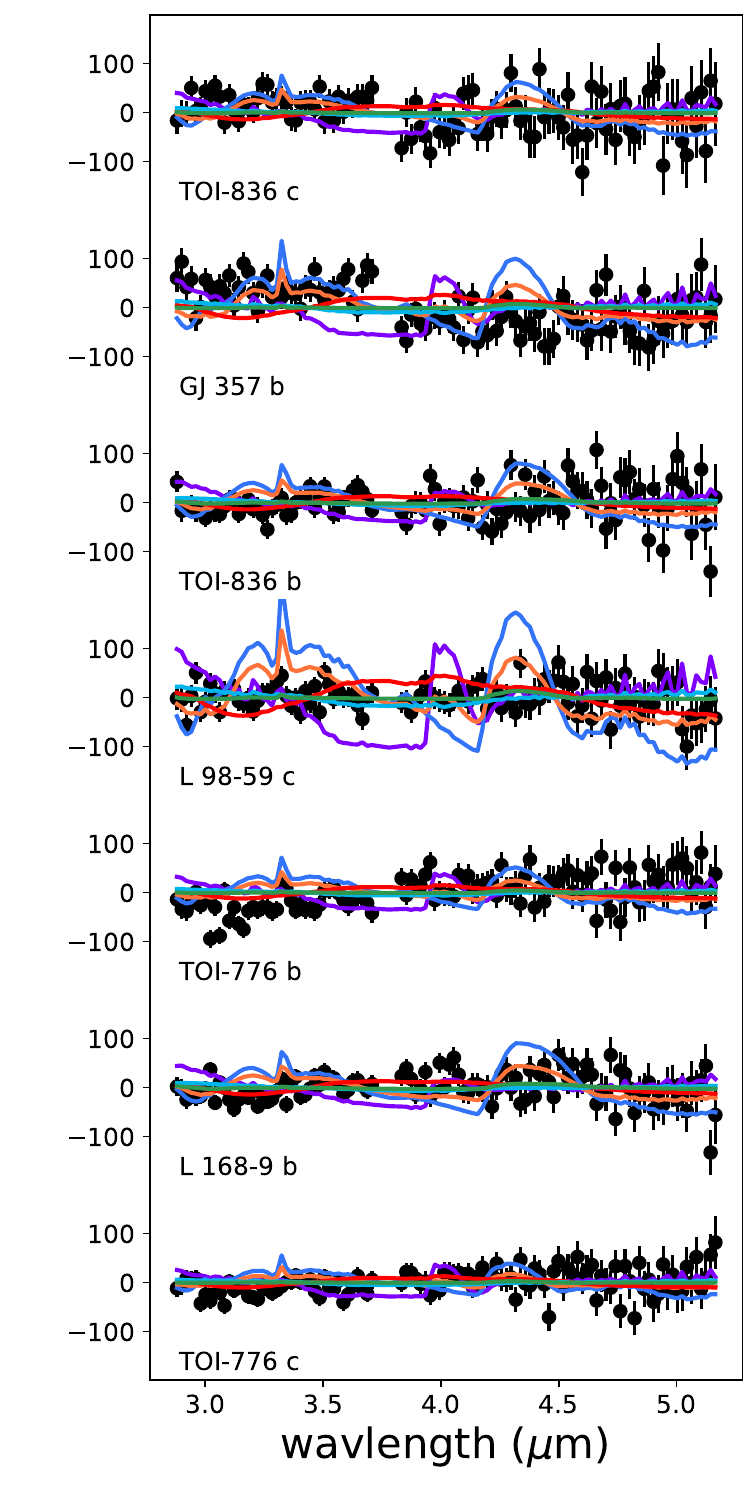} 
            \end{subfigure}
            \caption{\textbf{Left:} Spectra and approximate best-fit models for six sub-Neptunes observed with JWST. \textbf{Right: } Spectra for each of the seven COMPASS targets overplotted with spectra matching the composition of the four planets in the left plot but scaled to the masses, radii, and equilibrium temperatures of the COMPASS targets.}
            \label{fig:comparisons}
        \end{figure*}

\begin{figure}
            \centering
            \includegraphics[width=0.5\textwidth]{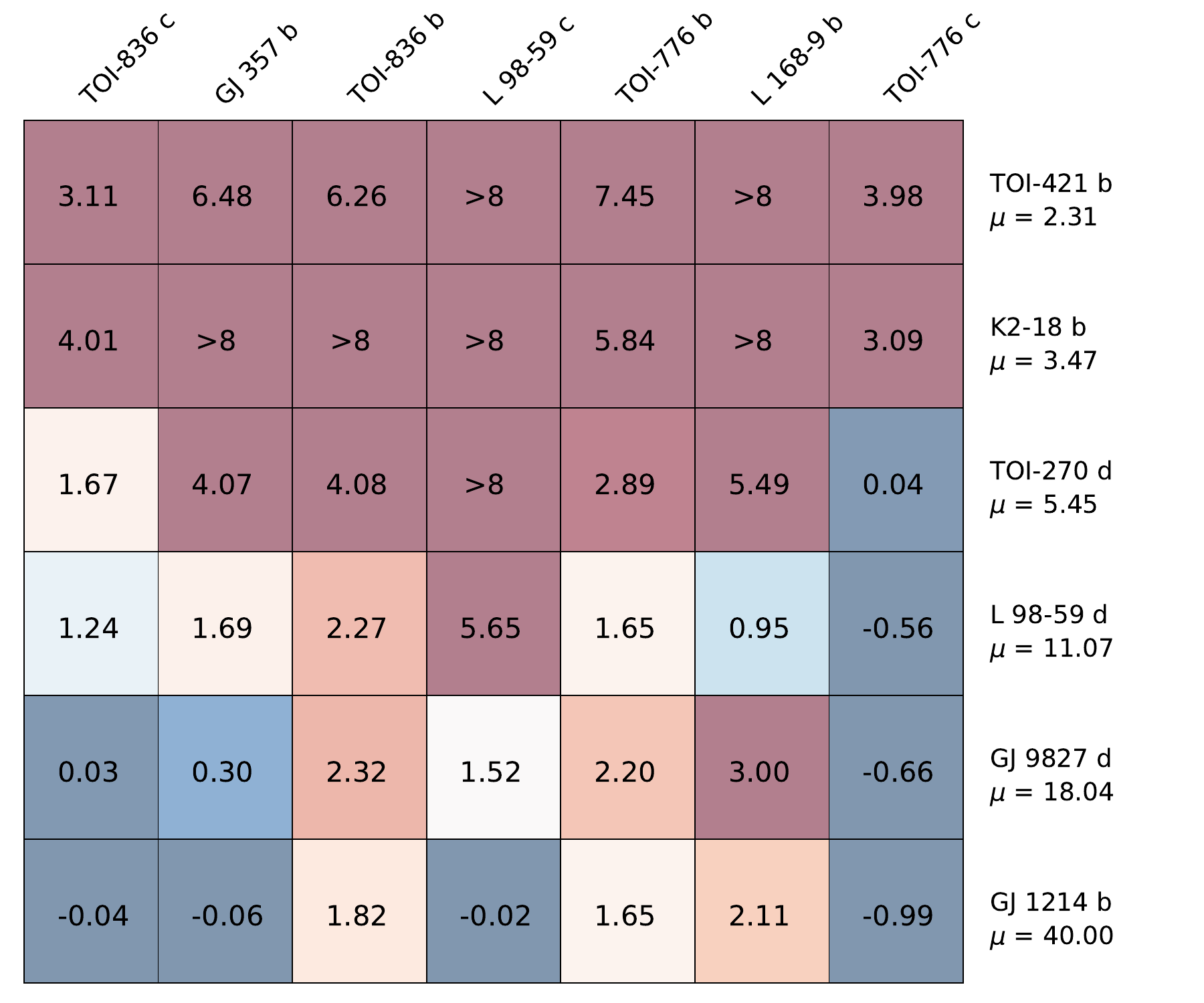}
            \caption{$\sigma_r$ (sigma-rejection threshold) values for model atmospheres with identical compositions to exoplanets with observed transmission features (rows) compared to the seven COMPASS spectra (columns). For each square we compute a forward model for the atmospheric composition corresponding to the row of the square for the mass, radius, and equilibrium temperature of the COMPASS target corresponding to the column of the square. We then compute $\chi^2/N$ between the model and the corresponding COMPASS planet spectrum, from which we compute the $\sigma_r$ value. We allow for an offset between the NRS1 and NRS2 detectors with a value that is chosen to minimize the sum of squares between the model atmosphere and the data.}
            \label{fig:xi2_grid}
        \end{figure}

        We examine the transmission spectra that would be produced by the atmospheres of the six planets with features if they were placed around each of the COMPASS planets. We take the retrieved molecular abundances from the original studies, for which best-fit spectra are shown in the left panel of Figure \ref{fig:comparisons}, and use these to compute model spectra for the COMPASS targets using \texttt{petitRADTRANS} \citep{Molliere2019}. These models are shown in the right panel of Figure \ref{fig:comparisons}, where they are plotted over the observed spectra of each atmosphere. In Figure \ref{fig:xi2_grid} we show the $\sigma$ rejection threshold values between the observed planet spectra (columns) and the model atmosphere (rows). For GJ\,1214\,b we used the CO$_2$ and CH$_4$ abundances from \cite{Schlawin2024}, who did not report a mean molecular weight. We found that in order to reproduce an approximate fit to the observations we had to assume a mean molecular weight of about 42 amu, which is roughly consistent with the reported abundances plus a background gas composed of N$_2$. We note that the composition of GJ\,1214\,b's atmosphere and haze properties are highly uncertain (see also \cite{Ohno2025} and \cite{Malsky2025}).

        We find that the low mean molecular weight compositions of K2-18\,b and TOI-421\,b are inconsistent with the flat spectra of all seven COMPASS targets at greater than 3-$\sigma$. This is because their low mean molecular weight atmospheres produce large transmission features, which are easy to rule out for the COMPASS spectra. TOI-270\,d has an intermediate mean molecular weight of $\mu$=5.82 amu, which results in muted spectral features. Its composition is ruled out for the super-Earths (except for TOI-776 b), but remains barely plausible for TOI-836 c at $\sigma_r=1.67$ and is fully consistent with the spectrum of TOI-776 c. The purported high mean molecular weight atmosphere of L\,98-59\,b is inconsistent with the flat spectrum of its sibling planet L\,98-59\,c, but is otherwise plausible for the remaining COMPASS targets. However, we note that there are uncertainties about the existence and composition of this atmosphere of L\,98-59\,b, with the observed JWST spectrum also being consistent with a bare rock \citep{Bello-Arufe2025}. The water-dominated atmosphere of GJ\,9827\,d is ruled out by the flat spectrum of the super-Earths L\,168-9\,b and disfavored for TOI-836\,b and TOI-776\,b, but is otherwise compatible with the remaining COMPASS planets. The very high mean molecular weight, hazy atmosphere of GJ\,1214\,b is flat enough to be plausible for all of the COMPASS planets. 
    
        Summarizing these observations, we see that only the low mean molecular weight atmospheres observed for TOI-421\,b and K2-18\,b can be reliably ruled out for the entire sample. Higher mean molecular weight atmospheres are largely consistent with the flat spectra observed for these COMPASS targets, with several exceptions in the cases of L\,98-59\,c and L\,168-9\,b. It is also worth noting that all of the planets in the left-hand panel of Figure \ref{fig:comparisons} have some short wavelength ($\lambda < 2.8$) coverage, including L 98-59\,d and GJ\,1214\,b which have HST observations that are not shown in the figure. These observations likely help to constrain their atmospheric properties beyond what would be possible with NIRSpec alone, especially due to the lack of strong water features between 2.8 and 5.3 $\mu$m. In general, the flatness of the COMPASS spectra does not appear all that surprising in light of the wavelength coverage of G395H and the compositional diversity and frequently high mean molecular weight atmospheres already observed among the collection of small planets seen by JWST. 

    \subsection{Future Observations}
        \label{sec:predictions}

        Sub-Neptunes with detections of molecules in their atmospheres tend to sit on the upper edge of the mass-radius diagram at large radii relative to their masses. Detections of transmission features for sub-Neptunes with less inflated radii have been less forthcoming, with the possible exception of K2-18\,b. Transmission features have been elusive for super-Earths, with only L\,98-59\,b showing tentative features \citep{Bello-Arufe2025}. This suggests the possibility that observers have been overly optimistic when estimating the number of transits necessary to detect atmospheres of all but the most inflated small planets. This may be due to a more frequent occurrence of low pressure aerosols and high metallicity atmospheres than was expected prior to the launch of JWST, the impact of stellar variability and surface inhomogeneities as well as other sources of correlated noise in JWST timeseries, and systematic overprediction of the transit precision from tools like \texttt{pandexo}. If we wish to begin building a picture of atmospheric composition across the mass-radius diagram, we will need to observe additional transits of those planets that have already been studied, and to allocate greater numbers of transits to new targets. 

        We investigate the prospects for detecting transmission features for our sample of seven COMPASS planets with additional observations using either NIRSpec/G395H or NIRISS/SOSS, under the assumption that these worlds possess clear, high metallicity atmospheres. We consider adding an additional $N_\mathrm{new}$ transit observations to the existing $N_\mathrm{existing}$ transits observed for the COMPASS program. Assuming a 300$\times$ Solar metallicity, C/O=1$\times$ Solar clear atmosphere and $N_\mathrm{new}$ transit observations for each planet, we estimate $\sigma_r$, the rejection threshold for a flat model assuming no detector offsets for the NIRSpec/G395H observations. In order to roughly marginalize over realizations of the white noise in the spectrum, we simulate 10,000 spectra with independent white noise realizations to obtain a 95\% confidence interval for $\sigma_r$. For the NIRSpec simulations we compute the uncertainty on the spectrum by scaling the measured uncertainty for the $N_\mathrm{existing}$ transits by a factor of $\sqrt{N_\mathrm{existing}/N_\mathrm{new}}$. For the NIRISS simulations we use \texttt{pandexo} to determine the expected precision of the spectrum. We then inflate the precision by a factor of 1.2 to account for the fact that \texttt{pandexo} underestimates transit depth uncertainties by assuming a flat-bottomed transit rather than a limb-darkened transit. The factor of 1.2 is chosen to match the observed discrepancy between the \texttt{pandexo} precision and the real precision for NIRSpec, according to \cite{Espinoza2023}. 

\begin{figure*}[t!]
            \centering
            \includegraphics[width=1.0\textwidth]{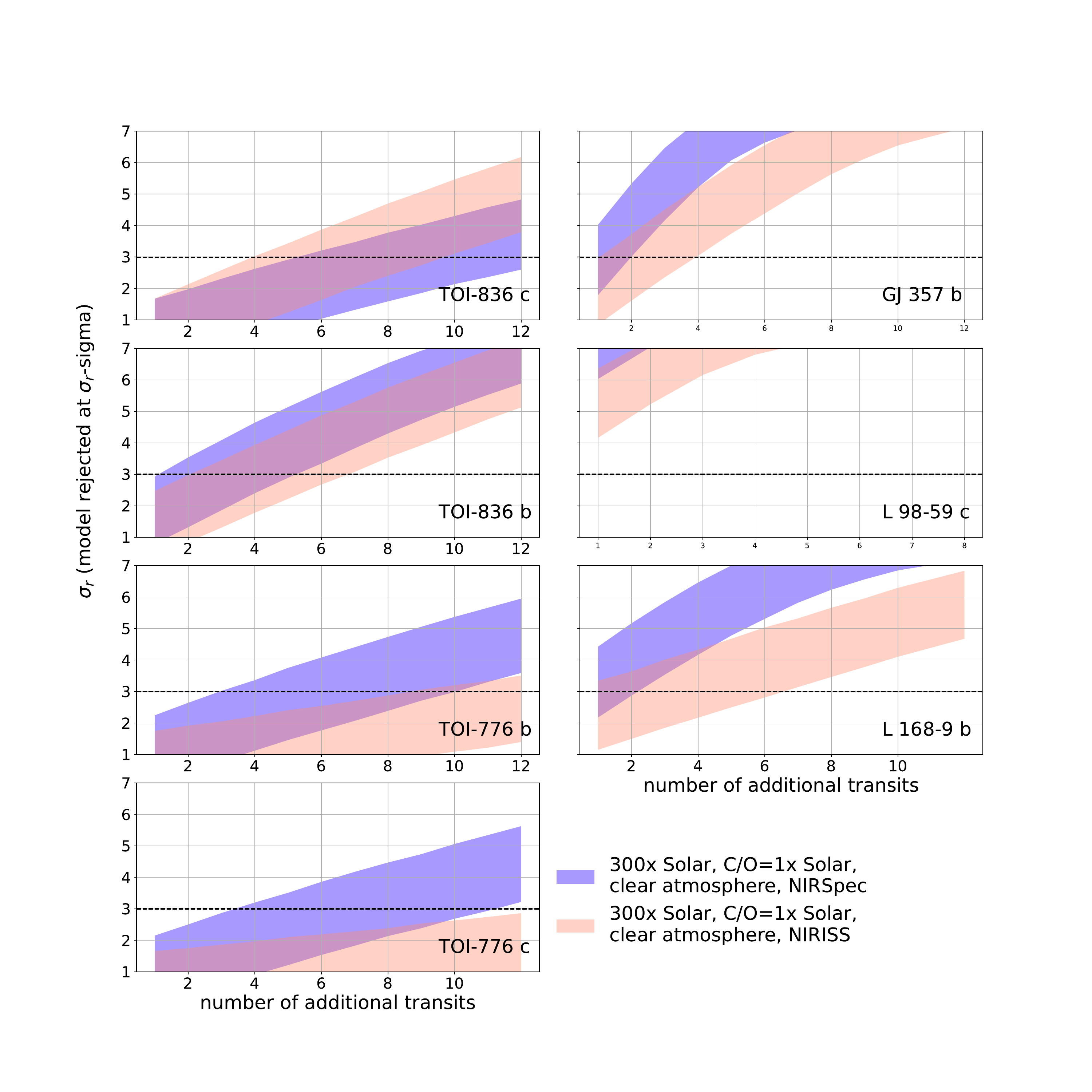}
            \caption{Predictions of the number of additional transit observations necessary to rule out a flat spectral model for an injected model representing a 300$\times$ Solar, C/O=1$\times$ Solar aerosol-free atmosphere for each planet. The shaded regions represent the 95\% confidence interval for the $\sigma$-rejection threshold for the flat spectral model when those transits are observed with either NIRISS/SOSS (orange region) or NIRSpec/G395H (blue region). An estimate of the number of transits required to detect the presence of an atmosphere at $\sigma_r=3$ can be read off the plot as the point on the x-axis where the upper edge of the shaded region first crosses $\sigma_r=3$ to the point where the lower edge clears the $\sigma_r=3$ line.}
            \label{fig:predictions_co1}
        \end{figure*}

        The results of these simulations are shown in Figure \ref{fig:predictions_co1}. In general, we find that NIRSpec is more useful than NIRISS for ruling out a flat spectrum in the case of a 300$\times$ Solar clear atmosphere. Our metallicity lower limits for L\,98-59\,c, GJ\,357\,b, and L\,168-9\,b are already above 300$\times$ Solar, making it unsurprising that few to no additional transits are necessary to rule out this atmosphere for those targets. For TOI-836\,c we find that between four and ten transits observed with NIRISS/SOSS would be sufficient to rule out this atmospheric composition, while five to 12 additional transits would be necessary with NIRSpec/G395H. For TOI-776\,b and TOI-776\,c, between three and ten additional NIRSpec transits are needed, and for TOI-836\,b, two to eight transits with either instrument would suffice. 

\begin{figure*}[t!]
            \centering
            \includegraphics[width=1.0\textwidth]{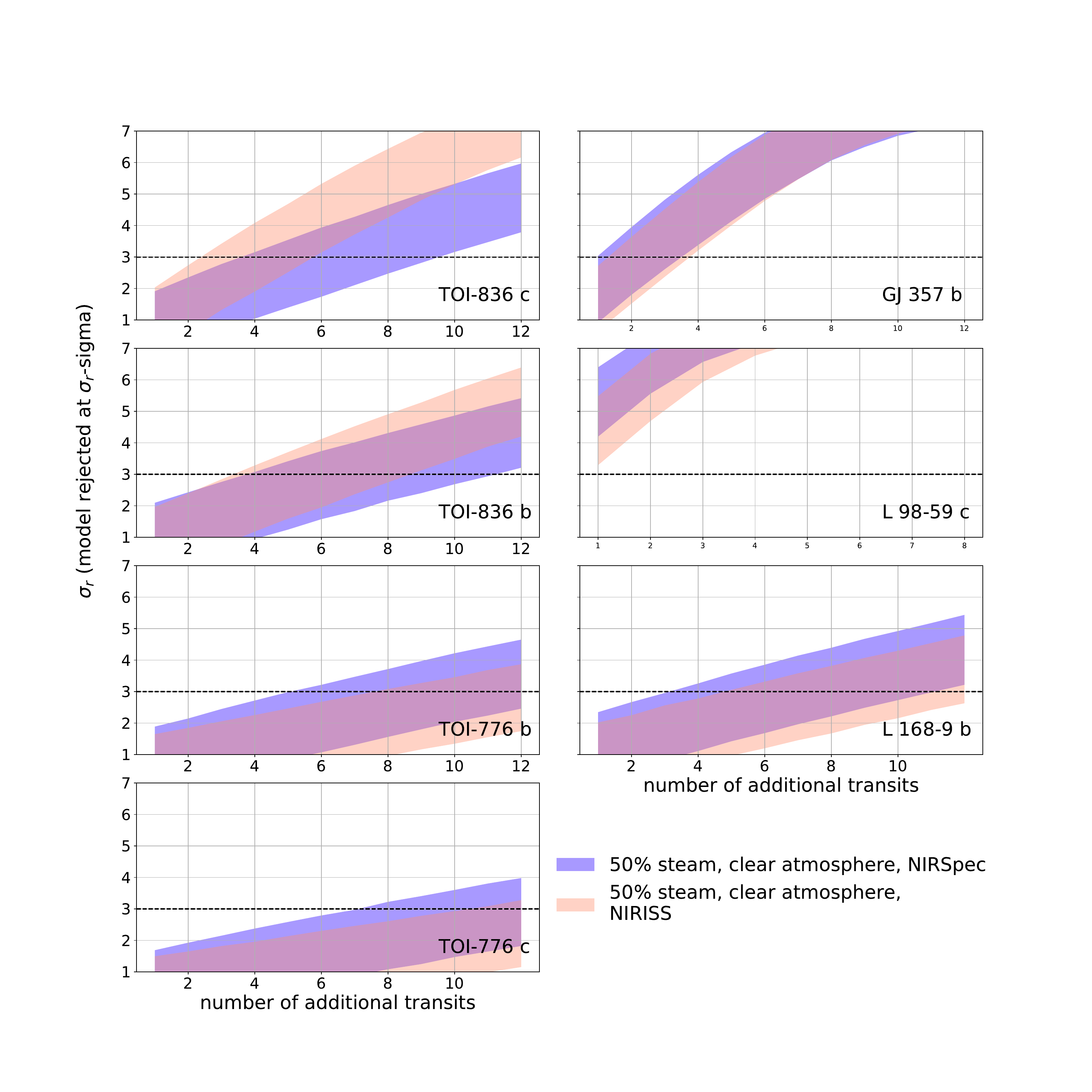}
            \caption{Same as Figures \ref{fig:predictions_co1} but for a 50\% steam/50\% H$_2$ clear atmosphere.}
            \label{fig:predictions_h2o}
        \end{figure*}

        As discussed in Section \ref{sec:results} and illustrated in Figure \ref{fig:comparisons}, a picture is beginning to emerge of the diversity of sub-Neptune atmospheres. In particular, the planets GJ\,9827\,d and TOI-421\,b both have spectra dominated by water absorption, with smaller relative abundances of CH$_4$ and CO$_2$ than for TOI-270\,d and K2-18\,b. In order to understand how this diversity impacts prospects for detecting atmospheres of these seven COMPASS targets, we carry out the same simulations described above for atmospheric compositions that are 50\% steam/50\% H$_2$. The results of these simulations are shown in Figure \ref{fig:predictions_h2o}. For these atmospheres NIRISS/SOSS and NIRSpec/G395H perform similarly, with the exception of TOI-836\,c and TOI-836\,b for which NIRISS requires approximately two fewer transits to detect the steam atmosphere than does NIRSpec. The improved performance of NIRISS over NIRSpec for the steam atmospheres over the equilibrium atmospheres in Figure \ref{fig:predictions_co1} is a result of the fact that NIRISS captures several prominent water features at wavelengths shorter than 2.8 $\mu$m which are missed by NIRSpec. 

        An additional complicating factor for NIRSpec/G395H observations of steam atmospheres is the degeneracy between the slope of the $\sim2.8\mu$m water feature extending across the NRS1 portion of the spectrum and the slopes that can arise from stellar contamination in this same wavelength range. This effect may further advantage NIRISS/SOSS over NIRSpec/G395H when it comes to steam atmospheres, particularly for M dwarfs where stellar contamination is expected to be a limiting factor. The presence of the detector gap and the associated offset between the NRS1 and NRS2 portions of the spectrum may also present a difficulty for constraining the presence of water vapor in an atmosphere: Because this gap overlaps with the edge of the water feature slope, it is possible for a degeneracy to arise between the amplitude of the water feature and the detector offset \citep[see, e.g.][]{Coulombe2025}.

        As discussed in previous COMPASS publications and illustrated in Figure \ref{fig:mcmc_individual}, the degeneracy between metallicity (or mean molecular weight) and opaque pressure level presents a challenge for interpreting flat transmission spectra. Potential means of breaking this degeneracy include atmospheric escape observations, which can indicate the presence of an atmosphere and provide some constraints on metallicity \citep[e.g.][]{Zhang2025}, emission spectroscopy through secondary eclipse observations or phase curves, which can also be used to demonstrate the presence of an atmosphere \citep{Kempton2024}, and additional transit observations to obtain more precise transmission spectroscopy. Emission spectroscopy is likely to be challenging for most off these small, temperate planets. The emission spectroscopy metric (ESM) put forward by \cite{Kempton2018}, which is a measure of the expected signal-to-noise ratio of the planet's secondary eclipse, ranges between 1.6 for TOI-776\,c and 13.8 for L\,168-9\,b. For reference, we compute the ESM for GJ\,1214\,b, which has been successfully observed in emission with a single transit observed by MIRI LRS \citep{Kempton2024}, to be 26.4 using the planetary and stellar parameters from \cite{Mahajan2024}. These low ESM values indicate that, while the hottest planets in our sample may permit emission spectroscopy in a manageable number of transits, transmission spectroscopy will be an essential tool if we wish to break the opaque pressure level/metallicity degeneracy and measure the atmospheric compositions of these planets. 

\begin{figure*}[t!]
            \centering
            \includegraphics[width=1.0\textwidth]{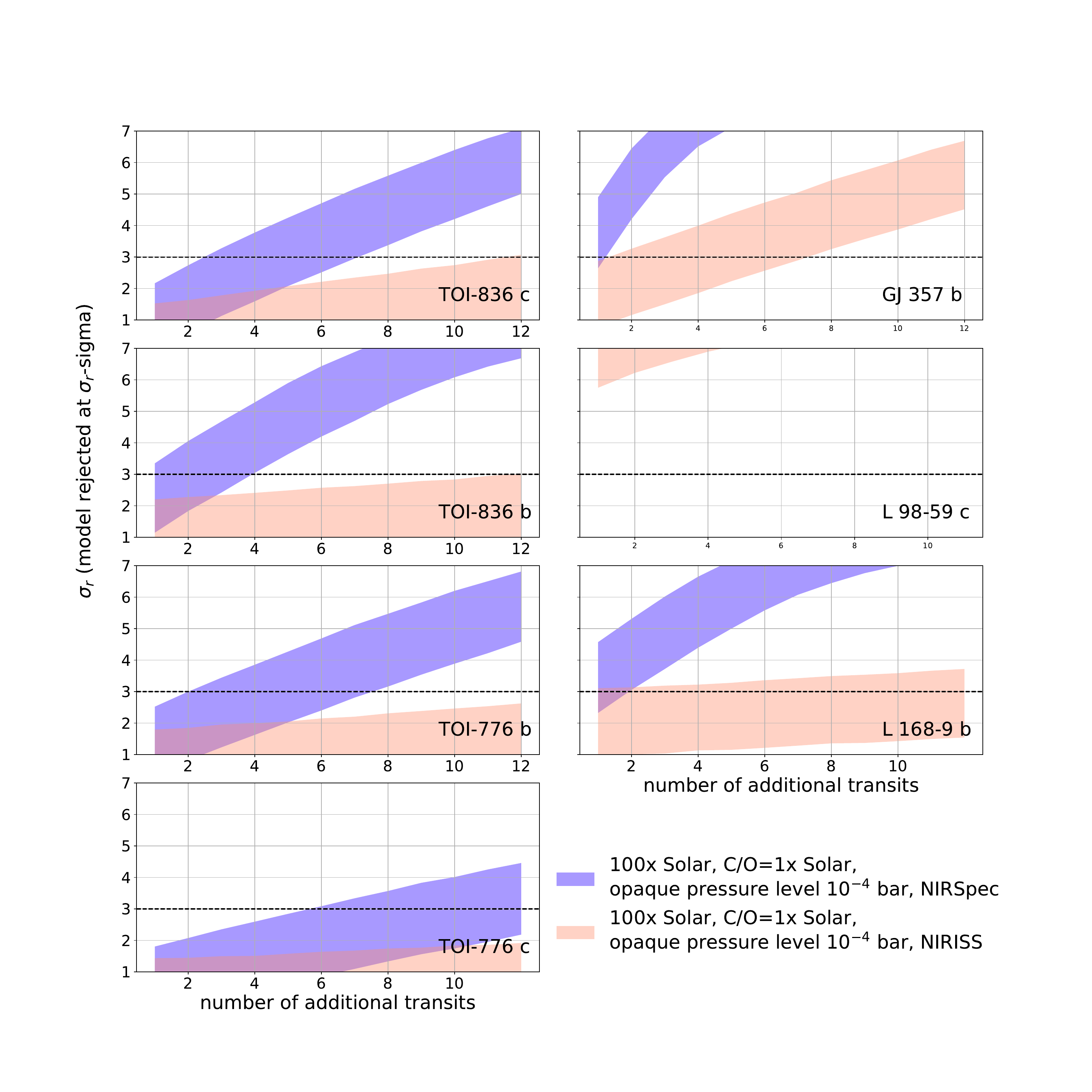}
            \caption{Same as Figures \ref{fig:predictions_co1} but for a 100$\times$ Solar, C/O=1$\times$ Solar atmosphere with an opaque pressure level of 10$^{-4}$ bar.}
            \label{fig:predictions_cloudy}
        \end{figure*}

\begin{figure}[t!]
            \centering
            \includegraphics[width=0.5
\textwidth]{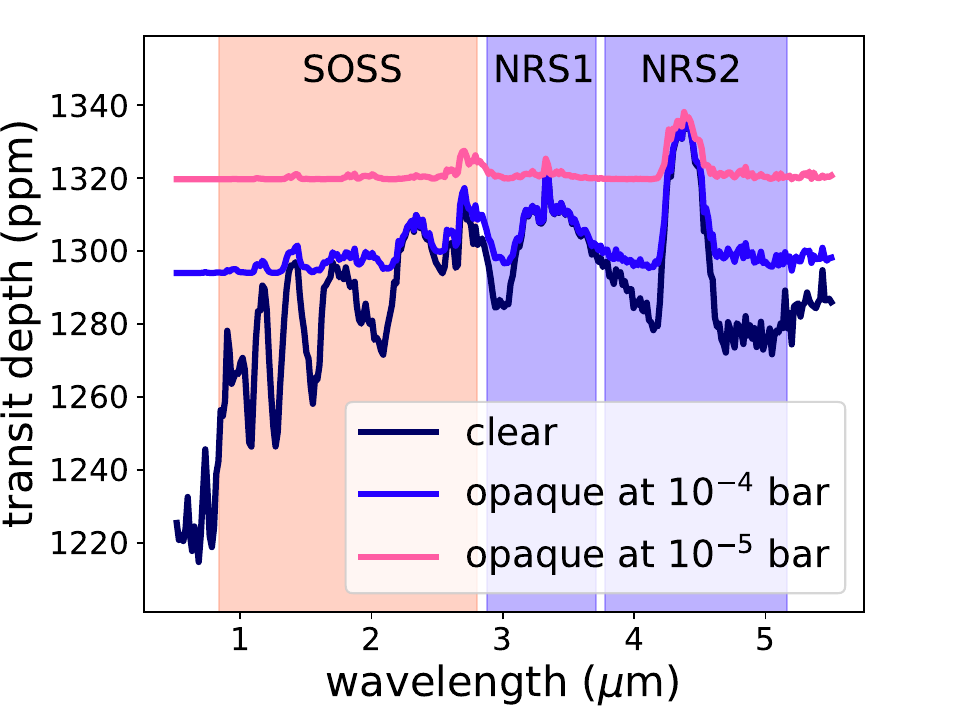}
            \caption{Model transmission spectra of TOI-836\,c for a clear atmosphere with a metallicity of 100$\times$ Solar, C/O=1$\times$ Solar, and opaque cloud pressures of 100 bar (clear atmosphere, dark blue line), 10$^{-4}$ bar (lighter blue line), and 10$^{-5}$ bar (pink line). The shaded regions show the wavelength range for NIRISS/SOSS (orange region) and NIRSpec/G395H (blue regions). This figure illustrates how the spectral features that are most prominent at NIRISS/SOSS wavelengths disappear in the presence of high-altitude aerosols, making NIRSpec a better choice when aerosols are expected to be present.}
            \label{fig:example_cloudy}
        \end{figure}

        To investigate observational prospects for breaking the opaque pressure level/metallicity degeneracy, we apply the same methodology as we used above for the clear atmosphere case, this time to 100$\times$ Solar metallicity atmospheres with opaque pressure levels of $10^{-4}$ bar. The results of these simulations are shown in Figure \ref{fig:predictions_cloudy}. The most notable difference between this case and the clear atmosphere cases is that NIRISS performs very poorly in the presence of a the high-altitude aerosol layer. The reason for this is illustrated in Figure \ref{fig:example_cloudy}: For high opaque pressure levels the H$_2$O features accessible to NIRISS are highly attenuated, while the CH$_4$ and particularly the CO$_2$ features at NIRSpec wavelengths remain prominent. 

        It should be noted that none of these simulations account for the possibility that NIRISS/SOSS observations could be impacted by stellar contamination via the transit light source effect \citep{Rackham2018}, which can alter the transmission spectrum at short wavelengths and potentially mimic a water vapor signature \citep[see, e.g.][]{Moran2023}. Another factor for NIRISS/SOSS observations not accounted for in our analysis is the possibility of a detection of the Rayleigh scattering slope of haze particulates in the atmosphere. Though potentially degenerate with both water features and stellar contamination, a successful detection of a Rayleigh scattering slope could confirm the presence of a haze, providing an additional path towards breaking the opaque pressure level/metallicity degeneracy. 

        Finally, we note that while NIRSpec/G395H has an edge over NIRISS/SOSS in the majority of cases, the relative performance of the two instruments is sensitive to the presence of aerosols and the expected amplitude of water features in the spectrum relative to CH$_4$ and CO$_2$, neither of which are generally known a-priori. This highlights the challenges inherent in optimizing observations for planets with unknown atmospheric compositions, and points towards the importance of modeling a wide range of possibilities before settling on an observing strategy.

\section{Conclusions}
    \label{sec:conclusions}

    We have carried out independent reductions of the first seven targets from the COMPASS program using the \texttt{ExoTiC-JEDI} code. We introduce a systematics model that accounts for variations in the trace morphology via the principal components of the relative pixel flux timeseries. We demonstrate that this systematics model reduces red noise in the residuals of the white light curves and improves the Akaike and Bayesian information criteria for low group-number observations. An analysis of the posteriors from the white light curve fitting confirms that the systematic effects are largest for the three-group observations. We also observe that the NRS1 detector suffers from systematics to a greater degree than NRS2. 

    We then carried out spectral light curve fitting. For the three-group observations we used the same 6-vector systematics model used on the white light curves to fit each spectral bin. For group numbers greater than three we found that the use of the full systematics model decreased the precision of the spectrum without substantially effecting its shape (see Figure \ref{fig:pca_comparison}), so we elected not to use the PCA-based systematics model for these observations, since the use of the more complex model slightly decreases the precision of the spectrum (see Figure \ref{fig:error_comparison}). For targets with multiple visits we simultaneously fit all the available transits to produce a single spectrum for each planet. 

    We derived new limits on metallicity and opaque pressure level assuming thermochemical equilibrium for each of the seven targets. We showed that the lower limits on metallicity are in broad agreement with those obtained in earlier studies with a few exceptions as noted in Section \ref{sec:results}. 

    In order to test the hypothesis that some fraction of the spectra in our sample have low-level transmission features in common which fail to attain statistical significance individually but which might be significant when the spectra are summed together to reduce the white noise scatter, we produce two composite spectra, one for the five super-Earths and one for the two sub-Neptunes in the sample. These composite spectra are produced by summing over the individual spectra while adding the error bars in quadrature. We find that both composite spectra are consistent with a flat spectrum including an offset between NRS1 and NRS2.

    By comparing best-fit models from the set of small planets with JWST observations of transmission features in the near-infrared to hypothetical chemically-identical atmospheres around each of the COMPASS targets, we demonstrate that only the lowest mean molecular weight sub-Neptune or super-Earth atmospheres known, those of TOI-421\,b and K2-18\,b, would be detectable around all of the planets in the sample. In other words, we confirm that none of the planets in this sample have atmospheres that are analogs to K2-18\,b or TOI-421\,b. Notably this includes TOI-836\,c, which is a close analog of TOI-421\,b in mass and radius, though it is substantially cooler with an equilibrium temperature of $\sim665 K$ compared to TOI-421\,b's $T_\mathrm{eq}\sim920 K$. The atmosphere of TOI-421\,b is ruled out for TOI-836\,c at $\sim3.6\sigma$. If these planets have similar atmospheres, it would therefore imply the presence of high-altitude aerosols in TOI-836\,c's atmosphere similar to those observed for GJ\,1214\,b. With two to five transits of TOI-836\,c observed with NIRISS/SOSS, a 50\% steam/50\% H$_2$ atmosphere could be ruled out, while $\sim6$ transits would be required to rule out a 100$\times$ Solar metallicity atmosphere with an opaque pressure level $>10^{-4}$ bar with NIRSpec/G395H. Using the same methods described in Section \ref{sec:predictions} we find that an atmosphere analogous to that of GJ-1214\,b for TOI-836\,c would be undetectable with fewer than several hundred NIRSpec/G395H transits due to its smaller transmission spectroscopy metric \citep{Kempton2018}. As a result, breaking the aerosol-metallicity degeneracy by detecting aerosols in the astmosphere of TOI-836\,c would likely require the use of NIRISS/SOSS to detect the Rayleigh scattering slope at wavelengths shorter than those accessible to G395H. 
    
    The heavier atmospheres of TOI-270\,d, L\,98-59\,b, and GJ\,9827\,d can all be ruled out for some of the seven COMPASS targets but are compatible with others, while the very high mean molecular weight, hazy atmosphere of GJ\,1214\,b is compatible with all seven of our spectra. In particular we note that TOI-270\,d's atmospheric composition is strongly ruled out for four of the seven targets ($\sigma_r > 3$), weakly ruled out for TOI-836\,c and TOI-776\,b ($\sigma_r > 2$) and consistent only with TOI-776\,c. \cite{Teske2025} also finds that TOI-776\,c's spectrum is compatible with TOI-270\,d's atmospheric composition, indicating the need for additional observations to rule out the hypothesis that these two planets, which have similar masses, radii, and equilibrium temperatures, have analogous atmospheres. 

    In order to break the degeneracy between high mean molecular weight atmospheres and aerosols in the atmospheres of sub-Neptunes, additional observations will be needed. To this end, we finish by providing predictions of the number of additional transits necessary to detect a 300$\times$ Solar metallicity clear atmosphere around each planet, and a 100x Solar metallicity atmosphere with an opaque pressure level of 10$^{-4}$ bar with either NIRISS/SOSS or NIRspec/G395H. We find that observing additional NIRspec/G395H transits is generally more productive than using NIRISS/SOSS, particularly in the case of a high opaque pressure level. However, our atmospheric models do not include Rayleigh scattering due to hazes, which may be detectable with NIRISS/SOSS. Further investigation is needed to determine whether NIRISS/SOSS is capable of detecting hazes and breaking the aerosol-metallicity degeneracy for any of these COMPASS targets. 

    We also find that fewer than three additional NIRSpec/G395H observations would be capable of ruling out a low-metallicity (100$\times$ Solar or less) atmosphere with an opaque pressure level of $10^{-4}$ bar for L\,98-59\,c, GJ\,357\,b, and L\,168-9\,b. Ruling out a 300$\times$ Solar clear atmosphere could also be accomplished for these planets with fewer than three additional NIRSpec/G395H transits, and a 50\% steam/50\% H$_2$ atmosphere could be ruled out for L\,98-59\,c, and GJ\,357\,b in $<3$ additional transits. 

\software{}

    In addition to the software cited in the text, we acknowledge the use of the following software in this work: \texttt{astropy} \citep{astropy}, \texttt{scipy} \citep{scipy}, \texttt{numpy} \citep{numpy}, \texttt{emcee} \citep{emcee}, \texttt{matplotlib} \citep{matplotlib}, \texttt{scikit-learn} \citep{scikit-learn}, \texttt{batman} \citep{Kreidberg2015}.

\begin{acknowledgements}
This work is based in part on observations made with the NASA/ESA/CSA James Webb Space Telescope. The data were obtained from the Mikulski Archive for Space Telescopes at the Space Telescope Science Institute, which is operated by the Association of Universities for Research in Astronomy, Inc., under NASA contract NAS 5-03127 for JWST. These observations are associated with program \#2512. Support for program \#2512 was provided by NASA through a grant from the Space Telescope Science Institute, which is operated by the Association of Universities for Research in Astronomy, Inc., under NASA contract NAS 5-03127.

Some of the data presented in this article were obtained from the Mikulski Archive for Space Telescopes (MAST) at the Space Telescope Science Institute. The specific observations analyzed can be accessed via \dataset[doi:10.17909/xzh5-zf02]{[https://doi.org/10.17909/xzh5-zf02](https://doi.org/10.17909/xzh5-zf02)}.

This research has made use of the NASA Exoplanet Archive, which is operated by the California Institute of Technology, under contract with the National Aeronautics and Space Administration under the Exoplanet Exploration Program.

T.G. gratefully acknowledges support from the Heising-Simons Foundation through grant No. 2021-3197. This work benefited from the 2022-2025 Exoplanet Summer Program in the Other Worlds Laboratory (OWL) at the University of California, Santa Cruz, a program funded by the Heising-Simons Foundation. This material is based upon work supported by NASA's Interdisciplinary Consortia for Astrobiology Research (NNH19ZDA001N-ICAR) under award number 19-ICAR19\_2-0041. Support for this work was provided by NASA through grant 80NSSC19K0290 to JT and NLW. We thank Lili Alderson for feedback on an initial draft of this work.
\end{acknowledgements}

\section*{Author Contributions}
Co-Author contributions are as follows: TAG carried out the data analysis and led the writeup. NMB advised TAG throughout the work. NEB provided independent lower limits for the atmospheric metallicity and contributed text. All co-authors provided feedback on paper drafts and/or gave feedback during team meetings.

\bibliography{main}

\end{document}